\documentclass[a4paper,11pt]{article}
  \pdfoutput=1 

\usepackage{jheppub} 

\usepackage[T1]{fontenc}

\usepackage[utf8]{inputenc}
\DeclareUnicodeCharacter{27E8}{\ensuremath{\langle}}
\DeclareUnicodeCharacter{27E9}{\ensuremath{\rangle}}
\usepackage{amsmath, amsthm, mathtools}
\usepackage{hyperref}
\usepackage{multicol,multirow}
\usepackage[sort&compress,numbers]{natbib}
\usepackage{subfigure}

\newcommand\fverb{\setbox\fverbbox=\hbox\bgroup\verb}
\newcommand\fverbdo{\egroup\medskip\noindent \fbox{\unhbox\fverbbox}\ }
\newcommand\fverbit{\egroup\item[\fbox{\unhbox\fverbbox}]}
\newbox\fverbbox

\def\half{\frac{1}{2}}
\def\as{\alpha_s}
\def\ep{\epsilon}
\def\bI{\boldsymbol{I}}
\def\bJ{\boldsymbol{J}}
\def\bZ{\boldsymbol{Z}}
\def\tauzero{{\cal T}_0}
\def\tauzerocut{{\cal T}_0^{cut}}
\def\taun{{\cal T}_N}
\def\tauncut{{\cal T}_N^{cut}}
     \def\LB{\left[}\def\RB{\right]}
\def\beq{\begin{equation}}
\def\eeq{\end{equation}}

\def\tr{\mathop{\rm tr}\nolimits}
\def\Lsnew{\mathop{\rm \widetilde {Ls}}\nolimits}

\def\Ls{\mathrm{Ls}}
\def\Li{\mathrm{Li}}
\def\Ll{\mathrm{L}}
\def\I33m{\mathrm{I}_3^{3{\mathrm m}}}

\def\be{\begin{equation}}
\def\ee{\end{equation}}
\def\beqn{\begin{eqnarray}}
\def\eeqn{\end{eqnarray}}
\def\bea{\begin{eqnarray}}
\def\eea{\end{eqnarray}}
\def\treenum{{(0)}}

\def\cg{c_\Gamma}

\def\spa#1.#2{\left\langle#1#2\right\rangle}
\def\spb#1.#2{\left[#1#2\right]}
\def\lor#1.#2{\left(#1#2\right)}
\def\sand#1.#2.#3{\left\langle\smash{#1}{\vphantom1}^{-}\right|{#2}\left|\smash{#3}{\vphantom1}^{-}\right\rangle}
\def\sandp#1.#2.#3{\left\langle\smash{#1}{\vphantom1}^{-}\right|{#2}\left|\smash{#3}{\vphantom1}^{+}\right\rangle}
\def\sandpp#1.#2.#3{\left\langle\smash{#1}{\vphantom1}^{+}\right|{#2}\left|\smash{#3}{\vphantom1}^{+}\right\rangle}
\def\sandpm#1.#2.#3{\left\langle\smash{#1}{\vphantom1}^{+}\right|{#2}\left|\smash{#3}{\vphantom1}^{-}\right\rangle}
\def\sandmp#1.#2.#3{\left\langle\smash{#1}{\vphantom1}^{-}\right|{#2}\left|\smash{#3}{\vphantom1}^{+}\right\rangle}
\def\spab#1.#2.#3{\langle#1|#2|#3]}
\def\spba#1.#2.#3{[#1|#2|#3\rangle}
\def\spaaold#1.#2.#3{\langle#1|#2|#3\rangle}
\def\spbbold#1.#2.#3{[#1|#2|#3]}
\def\spaa#1.#2.#3.#4{\left\langle#1|#2|#3|#4\right\rangle}
\def\spbb#1.#2.#3.#4{\left[#1|#2|#3|#4\right]}
\def\spaxa#1.#2.#3.#4{\langle#1|#2|#3|#4\rangle}
\def\spbxb#1.#2.#3.#4{[#1|#2|#3|#4]}
  
\def\Etmiss{E_T^{\rm{miss}}}

\def\aa{\mathcal{A}}

\def\lc{\mathrm{lc}}
\def\slc{\mathrm{sl}}
\def\dk{{dk}}
\def\Wrad{{W}}
\def\lrad{{e}}

\def\tree{\mathrm{tree}}

\def\prp34{\mathcal{P}_{34}}
\def\prpp345{\mathcal{P}_{345}}

\def\LsminOne#1#2#3{\mathrm{Ls}_{-1}\left(\frac{#1}{#3},\frac{#2}{#3}\right)} \def\LsminOneTME#1#2#3#4{\mathrm{Ls}_{-1}^{2\mathrm{me}}\left(#1,#2;#3,#4\right)} \def\LsminOneTMH#1#2#3#4{\mathrm{Ls}_{-1}^{2\mathrm{mh}}\left(#1,#2;#3,#4\right)} \def\LsminOneTMHT#1#2#3#4{\widetilde{\mathrm{Ls}}_{-1}^{2\mathrm{mh}}\left(#1,#2;#3,#4\right)} \def\Lzero#1#2{\mathrm{L}_0\left(\frac{#1}{#2}\right)} \def\Lone#1#2{\mathrm{L}_1\left(\frac{#1}{#2}\right)} \def\Lnrat#1#2{\ln\left(\frac{#1}{#2}\right)}
\def\iTm#1#2#3{I_3^{3m}\left(#1,#2,#3\right)}
\def\Vpole{V}
\def\DeltaTM#1#2#3#4{\Delta_{#1#2,#3#4}}

\def\deltaof{\delta_{15}}
\def\deltats{\delta_{26}}

\usepackage{slashed}
\allowdisplaybreaks

\preprint{\begin{minipage}[t]{8cm}\begin{flushright}FERMILAB-PUB-21-215-T,\\ FR-PHENO-2020-020,\, IPPP/20/96\end{flushright}\end{minipage}}
\title{The $pp\to W(\to l\nu)+\gamma$ process at next-to-next-to-leading order}

\author[a]{John M. Campbell,}
\emailAdd{johnmc@fnal.gov}

\author[b]{Giuseppe De Laurentis,}
\emailAdd{giuseppe.de.laurentis@physik.uni-freiburg.de}

\author[c]{R. Keith Ellis,}
\emailAdd{keith.ellis@durham.ac.uk}

\author[d]{Satyajit Seth}
\emailAdd{seth@prl.res.in}

\affiliation[a]{Fermilab, PO Box 500, Batavia IL 60510-5011, USA}
\affiliation[b]{Fakultät für Mathematik und Physik, Physikalisches Institut, D-79104 Freiburg, Germany}
\affiliation[c]{Institute for Particle Physics Phenomenology, Durham University, Durham, DH1 3LE, UK}
\affiliation[d]{Physical Research Laboratory, Navrangpura, Ahmedabad - 380009, India}
\date{\today}

\abstract{ We present details of the calculation of the $pp\to W(\to l\nu) \gamma$
  process at next-to-next-to-leading order in QCD, calculated
  using the jettiness slicing method.  The calculation is based
  entirely on analytic amplitudes. Because of the radiation zero, the NLO
  QCD contribution from the $gq$ channel is as important as the
  contribution from the Born $q\bar{q}$ process, disrupting the normal
  counting of leading and sub-leading contributions.  We also assess
  the importance of electroweak (EW) corrections, including the EW
  corrections to both the six-parton channel $0\to \bar{u} d \nu e^+\gamma g$
  and the five-parton channel $0\to \bar{u} d \nu e^+ \gamma$.
  Previous experimental results have been shown to agree
  with theoretical predictions, taking into account the large
  experimental errors.  With the advent of run II data from the LHC,
  the statistical errors on the data will decrease, and will be
  competitive with the error on theoretical predictions for the first
  time.  We present numerical results for $\sqrt{s}=7$ and $13$~TeV.
  Analytic results for the one-loop six-parton QCD amplitude and the
  tree-level seven-parton QCD amplitude are presented in appendices.  }


\begin{document} 
\setcounter{tocdepth}{2}
\maketitle

\section{Introduction}
The process $pp \to W(\to l\nu)+\gamma$ occupies a special place amongst the high-energy processes
sensitive to triple coupling of three vector bosons. Of all the vector-boson pair-production processes
which are sensitive to the triple gauge boson coupling, it has the largest cross section.
The discovery, some forty years ago, of the radiation zero~\cite{Mikaelian:1979nr,Brown:1979ux} in the leading-order
prediction for this process indicates a high amount of destructive interference between the contributing 
sub-amplitudes\footnote{The simplest explanation for the radiation zero in the $W\gamma$ process has been 
given in Ref.~\cite{Goebel:1980es}, exploiting an early precursor of BCJ relations~\cite{Bern:2008qj}.
For a recent discussion of the role of radiation zeroes and connections to the BCJ relations se Ref.~\cite{Brown:2016mrh}.}. 
This characteristic interference pattern, even though attenuated by higher-order corrections,
still implies that this process is a particularly sensitive test of the gauge structure, 
allowing incisive probes of the three-boson coupling.

On the theoretical side, early calculations of the QCD radiative corrections~\cite{Smith:1989xz,Ohnemus:1992jn,Baur:1993ir}
showed that the radiation zero, which manifests itself in $pp$ collisions as a dip in the centre-of-mass 
rapidity of the photon at $y_\gamma^*=0$, persists in the NLO theory, but with diminished importance.
With the advent of spinor techniques, compact analytic expressions became available for the one-loop $u \bar{d} \to W \gamma$ amplitudes~\cite{Dixon:1998py}.
For a review of the status at the dawn of the LHC era, see \cite{Campbell:2011bn} where results for NLO processes implemented in MCFM
are reported. More recently NLO $W\gamma$ production in hadronic collisions has been interfaced to a shower generator according to the POWHEG prescription 
in such a way that the contribution arising from hadron fragmentation into photons is fully modelled~\cite{Barze:2014zba}.

The large correction in passing from leading order to next-to-leading order,
caused by the radiation zero, hints at the potential importance of a NNLO calculation.
A NNLO calculation has been achieved in refs.~\cite{Grazzini:2016hai,Grazzini:2015nwa} 
using a $q_T$ slicing method. Subsequently this process has been treated in a unified framework by the 
MATRIX program~\cite{Grazzini:2017mhc}, where the matrix elements are calculated using the 
OpenLoops procedure~\cite{Cascioli:2011va}. A further development is the calculation of 
electroweak effects~\cite{Accomando:2001fn} and their combination with both 
NLO QCD calculations~\cite{Denner:2014bna} and NNLO QCD calculations~\cite{Kallweit:2019zez}.
The results presented in this paper are similar in spirit, but different in detail from the results
of refs.~\cite{Grazzini:2017mhc,Kallweit:2019zez}:
\begin{enumerate}
\item Instead of the $q_T$ slicing method, we use the $N$-jettiness
slicing method, which has been successfully implemented in MCFM for the following processes, 
$pp\to H+X$, $pp\to W^\pm$, $pp\to Z$, $pp\to WH$, $pp\to ZH$, $pp\to \gamma\gamma$~\cite{Boughezal:2016wmq},
$pp\to Z\gamma$~\cite{Campbell:2017aul},
$pp \to Z+$~jet~\cite{Boughezal:2015ded}, $pp \to H+$~jet~\cite{Campbell:2019gmd}.
\item
All amplitudes and hence matrix elements entering our NNLO QCD calculation are implemented using
analytic formulae. We believe this will have benefits for both the
stability and speed of the code.  Thus for example our NNLO
calculation needs the one-loop contribution to the six-parton process.
The first complete one-loop analytic result for this process is
presented in an appendix to this paper.  Part of this result is
derivable from Ref.~\cite{Bern:1997sc}.  As usual these one-loop
processes are calculated using analytic unitarity
methods~\cite{Britto:2004nc,Forde:2007mi,Mastrolia:2009dr}.  They
are further manipulated and simplified by means of high-precision
floating-point reconstruction~\cite{DeLaurentis:2019phz}.
\item
Our code can accommodate a non-diagonal 
CKM matrix. While this is known to give quite small modifications,
its inclusion eliminates an entirely avoidable theoretical error.
\item
We have studied the impact of electroweak corrections, including for the
first time the $O(\alpha_s)$ gluon-quark initiated process, which is numerically as important
as the quark-antiquark initiated process.
\end{enumerate}

On the experimental side, the $W\gamma$ process was first observed at
the Tevatron~\cite{Abe:1994fx,D0:1995mca} with a handful of events and
larger-statistics studies were later performed both by
CDF~\cite{Acosta:2004it} and D0
\cite{Abazov:2005ni,Abazov:2011rk}. However, even with the largest
statistics available at the Tevatron, the experimental errors were
larger than the theoretical errors presented in these
papers. Table~\ref{Experiments} presents a compilation of results from
both the Tevatron and the LHC at various energies and with various
accumulated integrated luminosities. Also indicated are the
predictions for the theoretical cross sections, presented in the 
papers cited by the experimental collaborations.  For the most part the
experimental results are fiducial cross sections for the process
$p\bar{p} \to \ell^\pm \nu \gamma$ or $pp \to \ell^\pm \nu \gamma$
with differing cuts, and as such they are not directly comparable,
even at the same energy.  The exception is Ref.~\cite{D0:1995mca}
which gives the $W\gamma$ cross section.  The experimental result~\cite{Sirunyan:2021zud} in
the penultimate row of Table~\ref{Experiments} reports the sum of the
four cross sections, ($e^-\nu\gamma$, $e^+\bar{\nu}\gamma$,
$\mu^-\nu\gamma$, $\mu^+\bar{\nu}\gamma$) where these event categories
include feed-down from $\tau$-decays.  In some cases the value of the cross
section is reported in an extended fiducial region beyond the actual
region of measurement by performing an acceptance correction.  The
experiments impose a separation between the lepton and the photon, $R_{l \gamma}>0.7$,
except for Ref.~\cite{Sirunyan:2021zud} which has $R_{l \gamma}>0.5$.  For the experimental measurements the errors
are statistical, followed by systematic error and in some cases the
luminosity error.  The column labelled theoretical cross section also
indicates the provenance of the theory prediction.

In all but the most recent measurement~\cite{Sirunyan:2021zud} the experimental
errors are bigger than the theoretical errors. However
in Ref.~\cite{Sirunyan:2021zud} where the errors in experiment and in theory
are commensurate, two results are presented for the theoretical
prediction which are only marginally consistent with one another, as
shown in Table~\ref{Experiments}.
\begin{table}
\begin{center}
\begin{tabular}{|l|l|l|l|l|l|}
\hline
Experiment                    & $\int L dt$ & $\sqrt{s}$ & $E_{Tmin}^\gamma$ & Experimental                       & Theoretical\\
                              & fb$^{-1}$   & [TeV]      & [GeV]             & cross section [pb]                 & cross section [pb]\\
\hline                                                                
CDF~\cite{Abe:1994fx}         & 0.020       & 1.8        & 7                 & $13.2 \pm 4.2 \pm 1.3$             & $18.6 \pm 2.8$~\cite{Baur:1989gk}\\     
D0~\cite{D0:1995mca}          & 0.0138      & 1.8        & 10                & $138^{+51}_{-38} \pm 21$           & $112\pm 10$~\cite{Baur:1989gh} \\     
CDF~\cite{Acosta:2004it}      & 0.200       & 1.96       & 7                 & $18.1 \pm 3.1$                     & $19.3 \pm 1.4$~\cite{Baur:1989gk,Baur:1992cd} \\     
D0~\cite{Abazov:2005ni}       & 0.162       & 1.96       & 8                 & $14.8 \pm 1.6\pm1.0 \pm 1.0$       & $16.0 \pm 0.4$~\cite{Baur:1989gk} \\
D0~\cite{Abazov:2011rk}       & 4.2         & 1.96       & 15                & $7.6\pm 0.4 \pm 0.6 $              & $7.6 \pm 0.2$~\cite{Baur:1989gk} \\     
\hline                                                                
CMS~\cite{Chatrchyan:2011rr}  & 0.036       & 7          & 10                & $56.3 \pm 5.0  \pm 5.0  \pm 2.3$   & $49.4 \pm 3.8$~\cite{Baur:1993ir} \\     
ATLAS~\cite{Aad:2011tc}       &  0.035      & 7          & 15                & $36.0\pm 3.6\pm 6.2 \pm 1.2$       & $36.0 \pm  2.3$~\cite{Baur:1993ir} \\     
ATLAS~\cite{Aad:2012mr}       & 1.02        & 7          & 15                & $4.60  \pm 0.11  \pm 0.64$         & $3.70\pm 0.28 $~\cite{Campbell:2010ff}\\     
CMS~\cite{Chatrchyan:2013fya} & 5           & 7          & 15                & $37.0 \pm 0.8 \pm 4.0 \pm 0.8$     & $31.8 \pm 1.8$~\cite{Campbell:2011bn,Campbell:1999ah}\\     
CMS~\cite{Sirunyan:2021zud}   & 137.1       & 13         & 25                & $15.58  \pm 0.05 \pm 0.73 \pm 0.15$& $15.4 \pm 1.2\pm 0.1$~\cite{Alwall:2014hca,Frederix:2012ps}\\     
                              &             &            &                   &                                    & $22.4 \pm 3.2\pm 0.1$~\cite{Barze:2014zba}\\
\hline
\end{tabular}
\caption{Experiments at various energies with $p \bar{p}$ and $pp$ on the $l^\pm\nu\gamma$ process.}
\label{Experiments}
\end{center}
\end{table}
It therefore seems opportune to us to re-examine the theoretical status of the $W\gamma$ process.
This is particularly important now that the statistical precision of the Run 2 data at $\sqrt{s}=13$~TeV 
approaches the precision of theoretical calculations. We encourage the LHC collaborations to 
perform the necessary analyses with the full run 2 data set.

In section 2 we outline the structure of our calculation of the $W(\to l \nu)+\gamma$-process.
In section 3 we provide a brief review of the $N$-jettiness method for calculating 
NNLO cross sections.
In section 4 we describe the setup which we will use for numerical results and take a first look at results
at $\sqrt{s}=13$~TeV. Section 5 describes how we have incorporated
the electroweak corrections. Section 6 presents our detailed numerical results for $\sqrt{s}=7$ and $13$~TeV.
Definitions of spinor products and analytic results for the 6- and 7-parton QCD processes are provided in the appendices. 

\section{Ingredients of the calculation}
In order to perform a calculation of the $pp\to W(\to l \nu) +\gamma$ process at NNLO
in QCD the following amplitudes are necessary ($g=\sqrt{4\pi\alpha_s}$ is the QCD coupling constant),
\begin{itemize}
\item 5-parton, $0 \rightarrow \bar u (p_1) + d(p_2) + \nu_e (p_3) + e^+ (p_4)  + \gamma (p_5)~\text{calculated at orders}~1,g^2,g^4$
\item 6-parton, $0 \rightarrow \bar u (p_1) + d(p_2) + \nu_e (p_3) + e^+ (p_4)  + \gamma (p_5)+g(p_6) ~\text{calculated at orders}~g,g^3$
\item 7-parton, $0 \rightarrow \bar u (p_1) + d(p_2) + \nu_e (p_3) + e^+ (p_4)  + \gamma (p_5)+g(p_6)+g(p_7) ~\text{calculated at order}~g^2$
\item 7-parton, $0 \rightarrow \bar u (p_1) + d(p_2) + \nu_e (p_3) + e^+ (p_4)  + \gamma (p_5)+q(p_6)+\bar{q}(p_7) ~\text{calculated at order}~g^2$
\end{itemize}
In the last 7-parton process in the above list, $q$ and $\bar{q}$ represent any of the five quarks $(d,u,s,c,b)$.
In addition, in all processes $\bar u$ can be replaced by $\bar c$, and $d$ by $s$ or $b$.

In this section we report on tree graph amplitudes. These are simple to calculate using spinor techniques 
and are included here to establish our notation.
\subsection{5-parton amplitude}
\subsubsection{Structure of the 5-parton amplitude}
Our calculation includes contributions where the photon is radiated off the positron, so it is really a
misnomer to call it $W+\gamma$ production.
However, it is a convenient way to refer to the Born-level 5-parton process which we shall employ throughout this paper.
The amplitude for the five-parton process is,
\begin{eqnarray}
\aa^\treenum ( 1^+_{\bar{u}},2^-_{d},3^-_{\nu},4^+_{\lrad},5^{h_5}_{\gamma}) 
&=&   i \, \sqrt{2} e g_W^2 \, \cdot A^\treenum  (5^{h_5}_{\gamma})\, .
\end{eqnarray}
In the naming of the amplitudes, the arguments $ 1^+_{\bar{u}},2^-_{d},3^-_{\nu},4^+_{\lrad}$ will be omitted,
except where they are needed (mainly for crossing relations).
\begin{figure}
\includegraphics[angle=270,width=\textwidth]{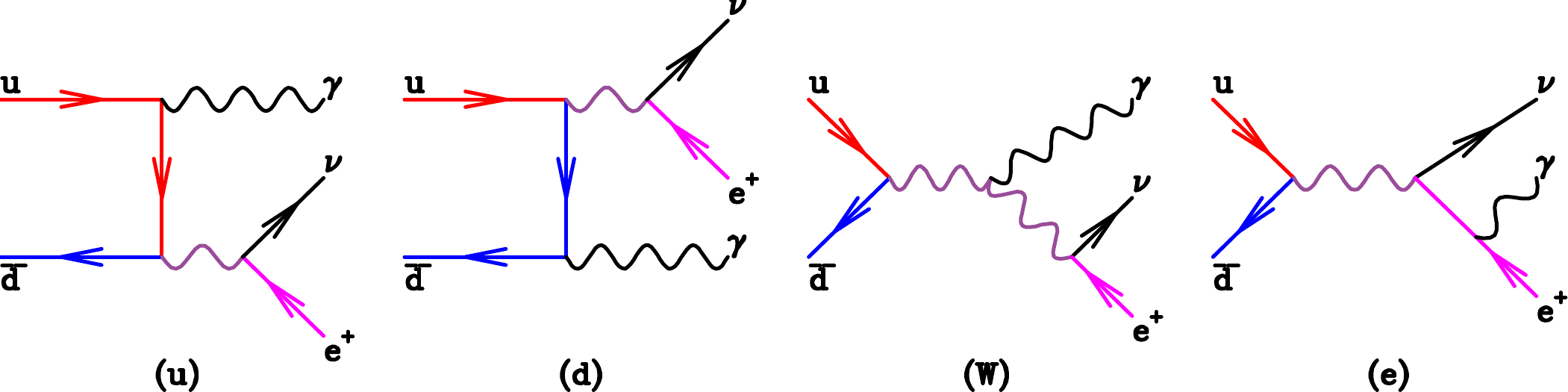}
\caption{Topologies of diagrams included at lowest order, shown for the specific case of 
$u \bar{d} \to \gamma \nu_e e^+$. Note that diagrams (u) and (d) are proportional to $Q_u$ and $Q_d$ respectively,
whereas diagrams, (\Wrad) and (e) are proportional to  $Q_u-Q_d$.}
\label{LowestOrderGraphs}
\end{figure}
The lowest-order graphs are shown in Fig.~\ref{LowestOrderGraphs}.
We separate the sub-amplitudes into contributions that are
sensitive to individual electric charges,
\begin{eqnarray}
A^{(0)}(5_{\gamma}^{h_5}) &=& 
  \left( Q_u \, A^u_\tree(5_{\gamma}^{h_5})
       + Q_d \, A^d_\tree (5_{\gamma}^{h_5}) \right) P(s_{34}) \\
&&  + (Q_u-Q_d) \, \left( A^\lrad_\tree(5_{\gamma}^{h_5}  ) 
                        + A^\Wrad_\tree(5_{\gamma}^{h_5}) P(s_{34}) \right)  P(s_{345})\, , \nonumber
\end{eqnarray}
where we have also pulled out factors of the $W$-boson propagator defined by,
\begin{equation}
P(s)=\frac{1}{s-m_W^2}=\frac{1}{s-M_W^2 +i M_W \Gamma_W},
\end{equation}
and $m_W^2=M_W^2-i M_W \Gamma_W$ indicates that we are working in the complex-mass scheme.
These sub-amplitudes are clearly not individually gauge invariant in the electroweak sector.

\subsubsection{Tree-level amplitudes}
The components of the tree-level Born amplitude are,
\begin{eqnarray}
A^u_\tree(5_{\gamma}^{-})&=&0,\;\;\;A^d_\tree(5_{\gamma}^{-})=\frac{\spb1.4^2 \spa4.3}{\spb1.5 \spb2.5},\;\;\;
A^\Wrad_\tree(5_{\gamma}^{-})=\frac{\spb1.4^2 \spa4.3 \spa2.5 }{\spb1.5},\nonumber \\
A^\lrad_\tree(5_{\gamma}^{-})&=&\frac{\spb1.4^2 \spa3.2 }{\spb1.5 \spb4.5} , \\[2mm]
A^u_\tree(5_{\gamma}^{+})&=&\frac{\spa2.3^2 \spb3.4}{\spa1.5 \spa2.5},\;\;\; A^d_\tree(5_{\gamma}^{+})=0,\;\;\;
A^\Wrad_\tree(5_{\gamma}^{+})= \frac{\spa2.3^2 \spb4.3 \spb1.5}{\spa2.5},\nonumber \\
A^\lrad_\tree(5_{\gamma}^{+})&=&\frac{\spa2.3^2 \spb3.1}{\spa2.5 \spa4.5}.
\end{eqnarray}

For all of the sub-amplitudes except for the one in which the photon is radiated from the positron in the $W$-boson decay, 
$A^\lrad$, there is a simple rule for flipping the helicities of the photon,
\begin{eqnarray}
  A^u_\tree  ( 1_{\bar u}^+,2_{d}^-,3_{\nu}^-, 4_{\lrad}^+, 5_{\gamma}^{-h_5})
&=& A^d_\tree  ( 2_{\bar u}^+,1_{d}^-,4_{\nu}^-, 3_{\lrad}^+, 5_{\gamma}^{h_5})
 \, \bigl\{ \langle \rangle \leftrightarrow [] \bigr\} \, ,\\
  A^\Wrad_\tree  ( 1_{\bar u}^+,2_{d}^-,3_{\nu}^-, 4_{\lrad}^+, 5_{\gamma}^{-h_5})
&=& - A^\Wrad_\tree  ( 2_{\bar u}^+,1_{d}^-,4_{\nu}^-, 3_{\lrad}^+, 5_{\gamma}^{h_5})
 \, \bigl\{ \langle \rangle \leftrightarrow [] \bigr\} \, .
\end{eqnarray}

Complete amplitudes for the charge-conjugate process,
\begin{equation}
0 \rightarrow u (p_1) + \bar{d}(p_2) + e^- (p_3) + \bar \nu_e (p_4)  + \gamma (p_5)
\end{equation}
are given by a transformation on the entire amplitude similar to the one given above for flipping helicities,
\begin{equation}
\aa^\treenum ( 1^+_{u},2^-_{\bar d},3^-_{\lrad},4^+_{\nu},5^{-h_5}_{\gamma})
 = \aa^\treenum ( 2^+_{\bar{u}},1^-_{d},4^-_{\nu},3^+_{\lrad},5^{h_5}_{\gamma})
 \, \bigl\{ \langle \rangle \leftrightarrow [] \bigr\} \, .
\end{equation}

\subsubsection{One-loop amplitude}
Five-parton results at one loop are taken from Ref.~\cite{Dixon:1998py} and supplemented with the contributions from radiation in 
decay.  They could equivalently be taken from Ref.~\cite{Gehrmann:2011ab} and we have checked explicitly that the two forms
are identical numerically.

\subsubsection{Two-loop amplitude}
Genuine two-loop contributions for Drell-Yan type processes were pioneered in Ref.~\cite{Matsuura:1988sm}.
The $W\gamma$ results that we use in our calculation are taken from Ref.~\cite{Gehrmann:2011ab}, where the two-loop amplitude
after UV renormalization at $\mu^2=s_{34}$, is presented as the finite term remaining after
extraction of the predicted IR singularity structure at two loops~\cite{Catani:1998bh}:
\beqn
\Omega
&=&\Omega^{(0)} + \left(\frac{\as(\mu^2)}{2 \pi}\right)\Omega^{(1)} + \left(\frac{\as(\mu^2)}{2 \pi}\right)^2 \Omega^{(2)}+O(\as^3) \nonumber \\
&=&\Omega^{(0)} +\left(\frac{\as(\mu^2)}{2 \pi}\right)(\Omega^{(1),f} +\Omega^{(0)} \bI_1(\ep))\nonumber \\
       &&+\left(\frac{\as(\mu^2)}{2 \pi}\right)^2 (\Omega^{(2),f}+\Omega^{(1),f} \bI_1(\ep)+\Omega^{(0)} (\bI_1(\ep)^2+\bI_2(\ep))) \, .
\eeqn
The finite parts $\Omega^{(1),f},\Omega^{(2),f}$ differ by finite terms from the hard function, $H$.
The construction of the hard function from the results for the one- and two-loop
amplitude is spelled out in refs.~\cite{Becher:2013vva,Campbell:2016yrh},
\begin{eqnarray}
H &=& H^{(0)} +\frac{\as(\mu^2)}{2 \pi}H^{(1)}+\left(\frac{\as(\mu^2)}{2 \pi}\right)^2 H^{(2)} \, ,\\
H^{(0)}&=& \Omega^{(0)}\nonumber \, , \\
H^{(1)}&=& \Omega^{(1),f}+ \bJ^{(1)} \Omega^{(0)} \, , \nonumber \\
H^{(2)}&=& \Omega^{(2),f}+\bJ^{(1)} \Omega^{(1),\text{f}} +\bJ^{(2)}\Omega^{(0)} \, ,
\end{eqnarray}
where the $\bJ$ are finite functions formed by combining the perturbative expansion of the Catani singularity structure
and the perturbative expansion of the inverse of the $\bZ$ matrix, 
\beq
\bZ^{-1}=1+\bZ^{(1)}(\ep) \frac{\as(\mu^2)}{2 \pi} + \bZ^{(2)}(\ep) \left(\frac{\as(\mu^2)}{2 \pi}\right)^2+\ldots 
\eeq
Explicitly,  
\beq
\bJ^{(1)}=\bI^{(1)}(\ep) +\bZ^{(1)}(\ep),\;\;\; \bJ^{(2)}=\bI^{(2)}(\ep) + \left(\bI^{(1)}(\ep) +\bZ^{(1)}(\ep)\right) \bI^{(1)}(\ep) +\bZ^{(2)}(\ep) \, ,
\eeq
resulting in,
\begin{eqnarray}
\bJ^{(1)}&=&-\frac{C_F}{2} (L^2  +3 L - \zeta_2) \, , \\
\bJ^{(2)}&=&\frac{C_F^2}{32}[4 L^4+ 24 L^3+ 36 L^2-8 \zeta_2 L^2- 24 \zeta_2 L+ 10\zeta_4] \nonumber  \\
&&+ \frac{C_F T_R n_f}{18}[2 L^3+19 L^2+30 L+3 \zeta_2 L +2\zeta_2+3 \zeta_3] \\
&&- \frac{C_F C_A}{144}[44 L^3+466 L^2-72 \zeta_2 L^2 +804 L-150 \zeta_2 L +32 \zeta_2+66 \zeta_3+45 \zeta_4] , \nonumber 
\end{eqnarray}
where $n_f$ is the number of active flavours and
\beq
C_F=\frac{4}{3},\;\;N=3,\;\;T_R=\frac{1}{2},\;\;\zeta_2=\frac{\pi^2}{6},\;\;\zeta_3\approx1.20206,\;\;\zeta_4=\frac{\pi^4}{90},\;\; L=\ln\frac{\mu^2}{-s_{12}-i0}\, .
\eeq
\subsection{6-parton amplitude}

\subsubsection{Structure of the 6-parton amplitude}
We calculate the one-loop helicity amplitudes for,
\begin{equation}
0 \rightarrow \bar u (p_1) + d(p_2) + \nu_e (p_3) + e^+ (p_4)  + \gamma (p_5) + g(p_6)\, .
\end{equation}
Our calculation includes contributions where the photon is radiated of the positron. The reduced amplitude, $A^\treenum$, is defined as,
\begin{eqnarray}
\aa^\treenum ( 1^+_{\bar{u}},2^-_{d},3^-_{\nu},4^+_{\lrad},5^{h_5}_{\gamma},6^{h_6}_g ) 
&=&   i \sqrt{2} \, e g_W^2 g_s \, t^{a_6}_{i_2 i_1} \cdot A^\treenum  (5^{h_5}_{\gamma},6^{h_6}_g ) \, ,
\end{eqnarray}
with $ \tr (t^a t^b)=\delta^{ab}$.
Furthermore, it is useful to further separate the sub-amplitudes into contributions that are
sensitive to individual electric charges,
\begin{eqnarray}
A^{(0)}(5_{\gamma}^{h_5} ,6_g^{h_6} ) &=& 
  \left( Q_u \, A^u_\tree(5_{\gamma}^{h_5} ,6_g^{h_6} )
       + Q_d \, A^d_\tree (5_{\gamma}^{h_5} ,6_g^{h_6} ) \right) P(s_{34}) \\
&&  + (Q_u-Q_d) \, \left( A^\lrad_\tree(5_{\gamma}^{h_5} ,6_g^{h_6} ) 
                        + A^\Wrad_\tree(5_{\gamma}^{h_5} ,6_g^{h_6} ) P(s_{34}) \right)  P(s_{345})\, . \nonumber
\end{eqnarray}

\subsubsection{Tree-level amplitudes}

The components of the tree-level amplitude for a photon and a gluon both of positive helicity are,
\begin{eqnarray}
A^u_\tree(5_\gamma^+,6_g^+)&=&
  \frac{\spa1.2 \spa2.3^2 \spb4.3}{\spa1.5 \spa1.6 \spa2.5 \spa2.6} \, , \label{eq:LOpp-Qu} \\
A^d_\tree(5_\gamma^+,6_g^+)&=& 0 \, , \label{eq:LOpp-Qd} \\
A^\lrad_\tree(5_\gamma^+,6_g^+)&=&
  \frac{\spa2.3^2 \spab2.(4+5).3}{\spa1.6 \spa2.5 \spa2.6 \spa4.5} \, , \label{eq:LOpp-Wnorad} \\
A^\Wrad_\tree(5_\gamma^+,6_g^+)&=&
  \frac{\spa2.3^2 \spb4.3 \spab2.(1+6).5}{\spa1.6 \spa2.5 \spa2.6} \, . \label{eq:LOpp-Wrad}
\end{eqnarray}

For the case where the photon has negative helicity and the gluon positive helicity we have,
\begin{eqnarray}
A^u_\tree(5_\gamma^-,6_g^+)&=& - \frac{\spa2.3 \spb1.6 \spab5.(2+3).4)}{\spb1.5 \spa1.6 s_{234}} \, , \label{eq:LOmp-Qu} \\
A^d_\tree(5_\gamma^-,6_g^+)&=&
     - \Big[\frac{\spab2.(1+6).4 \spab3.(2+5).1}{\spb1.5 \spa1.6 \spb2.5 \spa2.6}
     +\frac{\spb1.4 \spa2.5\spab3.(1+4).6}{\spb2.5 \spa2.6 s_{134}} \Big] \, , \label{eq:LOmp-Qd} \\
A^\lrad_\tree(5_\gamma^-,6_g^+)&=&
     -  \frac{ \spb1.4 \spab2.(1+6).4 \spa2.3}{\spa1.6 \spa2.6 \spb1.5 \spb4.5} \, , \label{eq:LOmp-Wnorad} \\
A^\Wrad_\tree(5_\gamma^-,6_g^+)&=&
     - \frac{\spab2.(1+6).4 (\spa2.5\spa3.4\spb1.4+\spa2.6\spa3.5\spb1.6)}{\spa1.6 \spa2.6 \spb1.5 } \, . \label{eq:LOmp-Wrad}
\end{eqnarray}

For all of the sub-amplitudes except for the one in which the photon is radiated from the positron in the $W$-boson decay, 
$A^\lrad$, there is a simple rule for flipping the helicities of the photon and gluon,
\begin{eqnarray}
  A^u_\tree  ( 1_{\bar u}^+,2_{d}^-,3_{\nu}^-, 4_{\lrad}^+, 5_{\gamma}^{-h_5}, 6_g^{-h_6} )
&=& A^d_\tree  ( 2_{\bar u}^+,1_{d}^-,4_{\nu}^-, 3_{\lrad}^+, 5_{\gamma}^{h_5}, 6_g^{h_6} )
 \, \bigl\{ \langle \rangle \leftrightarrow [] \bigr\} \, , \\
  A^\Wrad_\tree  ( 1_{\bar u}^+,2_{d}^-,3_{\nu}^-, 4_{\lrad}^+, 5_{\gamma}^{-h_5}, 6_g^{-h_6} )
&=& - A^\Wrad_\tree  ( 2_{\bar u}^+,1_{d}^-,4_{\nu}^-, 3_{\lrad}^+, 5_{\gamma}^{h_5}, 6_g^{h_6} )
 \, \bigl\{ \langle \rangle \leftrightarrow [] \bigr\} \, .
\end{eqnarray}
For the remaining amplitudes we have,
\begin{eqnarray}
A^\lrad_\tree(5_\gamma^-,6_g^-)&=&
    \frac{\spb1.4^2 \spab3.4+5.1}{\spb2.6 \spb1.5 \spb1.6 \spb4.5} \label{eq:LOmm-Wnorad} \, , \end{eqnarray}
and,
\begin{eqnarray}
A^\lrad_\tree(5_\gamma^+,6_g^-)&=&
      - \frac{\spab2.(4+5).1 \spab3.(2+6).1}{\spb2.6 \spb1.6 \spa2.5 \spa4.5} \label{eq:LOpm-Wnorad} \, . \end{eqnarray}

Complete amplitudes for the charge-conjugate process,
\begin{equation}
0 \rightarrow u (p_1) + \bar{d}(p_2) + e^- (p_3) + \bar \nu_e (p_4)  + \gamma (p_5) + g(p_6)
\end{equation}
are given by a transformation on the entire amplitude similar to the one given above for flipping helicities,
\begin{equation}
\aa^\treenum ( 1^+_{u},2^-_{\bar d},3^-_{\lrad},4^+_{\nu},5^{-h_5}_{\gamma},6^{-h_6}_g )
 = \aa^\treenum ( 2^+_{\bar{u}},1^-_{d},4^-_{\nu},3^+_{\lrad},5^{h_5}_{\gamma},6^{h_6}_g )
 \, \bigl\{ \langle \rangle \leftrightarrow [] \bigr\} \, .
\end{equation} 
The $W$-boson component of this amplitude can be isolated by partial fractioning
\begin{equation}\label{eq:Wpropagatorspartialfraction}
P(s_{34}) P(s_{345})=\frac{1}{s_{345}-s_{34}}\, \Big[ P(s_{34})-P(s_{345})\Big]
\end{equation}
and dropping all terms not containing $P(s_{34})$. The result in this limit is given in Ref.~\cite{Dixon:1998py}.

\subsubsection{One-loop amplitude}

We extract a similar factor at the one-loop level,
\begin{equation}
 \aa^{(1)}(5^{h_5} _{\gamma},6_g^{h_6} ) = 
 \quad i \sqrt{2} \, e g_W^2 g_s^3 \, \cg \; t^{a_6}_{i_2 i_1}   \bigg\lbrace  N_c \, A^{(1)}_\lc  (5^{h_5}_{\gamma},6^{h_6}_g )
   + \frac{1}{N_c} \, A^{(1)}_\slc (5_{\gamma}^{h_5} ,6_g^{h_6} ) \bigg\rbrace \, ,
\end{equation}
where this time, in addition, we have performed a decomposition into
leading- (`$\lc$') and subleading-colour (`$\slc$') components.
We work in dimensional regularization with $d= 4 - 2 \ep$, resulting in the
overall factor $\cg$ given by,
\beq
\cg =\frac{1}{(4\pi)^{2-\ep}} \frac{\Gamma(1+\ep)\Gamma^2(1-\ep)}{\Gamma(1-2\ep)}=\frac{1}{(4\pi)^{2-\ep}} \frac{1}{\Gamma(1-\ep)}+O(\ep^3)\, .
\label{cgammadef}
\eeq

Furthermore, it is useful to further separate the sub-amplitudes into contributions that are
sensitive to individual electric charges,
\begin{eqnarray}
A^{(1)}_\lc(5_{\gamma}^{h_5} ,6_g^{h_6} ) &=& 
  \left( Q_u \, A^u_\lc(5_{\gamma}^{h_5} ,6_g^{h_6} )
       + Q_d \, A^d_\lc (5_{\gamma}^{h_5} ,6_g^{h_6} ) \right) P(s_{34}) \\
&&  + (Q_u-Q_d) \, \left( A^\lrad_\lc(5_{\gamma}^{h_5} ,6_g^{h_6} ) 
                        + A^\Wrad_\lc(5_{\gamma}^{h_5} ,6_g^{h_6} ) P(s_{34}) \right)  P(s_{345})\, , \nonumber
\end{eqnarray}
(and similarly for $A^{(1)}_\slc$).
Note that the individual terms in this separation are not invariant under electroweak or QCD gauge transformations. 
Full analytic results for $A_\lc$ and $A_\slc$ are given in Appendix~\ref{Oneloopappendix}. 
We have checked numerically that our results agree perfectly with those provided by OpenLoops~\cite{Cascioli:2011va}, but
are faster by at least a factor of four.

\subsection{7-parton amplitude}
\subsubsection{Two-quark two-gluon processes}
We can decompose the amplitude for the two-gluon process 
$0 \rightarrow \bar u (p_1) + d(p_2) + \nu_e (p_3) + e^+ (p_4)  + \gamma (p_5)+g(p_6)+g(p_7)$
into two colour-ordered sub-amplitudes,
\begin{eqnarray}
&&\aa^\treenum ( 1^+_{\bar{u}},2^-_{d},3^-_{\nu},4^+_{\lrad},5^{h_5}_{\gamma},6^{h_6}_g,7^{h_7}_g) 
=   i\,\sqrt{2} \, e g_W^2 g_s^2  \nonumber \\
&\times& \, \left[
 \left( t^{a_6} t^{a_7}\right)_{i_2 i_1} \cdot A^{67}  ( 5^{h_5}_{\gamma},6^{h_6}_g,7^{h_7}_g) 
+\left( t^{a_7} t^{a_6}\right)_{i_2 i_1} \cdot A^{76}  ( 5^{h_5}_{\gamma},6^{h_6}_g,7^{h_7}_g) \right]\, .
\end{eqnarray}
Full analytic results for the colour-ordered trees are presented in subsection~(\ref{Sevenpointtrees:gluons}).
Squaring the amplitude and summing over the colours of the gluons, we obtain,
\begin{eqnarray}
&&\sum_{a_6,a_7} \sum_{h_5,h_6,h_7} \big|\aa^\treenum ( 1^+_{\bar{u}},2^-_{d},3^-_{\nu},4^+_{\lrad},5^{h_5}_{\gamma},6^{h_6}_g,7^{h_7}_g)\big|^2 = 2 e^2 g_W^4 g_s^4 (N^2-1) \nonumber \\
&\times &\sum_{h_5,h_6,h_7}  \Big[ 
N \left(\big|A^{67} ( 5^{h_5}_{\gamma},6^{h_6}_g,7^{h_7}_g)\big|^2+\big|A^{76} ( 5^{h_5}_{\gamma},6^{h_6}_g,7^{h_7}_g)\big|^2\right) \nonumber \\
&-&\frac{1}{N} \big|A^{67} ( 5^{h_5}_{\gamma},6^{h_6}_g,7^{h_7}_g)+A^{76} ( 5^{h_5}_{\gamma},6^{h_6}_g,7^{h_7}_g)\big|^2 \Big] \, .
\end{eqnarray}

\subsubsection{Four-quark processes}
In a similar fashion the amplitude for the process,
$0 \rightarrow \bar u (p_1) + d(p_2) + \nu_e (p_3) + e^+ (p_4)  + \gamma (p_5)+q(p_6)+\bar{q}(p_7)$
for the simple case of $q$ and $\bar{q}$ consisting of non-identical flavours can be written as,
\beq
\label{fourquark}
\aa^\treenum ( 1^+_{\bar{u}},2^-_{d},3^-_{\nu},4^+_{\lrad},5^{h_5}_{\gamma},6^{h_6}_q,7^{h_7}_{\bar{q}})
=i\,\sqrt{2}\, e g_W^2 g_s^2   \, \sum_{a}  t^a_{i_7 i_1} \, t^a_{i_8,i_2} \, \aa^\treenum (5^{h_5}_{\gamma},6^{h_6}_q,7^{h_7}_{\bar{q}}) \, .
\eeq
Many other needed amplitudes are related to the amplitude in Eq.~(\ref{fourquark}) by crossing.
Full analytic results are presented in subsection~(\ref{Sevenpointtrees:quarks}).
Performing the colour sums, the squared amplitude is written as,
\beq
\sum_{h_5} \Big| \aa^\treenum ( 1^+_{\bar{u}},2^-_{d},3^-_{\nu},4^+_{\lrad},5^{h_5}_{\gamma},6^{+}_q,7^{-}_{\bar{q}})\Big|^2 = 
2 e^2 g_W^4 g_s^4 (N^2-1)\sum_{h_5} \Big| \aa^\treenum (5^{h_5}_{\gamma},6^{+}_q,7^{-}_{\bar{q}})\Big|^2 \, .
\eeq
The case for identical quarks is slightly more complicated, but can be easily derived from the results presented.
\section{$N$-jettiness method for NNLO cross sections}
\subsection{Jettiness}
A collision of partons $a$ and $b$ with momentum fractions $x_{a,b}$,
originating from the incoming-beam protons with momenta  $p_{a,b}$,
can potentially produce a final state including $N$ jets with momenta $\{p_i\}$. The
jettiness of parton $j$ with momentum $q_j$ is defined as
\beq
\taun(q_j)=\min_{i=a,b,1,\ldots,N}\left\{\frac{2\,p_i\cdot q_j}{{P_i}}\right\}\ .
\eeq
We denote by $E_i$ the jet or beam energy.
$P_i$ is a measure of the jet/beam hardness.
In our numerical results we set this equal to twice the jet/beam energy,
$P_i=2 E_i$~\cite{Stewart:2010tn}.
We can now define the event jettiness, or $N$-jettiness, as the sum
over all the $M$ final-state parton-jettiness values
\beq
\taun=\sum_{j=1}^M\taun(q_j)=\sum_{j=1}^M
\min_{i=a,b,1,\ldots,N}\left\{\frac{2\,p_i\cdot q_j}{ { P_i}}\right\}\ .
\eeq
For Leading Order (LO) events we have $\{q_i\}=\{p_i\}$ and the event jettiness is zero.
Beyond LO extra particles are emitted ($M>N$), the event jettiness goes to zero only in the soft/collinear limit. 
The event $N$-jettiness can be used in a non-local subtraction approach where we can isolate
the doubly unresolved region of the phase space by demanding $\taun<\tauncut$~\cite{Boughezal:2011jf,Gaunt:2015pea}.

\subsection{Colour singlet final states}
For the case at hand we have no coloured final-state partons at leading order.
We can therefore use the event shape variable $\tauzero$ to regulate the initial-state radiation.
By demanding $\tauzero<\tauzerocut$ one isolates the doubly unresolved
regions of phase space.  In this region the jettiness has simple factorization properties 
derivable using soft-collinear effective theory.
We exploit the fact that the matrix elements in the soft/collinear
approximation can be analytically integrated over this region and
added to the virtual contributions.  The regions of phase space where
$\tauzero>\tauzerocut$ are integrated over numerically.

In the context of MCFM, the application of the $N$-jettiness method to the particular case of 
processes involving the production of a colour-singlet final state 
at the Born level has been described in a series of 
papers~\cite{Boughezal:2016wmq,Campbell:2016yrh,Campbell:2017aul}.  
In particular refs.~\cite{Boughezal:2016wmq,Campbell:2017aul} contain 
details of the construction of the soft function and of the modifications to the two-loop
matrix elements needed to construct the hard function.

For the process studied in this paper the cut that defines the below-cut and
above-cut contributions is expressed in terms of a dimensionless variable $\epsilon$
that is defined by,
\begin{equation}
\label{epsilondef}
\tauzerocut = \epsilon \times m_{\ell \nu \gamma} \,,
\end{equation}
where $m_{\ell \nu \gamma}^2 = (p_\ell + p_\nu + p_\gamma)^2$.
Compared to a fixed (dimensionful) value of the cut this yields better numerical stability
and aids an automatic fitting of the $\tauzerocut$ dependence that
can be used to extrapolate the $\tauzerocut \to 0$ result.

\section{Setup of numerical results for $W(\to e \nu)+\gamma$ cross section}
\subsection{Parameter setup}
We investigate the processes $pp \to  e^- \bar{\nu}_e \gamma$ and $pp \to  e^+  \nu_e \gamma  $,
using the parameters shown in Table~\ref{parameters}. The parton distributions sets used
are the sets, 'NNPDF30\_xx\_as\_0118' with xx=lo,nlo,nnlo according to
the order calculated~\cite{Buckley:2014ana}. In all three cases the value of the strong coupling is taken to be $\alpha_s(M_z)=0.118$. 
The electromagnetic coupling $\alpha$ is a derived parameter, calculated with the definition shown in Table~\ref{parameters}.
The complex-mass scheme is used, so that the Weinberg angle is also complex.
In this section our parameter choices are made so as to agree with the choices made in Ref.~\cite{Grazzini:2017mhc}.

\renewcommand{\baselinestretch}{1.4}
\begin{table}
\begin{center}
\begin{tabular}{|l|l|l|l|}
\hline
$M_W$ & 80.385~GeV                & $\Gamma_W$ & 2.0854~GeV \\[-1.5mm]
$M_Z$ & 91.1876~GeV               & $\Gamma_Z$ & 2.4952~GeV \\[-1.5mm]
$G_\mu$ & $1.166390\times10^{-5}$~GeV$^{-2}$   &            & \\[-.5mm]
\hline
\hline
$m_W^2 = M_W^2-i M_W \Gamma_W $ & \multicolumn{3}{l|}{$(6461.748225 - 167.634879\, i) $~GeV$^2$} \\
$m_Z^2 = M_Z^2-i M_Z \Gamma_Z $ & \multicolumn{3}{l|}{$(8315.17839376 - 227.53129952\, i)$~GeV$^2$} \\
$\cos^2\theta_W={m_W^2}/{m_Z^2}$          & \multicolumn{3}{l|}{$(0.7770725897054007 + 0.001103218322282256\, i)$}\\
$\alpha = \frac{\sqrt{2}G_\mu}{\pi} M_W^2 (1-\frac{M_W^2}{M_Z^2})$ & \multicolumn{3}{l|}{$7.56246890198475\times10^{-3}$ giving $1/\alpha\approx 132.23\ldots$}\\
\hline
\end{tabular}
\caption{Input and derived parameters used for our numerical estimates.\label{parameters}}
\end{center}
\end{table}
\renewcommand{\baselinestretch}{1}

We construct an explicitly unitary, CP-conserving CKM matrix in the standard form~\cite{Zyla:2020zbs} 
with $(c_{ij}=\cos\theta_{ij},s_{ij}=\sin\theta_{ij}$)
\beq
V_{CKM}=
\left(\begin{matrix}
c_{12} c_{13}          &          c_{13} s_{12}     &        s_{13}\\  
- c_{23} s_{12} - c_{12} s_{13} s_{23} &  c_{12} c_{23} - s_{12} s_{13} s_{23} &   c_{13} s_{23} \\
 s_{12} s_{23} - c_{12} c_{23} s_{13}  &  - c_{12} s_{23} - c_{23} s_{12} s_{13} & c_{13} c_{23}
\end{matrix}\right) \, .
\eeq
Starting with the following values for four of the measurements $V_{ud}=0.97417$, $V_{us}=0.22480$, $V_{ub}=0.00409$, $V_{cb}=0.04050$,
and the following definitions for the angles
\beqn
s_{12}&=&\frac{V_{us}}{\sqrt{V_{ud}^2+V_{us}^2}},\;\;\;
s_{23}=\frac{V_{cb}}{V_{us}}{s_{12}},\;\;\;
s_{13}=V_{ub} \, ,
\eeqn
we obtain a unitary CKM matrix of the following form,
\beq
V_{CKM}=
\left(\begin{matrix}
0.97438  &  0.22485  &       0.00409\\       
 - 0.22483 &  0.97356  & 0.04051 \\
  0.00513  & - 0.04039 &  0.99917
\end{matrix}\right) \, .
\label{CKMnumerical}
\eeq
The third row in Eq.~(\ref{CKMnumerical})
involving couplings to the top is irrelevant for the current calculation.

We estimate the scale variation by varying the renormalization $(\mu_R=\kappa_R\mu_0)$ and factorization $(\mu_F=\kappa_F\mu_0)$ 
scales independently by a factor $\kappa_{R,F}$ about the central scale $\mu_0$,
\beq \label{Centralscale}
\mu_0 = \sqrt{M_W^2 + (p_T^\gamma)^2}.
\eeq
We choose $\kappa_{R,F}=\half$ or $2$.
This gives us eight possible scale variations about the central scale $(1,1)$, or six variations if we drop the choices
where $\kappa_F$ and $\kappa_R$ differ by a factor of 4,
\beqn
\text{7-point}:(\kappa_R,\kappa_F)&=&(1,1),(\half,1),(1,\half),(2,1),(1,2),(\half,\half),(2,2) \, ;\nonumber \\
\text{9-point}:(\kappa_R,\kappa_F)&=&(1,1),(\half,1),(1,\half),(2,1),(1,2),(\half,\half),(2,2),(\half,2),(2,\half) \, .
\eeqn
The assigned error is the maximum of the deviation from the value at the central scale $(1,1)$ in both the up and down directions. 
We note that although the 7-point variation has become a somewhat standard procedure, the extension to the 9-point
variation for our process is motivated by an accidental cancellation between renormalization and factorization scale
dependence observed at NLO~\cite{Campbell:2011bn}.

\subsection{Photon isolation}
\label{sec:photoniso}

Rather than performing a calculation that implements the effects of photon fragmentation,
necessitating the use of data-derived fragmentation functions, we pursue a simpler approach
that is readily applied in the NNLO case.  We use a ``smooth cone''
isolation procedure~\cite{Frixione:1998jh} to avoid infrared singularities
arising from the emission of photons from partons.  In this method one defines 
a cone of radius $R=\sqrt{(\Delta\eta)^2 + (\Delta\phi)^2}$
around the photon, where $\Delta\eta$ and $\Delta\phi$ are the pseudorapidity and azimuthal angle
difference between the photon and any parton. 
The total partonic transverse energy inside
a cone with radius $R$ is then constrained according to,
\begin{equation} \label{Frixione}
E_T^{had}(R) < \epsilon_s \, E_T^\gamma \left( \frac{1-\cos R}{1-\cos R_s} \right)^n\,,
\end{equation}
for all cones $R<R_s$, where $E_T^\gamma$ is the transverse photon energy, and $\epsilon_s$, $R_s$
and $n$ are parameters.

In addition, one can also choose to further impose an additional
fixed-cone isolation that more closely mimics the experimental sensitivity to photon isolation effects, resulting
in a so-called ``hybrid isolation'' scheme~\cite{Siegert:2016bre,Chen:2019zmr}.  The form of this
additional cut is,
\begin{eqnarray} 
\label{hybrid}
E_T^{had}(R)< \epsilon_f \, E_T^\gamma + E_T^f \qquad \mbox{for~}R<R_f\, .
\end{eqnarray} 
We shall make use of an additional cut of this form in Section~\ref{sec:7and13}.

\subsection{A first look at results at $\sqrt{s}=13$~TeV} 
\label{First_look}
\begin{figure}[t]
\begin{center}
\includegraphics[angle=90,width=\textwidth]{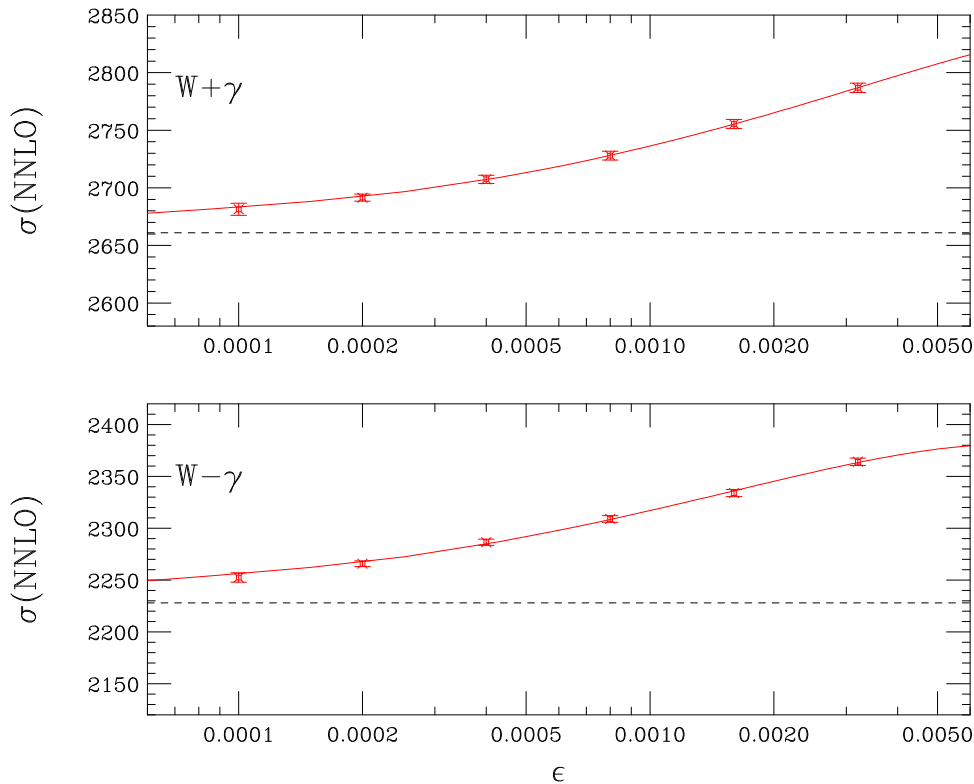}
\vspace{-14mm}
\caption{$\tau$-cut dependence of the NNLO cross sections for $W^+\gamma$ (upper) and
$W^-\gamma$ (lower) production using our setup.
The solid lines indicate the results of a fit to the points using the expected form
of power corrections given in Eq.~(\ref{eq:fitform}).
The asymptotic value of the NNLO cross section as $\epsilon \to 0$ (i.e. $\sigma_0$) is
indicated by the dashed line.}
\vspace{-1mm}
\label{fig:taudep1}
\end{center}
\end{figure}

We first perform a comparison with the MATRIX results given in Ref.~\cite{Grazzini:2017mhc},
that are computed at $\sqrt s=13$~TeV and using the cuts shown in Table~\ref{cuts}.
\renewcommand{\baselinestretch}{1.15}
\begin{table}
\begin{center}
\begin{tabular}{|l|l|}
\hline
Electron cuts & $p_{T,e} > 25$~GeV, $|\eta_e| < 2.47$ \\
Neutrino cuts & $p_{T}^{\rm missing} > 35$~GeV \\
Photon cuts & $p_{T,\gamma} > 15$~GeV, $|\eta_\gamma| < 2.47$ \\
Separation cuts & $\Delta_{ej}>0.3,\Delta_{\gamma j}>0.3,\Delta_{e\gamma}>0.7$\\
Photon Isolation & Isolation with $n=1$, $\epsilon_s=0.5$, $R_s=0.4$, c.f.~Eq.~(\ref{Frixione})\\
Jet definition & Anti-$k_T$ algorithm with R=0.4, $p_{T,j}>30$~GeV, $|\eta_j|<4.4$\\
\hline
\end{tabular}
\caption{Cuts used for the cross-section results in section~\ref{First_look}.}
\label{cuts}
\end{center}
\end{table}
The results of the MCFM calculation in this setup are shown in
Table~\ref{cross-sections}, indicating the cross sections obtained at
each order of perturbation theory up to NNLO, both with and without
the effects of the CKM matrix.  We have run the NNLO calculation to
ensure that the Monte Carlo uncertainty on the result at
$\epsilon=10^{-4}$, as defined in Eq.~(\ref{epsilondef}), is at the 2
per-mille level. An automated fit to the $\tau$ dependence is
performed using the known form of the residual power-corrections to
the SCET factorization formula~\cite{Campbell:2019dru},
\begin{equation}
\sigma = \sigma_0 + a\epsilon \log^3\epsilon +  b\epsilon \log^2\epsilon \,.
\label{eq:fitform}
\end{equation}
The result
(with the CKM matrix included) is illustrated in
Fig.~\ref{fig:taudep1}, yielding a correction to the
$\epsilon=10^{-4}$ result of around 1\%, with a corresponding
uncertainty of about 3 per-mille.  The difference between the result
using a diagonal CKM matrix, and the result with the CKM matrix of Eq.~(\ref{CKMnumerical}) included, is about
$0.8\%$ at LO but decreases to about $0.3\%$ at NLO and NNLO. Note
that because of the unitarity of the CKM matrix, parton processes
involving $gq$ or $g\bar{q}$ are approximately unchanged by including a non-diagonal
CKM matrix.  At NLO $gq$ and $g\bar{q}$ initial states contribute
$39\%$ of the cross section, which explains the reduced effect of a
non-diagonal CKM matrix.

Our results with a diagonal CKM matrix are in perfect agreement within mutual uncertainties
with those from MATRIX (reported in Table 6 of Ref.~\cite{Grazzini:2017mhc}) which are,
\begin{eqnarray}
\sigma^{\rm extrap}_{{\rm MATRIX}}(p p \to e^+ \nu_e \gamma ) &=& 2671(35)^{+3.8\%}_{-3.6\%} \, {\rm fb}, \\
\sigma^{\rm extrap}_{{\rm MATRIX}}(p p \to e^- \bar{\nu}_e \gamma ) &=& 2256(15)^{+3.7\%}_{-3.5\%} \, {\rm fb} .
\end{eqnarray}
We note that the uncertainty in the result, stemming
from the fit (or, equivalently, the extrapolation performed in MATRIX) is smaller in our case.

\renewcommand{\baselinestretch}{1.4}
\begin{table}
\begin{center}
\begin{tabular}{|l|l|l|l|l|}
\hline
Process                                      & $\sigma_{\text{LO}}$ [fb] & $\sigma_{\text{NLO}}$ [fb] & $\sigma_{\text{NNLO}}^{\epsilon=10^{-4}}$ [fb] & $\sigma_{\text{NNLO}}^{\rm
fit}$ [fb]\\
\hline
$p p \to e^+ \nu_e \gamma $ (no CKM)         & 861.6             & $2187^{+6.6\%}_{-5.3\%}$  & 2689(5)  & $2668(8)^{+3.9\%}_{-3.7\%}$ \\
$p p \to e^+ \nu_e \gamma $ (with CKM)       & 854.6             & $2181^{+6.6\%}_{-5.3\%}$  & 2681(5)  & $2661(8)^{+3.9\%}_{-3.7\%}$ \\
$p p \to e^- \bar{\nu}_e \gamma $ (no CKM)   & 726.2             & $1849^{+6.6\%}_{-5.3\%}$  & 2260(4)  & $2240(7)^{+3.7\%}_{-3.5\%}$ \\
$p p \to e^- \bar{\nu}_e \gamma $ (with CKM) & 720.1             & $1843^{+6.6\%}_{-5.3\%}$  & 2252(4)  & $2228(7)^{+3.7\%}_{-3.5\%}$ \\
\hline
\end{tabular}
\renewcommand{\baselinestretch}{1.2}
\caption{Cross-section results with the cuts of Table~\ref{cuts}. The theoretical error is 
estimated by a 7-point scale variation.  Parentheses indicate the residual error
resulting from numerical integration of the NNLO result (this error is beyond the
indicated number of digits at NLO).}
\label{cross-sections}
\end{center}
\end{table}
\renewcommand{\baselinestretch}{1.2}

Finally, we comment on the size of the higher-order corrections reported in
Table~\ref{cross-sections}, where the NLO cross-sections are larger than the
LO ones by a factor of about 2.5.  This is due in part to the filling
of the radiation zero that is present at LO, but is also the result of
significant contributions from corrections in which a gluon is present in the
initial state, as discussed above.  Further corrections at NNLO are more modest,
reflecting the fact that all important partonic channels have already been
opened at the preceding order.  Nevertheless, the importance of the gluon-quark
initiated contributions will be an important consideration in the discussion 
of electroweak corrections to this process.

 \section{Electroweak corrections}

As we have seen, the radiation zero, present in the lowest-order
process for $W\gamma$ production, suppresses the lowest-order
cross section so that the $O(1)$ and $O(\alpha_s)$ contributions
to the cross section are of similar size. Consequently the
$O(\alpha_s^2)$ QCD corrections are of great importance, effectively
playing the role of NLO corrections for the $gq$-induced part of 
the $O(\alpha_s)$ result. For the same reason, when considering
electroweak corrections, it will be important to
consider corrections to both the $O(1)$ and $O(\alpha_s)$ processes,
\begin{eqnarray}
  \label{udbarprocess}
  u + \bar{d} &\to& \nu + e^+ + \gamma \, , \\
  \label{qgprocess}
  u + g &\to& \nu + e^+ + \gamma + d \, .
\end{eqnarray}
The processes in Eqs.~(\ref{udbarprocess},\ref{qgprocess}) should be understood to include all
related $q\bar{q}$ and $qg$-initiated processes.

Because of the disruption of the normal hierarchy of $O(1)$ and $O(\alpha_s)$ contributions,
the $W\gamma$ process would be a prime candidate
for a complete mixed electroweak-QCD calculation, such as was
recently completed for the single $W$ process~\cite{Behring:2020cqi,Buonocore:2021rxx}.
The two-loop component of such a calculation will present a considerable
challenge, so for the moment we limit our discussion to the
electroweak corrections to the lowest-order processes shown
in Eqs.~(\ref{udbarprocess},\ref{qgprocess}).
For both of these processes we can identify two distinct types of
contributions,
\begin{itemize}
\item Processes involving initial-state photons, and associated
 terms needed to remove initial state collinear singularities;
\item Virtual electroweak corrections to the basic processes
  and real corrections associated with the emission of extra photons,
  together with the counterterms needed to remove singularities from soft and collinear 
  photon emission.
\end{itemize}
The electroweak corrections to the $W\gamma$ process have previously been considered
in refs.~\cite{Accomando:2005ra,Denner:2014bna}, but without the process in Eq.~(\ref{qgprocess}).
In the absence of the two-loop corrections mentioned above, we shall treat the
process in Eq.~(\ref{qgprocess}) by demanding an observed jet in the final state and will investigate
the sensitivity of the corrections to the value of the jet transverse momentum cut.  For simplicity,
in our calculations of electroweak corrections we assume a diagonal CKM matrix.

Note that in the case of real-radiation contributions to the electroweak
corrections we combine a photon and a charged lepton if they become collinear,
$\Delta R^{\ell\gamma} < 0.1$.  We subsequently demand that at least one photon 
satisfying $\Delta R^{\ell\gamma} > \Delta R^{\ell\gamma}_{\rm min}$ is observed
according to our smooth-cone isolation procedure (c.f. Section~\ref{sec:photoniso}),
with both $\Delta R^{\ell\gamma}_{\rm min}$ and the isolation parameters depending on the
analysis at hand, as described in Section~\ref{sec:7and13}.
Strictly speaking the recombination procedure is only
appropriate for observed electrons (not muons), although the difference
between this approach and retaining mass effects that lead to contributions proportional to
$\alpha \log(m_\mu)$ is small for all the observables we will consider
in this paper~\cite{Denner:2014bna}. 

\subsection{Effects of incoming photons}
It is opportune to re-examine the electroweak effects due to incoming
photons, in the light of updated distributions for the photon
structure of the proton~\cite{Manohar:2016nzj,Manohar:2017eqh}.  We
can assess the impact of photon-induced corrections in a
straightforward manner by evaluating contributions from the diagrams
representing the processes,
\begin{eqnarray}
  \label{gammauprocess}
  \gamma + u &\to& \nu + e^+ + \gamma + d \, , \\
  \label{gammagprocess}
  \gamma + g &\to& \nu + e^+ + \gamma + d + \bar u \, ,
\end{eqnarray}
\begin{figure}
\includegraphics[angle=270,width=\textwidth]{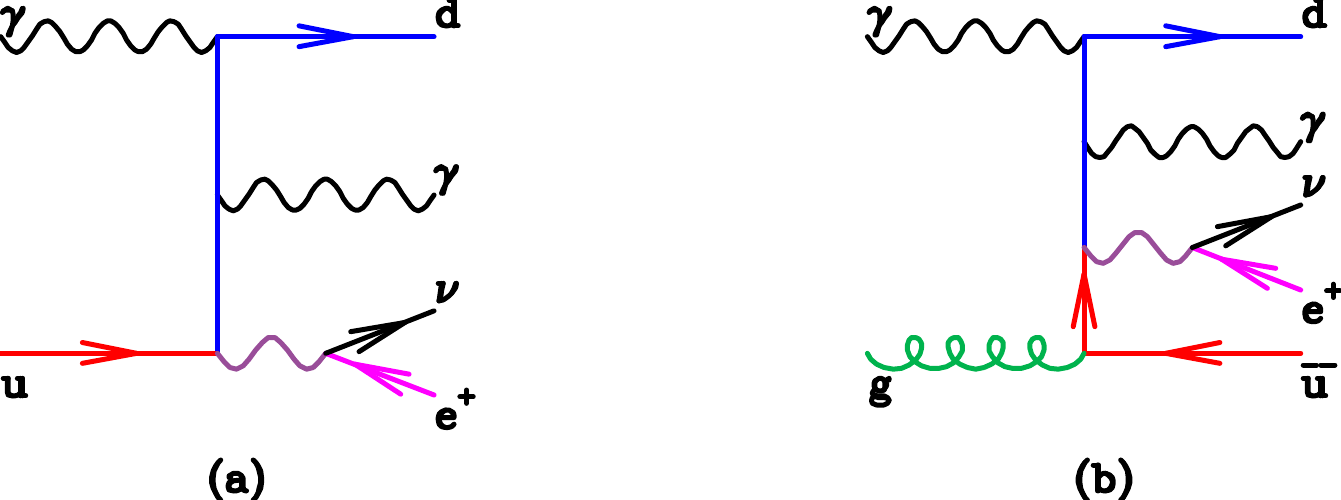}
\renewcommand{\baselinestretch}{1.2}
\caption{Representative photon-induced diagrams (a) for the process in Eq.~(\ref{gammauprocess}) and (b)
  for the process in Eq.~(\ref{gammagprocess}).}
\vspace{2mm}
\label{photoninduced}
\end{figure}
that are shown in Fig.~\ref{photoninduced}.
The singularity associated with the initial-state
splitting $\gamma \to q \bar{q}$ is absorbed into the quark
parton distribution function (pdf) in the $\overline{MS}$ scheme, in
accordance with the treatment of the photon pdf in the LUX
determination~\cite{Manohar:2016nzj,Manohar:2017eqh}.
For the process shown in Eq.~(\ref{gammagprocess}) initial-state singularities
for the splitting $g \to q \bar{q}$ are also absorbed in the same way.
Our numerical results for processes with
incoming photons are given in section~\ref{sec:7and13}.

\subsection{Electroweak virtual corrections}
The principal ingredient needed for the next step in evaluating the
radiative corrections are the one-loop corrections to the Born-level
process due to the exchange of $W, Z$ and $\gamma$. The numerical 
results for the electroweak corrections to the processes in Eqs.~(\ref{udbarprocess},\ref{qgprocess})
have been obtained using the Recola library~\cite{Actis:2016mpe,Denner:2017wsf}.
The Recola library supplies results in three different renormalization schemes,
\begin{itemize}
\item the $\alpha(G_\mu)$ scheme, where $\alpha=\sqrt{2}G_\mu/\pi M_W^2(1-M_W^2/M_Z^2)$, which includes universal terms
  associated with the renormalization of the weak mixing angle;
\item the $\alpha(0)$ scheme, where $\alpha$ is fixed by the measured value at $p^2=0$;
\item the $\alpha(M_Z)$ scheme, where $\alpha$ is fixed by the value at $p^2=M_Z^2$, taking into account
the running from $p^2=0$ to $p^2=M_Z^2$, which at low $p^2$ is inadequately treated in perturbation theory. 
\end{itemize}
Since the $W \gamma$ process involves both a real photon and $W$ bosons
we use a hybrid scheme in which the photon is treated in the $\alpha(0)$
scheme but all remaining powers of $\alpha$ (including factors associated
with the radiation of additional real or virtual photons) are considered
in the $\alpha(G_\mu)$-scheme.  Since we work in dimensional regularization, we note
that the translation between $\alpha(0)$ and $\alpha(G_\mu)$ schemes involves
the addition of singular terms -- single poles in $\epsilon$ -- as well as
a finite difference.
Clearly, we must also modify our earlier specification of the parameter setup
by replacing one power of $\alpha \equiv \alpha(G_\mu)$ by $\alpha(0) =
1/137.036$, resulting in all cross-sections being reduced by a factor
$\alpha(0)/\alpha(G_\mu) = 0.965$.
All the cross sections quoted in section~\ref{sec:7and13} have this rescaling already applied.

 \section{Numerical results for 7 and 13 TeV}
\label{sec:7and13}
\subsection{Comparison with CMS at 7 TeV}

In this section we compare the CMS results~\cite{Chatrchyan:2013fya} based on 5.0~fb$^{-1}$
of $\sqrt{s}=7$~TeV data with our predictions. The CMS results are not cross sections in the fiducial 
region, but rather have been corrected for the geometric and kinematic acceptance of the detector. 

Note that, compared to the earlier NLO predictions from MCFM reported in Ref.~\cite{Chatrchyan:2013fya},
the ones here differ in multiple respects.  As detailed earlier, here we have used 
the complex-mass scheme and slightly different electroweak parameters (including a non-diagonal CKM matrix),
updated the PDF set (to NNPDF3.0, instead of CTEQ6.6), and used a different
central scale choice ($\sqrt{M_W^2 + (p_T^\gamma)^2}$, cf.~Eq.~(\ref{Centralscale}),  instead of $M_W$),
and scale variation (a 9-point variation about the central choice instead of previously a common variation in opposite directions only).
Furthermore, anticipating the inclusion of electroweak corrections we replace a single power of $\alpha(G_\mu)$ in all
predictions by $\alpha(0)$, as discussed in the previous section.
In addition, our photon isolation is rendered theoretically viable by the hybrid isolation scheme, rather than having recourse to a
non-perturbative fragmentation function.
The parameters for the hybrid scheme defined in Eqs.~(\ref{Frixione}, \ref{hybrid}) are,
\begin{equation}
\epsilon_s =0.5, \;\; n=1, \;\; R_s= 0.1, \;\;
\epsilon_f = 0.0025, \;\; E_T^f = 2.2~\mbox{GeV}, \;\; R_f=0.4 \,,
\end{equation}
in order to mimic the HCAL photon isolation cut in Ref.~\cite{Chatrchyan:2013fya}.
Our theoretical results with the input parameters as described in this paragraph are given in Table~\ref{7TeVtheory}.

\renewcommand{\baselinestretch}{1.4}
\begin{table}
  \begin{center}
    \begin{tabular}{|l|l|l|l|l|}
    \hline 
    Process                     & $p_T^\gamma(min)$  & $\sigma_{\text{NLO}}$ [pb]      & $\sigma_{\text{NNLO}}$ [pb]   & $\sigma_{\text{NNLO}}/\sigma_{\text{NLO}}$  \\
    \hline
    $pp \to e^+\nu \gamma$      &$ 15$~[GeV]         & $17.98^{+0.77}_{-1.00}$   & $19.23^{+0.14}_{-0.41}$ & 1.070\\
    $pp \to e^-\bar{\nu}\gamma$ &$ 15$~[GeV]         & $12.43^{+0.54}_{-0.71}$   & $13.35^{+0.10}_{-0.28}$ & 1.074\\
    $pp \to e^+\nu \gamma$      &$ 60$~[GeV]         & $0.369^{+0.032}_{-0.026}$ &$0.451^{+0.021}_{-0.019}$& 1.22\\
    $pp \to e^-\bar{\nu}\gamma$ &$ 60$~[GeV]         & $0.247^{+0.023}_{-0.019}$ &$0.314^{+0.017}_{-0.014}$& 1.27\\
    $pp \to e^+\nu \gamma$      &$ 90$~[GeV]         & $0.117^{+0.013}_{-0.010}$ &$0.144^{+0.008}_{-0.007}$& 1.23\\
    $pp \to e^-\bar{\nu}\gamma$ &$ 90$~[GeV]         & $0.072^{+0.009}_{-0.007}$ &$0.095^{+0.006}_{-0.006}$& 1.32\\
    \hline
  \end{tabular}  
    \end{center}
\renewcommand{\baselinestretch}{1.2}
  \caption{Our theoretical predictions for $e^+\nu \gamma$ and $e^-\bar{\nu}\gamma$ at $\sqrt{s}=7$~TeV.
    The ratio $\sigma_{\text{NNLO}}/\sigma_{\text{NLO}}$ is given without uncertainty,
    to illustrate the trend with $p_T^\gamma(min)\,$. \label{7TeVtheory}}
    \end{table}  

A comparison between the theoretical predictions and CMS measurements, for these three different values of the
minimum photon $p_T$, is shown in Table~\ref{comparison-7tev}.  Since the CMS measurement is not in a fiducial region
and suffers from rather large systematic errors we do not present the effect of electroweak corrections here, but
postpone such a discussion until the following section.
\renewcommand\baselinestretch{1.5}
\begin{table}
\begin{center}
\begin{tabular}{|l|l|l|l|}
\hline
$\sigma(pp\to e^+\nu\gamma)+\sigma(pp \to e^-\bar{\nu}\gamma)$  & NLO [pb] &NNLO [pb] & CMS Experiment [pb]\\
\hline
$\sigma(p_T^\gamma>15~{\rm GeV})$ &  $30.41^{+1.31}_{-1.72}$   & $32.58^{+0.24}_{-0.69}$   & $ 37.0 \pm 0.8\pm 4.0 \pm 0.8$ \\
$\sigma(p_T^\gamma>60~{\rm GeV})$ &  $0.616^{+0.055}_{-0.045}$ & $0.765^{+0.039}_{-0.035}$ & $ 0.76\pm 0.05\pm 0.08\pm 0.02$\\
$\sigma(p_T^\gamma>90~{\rm GeV})$ &  $0.189^{+0.021}_{-0.016}$ & $0.238^{+0.014}_{-0.013}$ & $ 0.20\pm 0.03\pm 0.04\pm 0.01$\\
\hline
\end{tabular}
\renewcommand\baselinestretch{1.2}
\caption{NLO and NNLO predictions for cross sections at $\sqrt{s}=7$~TeV, for comparison with the
values measured by the CMS experiment~\cite{Chatrchyan:2013fya}. The experimental result is the 
average of the cross section to electrons and the cross section to muons.
The errors on the experimental results are statistical, systematic and luminosity respectively.
All results have a cut $\Delta R^{l\gamma}>0.7$.}
\label{comparison-7tev}
\end{center}
\end{table}

In order to probe the radiation zero that occurs at $y_\gamma^\star = 0$ (centre-of-mass frame)
it is easiest to construct a boost-invariant difference of rapidities between the lepton and the photon.
Weighting this by the charge of the lepton results in the ``signed rapidity difference'', the quantity
used in the original Tevatron probes of this phenomenon.  This quantity has also been measured in
Ref.~\cite{Chatrchyan:2013fya} in the fiducial region defined by additional acceptance cuts on the
leptons and photon.  For the sake of comparison, we show corresponding predictions for this quantity
based on the electron-channel cuts shown in Table~\ref{CMSrapzerocuts}, and after the application
of a veto on any jets observed in the region $p_T>30$~GeV, $|\eta|<4.4$ using the anti-$k_T$ clustering
algorithm with $R=0.4$.   The transverse cluster mass of the photon, lepton and missing $E_T$ (neutrino)
system ($M_T(\ell\gamma\Etmiss)$) is defined by,
\begin{equation}
M_T(\ell\gamma\Etmiss) = \left[
\left(M_{\ell\gamma}^2 + |\vec p_T(\gamma) + \vec p_T(\ell)|^2 \right)^{\frac{1}{2}}
 + \Etmiss \right]^2 - \left|\vec p_T(\gamma) + \vec p_T(\ell) + \vec \Etmiss \right|^2 \,.
\label{eq:TCM}
\end{equation}

\begin{table}
\begin{center}
\begin{tabular}{|l|l|}
\hline
$|\eta^\gamma| < 2.5$   & $p^\gamma_T >25$ GeV \\
$|\eta^\ell | < 2.5$, excluding $1.44 < |\eta^\ell | < 1.57$       & $p^\ell >35$ GeV \\
$\Delta R^{\ell\gamma} > 0.7$ & $M_T(\ell\gamma\Etmiss) > 110$~GeV \\
\hline
\end{tabular}
\caption{Fiducial cuts for theoretical predictions of the signed
rapidity difference at 7 TeV.}
\label{CMSrapzerocuts}
\end{center}
\end{table}

Our results are shown in Figure~\ref{veto7ydiff_cuts}, where the NLO and NNLO predictions also indicate
the uncertainties obtained by scale variation.  After the large correction from LO to NLO, for this set
of cuts there do not appear to be significant further corrections at NNLO and the scale
uncertainties somewhat overlap.  However, as is clear from the lower panel of
Figure~\ref{veto7ydiff_cuts}, the shape of the NNLO prediction does differ slightly from
the NLO one.  These predictions appear to be in reasonable agreement with the
experimental results of Figure 7 in Ref.~\cite{Chatrchyan:2013fya}.

\renewcommand\baselinestretch{1}
\begin{figure}
\begin{center}
\includegraphics[angle=90,width=0.7\textwidth]{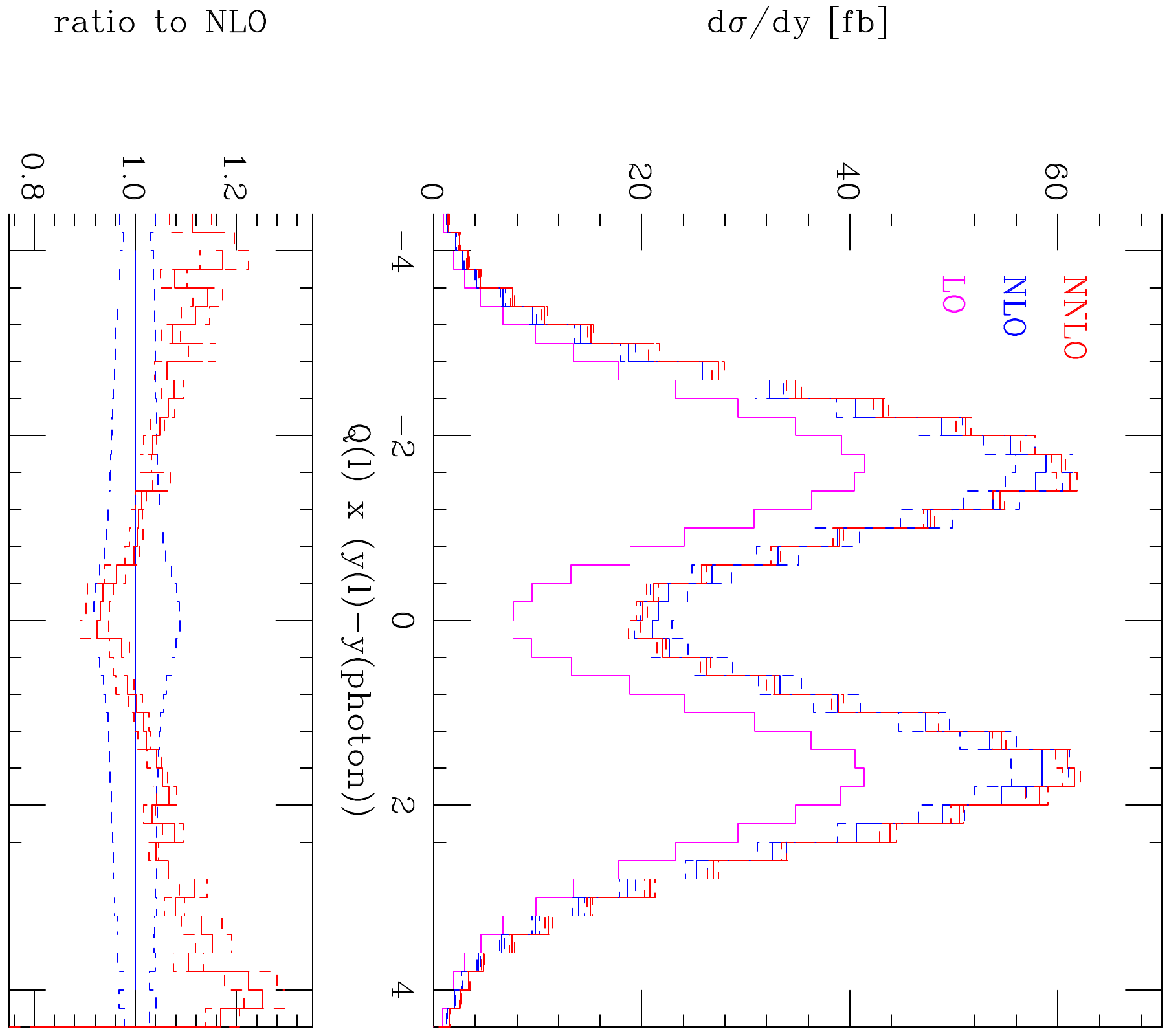}  
\renewcommand\baselinestretch{1.2}
\caption{Signed rapidity difference between the lepton and the photon,
after the application of the cuts detailed in Table~\ref{CMSrapzerocuts} and the
jet veto described in the text. The scale variation uncertainties are shown at NLO and NNLO.
The lower panel shows the ratio of the NLO and NNLO results, including the uncertainty bands,
to the central NLO result.}
\label{veto7ydiff_cuts}
\end{center}
\end{figure}

 \subsection{MCFM projections for $\sqrt{s}=13$~TeV}

In this section we make projections for 13 TeV.
We first consider the overall cross section in a  fiducial region using the cuts shown in Table~\ref{CMSprojectioncuts}
taken from Ref.~\cite{Sirunyan:2021zud}. The fiducial region is further reduced by demanding that the leptons and photons are isolated.
In Ref.~\cite{Sirunyan:2021zud} a lepton or photon is considered
isolated if the sum of the $p_T$ of all stable particles within $\Delta R = 0.4$, divided by the $p_T$ of the
lepton or photon, is less than 0.5.
\begin{table}
\begin{center}
\begin{tabular}{|l|l|}
\hline
$|\eta^\gamma| < 2.5$   & $p^\gamma_T >25$ GeV \\
$|\eta^\ell | < 2.5$       & $p^\ell >25$ GeV \\
$\Delta R^{\ell\gamma} > 0.5$ & \\
\hline
\end{tabular}
\caption{Fiducial cuts imposed for calculations of the total rate at 13 TeV.}
\label{CMSprojectioncuts}
\end{center}
\end{table}
In our numerical work we follow a procedure that is close to the procedure used in experiments.
We compute the inclusive cross section with the anti-$k_T$ jet algorithm and identify jets by clustering with
$R=0.4$ and demanding $p_T({\rm jet})$ > 12.5 GeV. We subsequently check to see whether any jet axes lie within
a cone of $\Delta R$=0.4 of the lepton and then reject the event if the scalar sum of any such jet $p_T$'s
is greater than $0.5 \times p_T^l$.
For isolation of the photon we again use a hybrid procedure, with smooth cone parameters (c.f. Eq.~(\ref{Frixione})),
\begin{eqnarray}
\epsilon_s =0.5, \qquad n=1, \qquad R_s= 0.1 \,,
\end{eqnarray}
and fixed-cone isolation parameters (c.f. Eq.~(\ref{hybrid})) taken from Ref.~\cite{Sirunyan:2021zud},
\begin{equation}
\epsilon_f = 0.5, \qquad E_T^f = 0, \qquad R_f=0.4 \,.
\end{equation}
Under these cuts our results are shown in Table~\ref{13TeVprojection}.
We note that the NNLO results lie outside the band of values predicted on the basis of scale variation in the NLO result.
Summing both charges in Table \ref{13TeVprojection} we obtain predictions 
for the cross section in the fiducial region
for both electrons and muons (ignoring any feed-down from $\tau$ decays),
\renewcommand{\baselinestretch}{1.4}
\begin{table}
  \begin{center}
    \begin{tabular}{|l|l|l|l|}
    \hline 
    Process                                               & $\sigma_{\text{NLO}}$ [pb]      & $\sigma_{\text{NNLO}}$ [pb]   & $\sigma_{\text{NNLO}}/\sigma_{\text{NLO}}$  \\
    \hline
    $pp\to \nu e^+\gamma   +pp \to \nu \mu^+ \gamma      $& $5.40^{+0.56}_{-0.56}$    &$6.73^{+0.31}_{-0.36}$   & 1.25 \\
    $pp\to e^-\bar\nu\gamma+pp \to \mu^- \bar\nu\gamma   $& $4.81^{+0.52}_{-0.52}$    &$5.96^{+0.28}_{-0.33}$   & 1.24  \\
    \hline
  \end{tabular}  
    \end{center}
\renewcommand\baselinestretch{1.2}
  \caption{Theoretical predictions for $\sqrt{s}=13$~TeV using cuts of Table~\ref{CMSprojectioncuts} and the isolation cuts described in
    the text. The ratio $\sigma_{\text{NNLO}}/\sigma_{\text{NLO}}$, given without an uncertainty, indicates that the corrections are still
    substantial at NNLO. \label{13TeVprojection} }
\end{table}  
\begin{eqnarray}
\sigma_{{\rm NLO}}(e^{\pm} \nu \gamma + \mu^{\pm} \nu \gamma) &=& 10.21^{+1.08}_{-1.08}~\text{pb}, \\
\sigma_{{\rm NNLO}}(e^{\pm} \nu \gamma+ \mu^{\pm} \nu \gamma)&=& 12.69^{+0.59}_{-0.69}~\text{pb}.
\label{NNLOxsec}
\end{eqnarray}
The only experimental result at $\sqrt{s}=13$~TeV reported so far is 
the sum of the cross sections for electrons and muons of both charges, 
\beq
\label{CMS}
\sigma=15.58 \pm 0.05 \pm 0.73 \pm 0.15~\text{pb} 
\eeq
from Ref.~\cite{Sirunyan:2021zud}, where the uncertainties are, respectively,
statistical, systematic and related to theoretical inputs.

The NNLO prediction given above is smaller than the experimental measurement
by about $20\%$.  However a comparison with this data requires a correction for the feed down from the $W \to \tau \nu$
channel, which is present in the result in Eq.~(\ref{CMS}) but not included in our theoretical rates. 
We think that this correction is best performed by the experimental collaborations, based on the
well-modelled properties of $\tau$ decay. 

However in order to get an idea of the order of magnitude of this correction we have studied the number of additional
events that may be produced from $\tau$-lepton decays using the Herwig Monte Carlo~\cite{Bahr:2008pv}.
The $17\%$ branching ratio of the $\tau$-lepton to electrons and muons sets an upper limit on the size of this 
correction. However, the leptons of the first two generations coming from $\tau$ decay are much softer than primary leptons,
and less frequently isolated. Our studies with Herwig, combining a NLO calculation with the effects of the parton shower,
indicate that less than $2\%$ of the produced $\tau$-leptons end up in the CMS event sample.
This source of additional events therefore seems unlikely to account for the difference between the
theoretical prediction in Eq.~(\ref{NNLOxsec}) and the measured cross-section in Eq.~(\ref{CMS}).

Finally we turn to the inclusion of electroweak corrections.  For the 
EW corrections to the LO process in Eq.~(\ref{udbarprocess}) 
we summarize their effect by using a relative factor $\delta_{EW}$ that is defined by,
\begin{equation}
\label{eq:EWdelta}
\delta_{EW}^{q\bar q} = \frac{\sigma_{EW}}{\sigma_{LO}} \,,
\end{equation}
where both numerator and denominator are computed using the same (LUX) pdf set.
This form allows electroweak effects to be incorporated in a QCD-corrected calculation
of any order in a straightforward manner.
We find the relative correction factors,
\begin{equation}
\label{eq:dEWqqbar}
\delta_{EW}^{q\bar q}(\nu\ell^+\gamma) = -0.013 \,, \;\;
\delta_{EW}^{q\bar q}(\ell^-\bar\nu\gamma) = -0.012 \,.
\end{equation}
Since the photon-initiated process shown in Eq.~(\ref{gammauprocess}) represents a new
partonic channel we do not expect its effects to factorize in the same way. 
We therefore present the corrections from this channel normalized to the NNLO cross section,
\begin{equation}
\label{eq:EWdeltaqgamma}
\delta_{EW}^{q\gamma} = \frac{\sigma_{EW}}{\sigma_{\text{NNLO}}} \,.
\end{equation}
In this way we find,
\begin{equation}
\label{eq:dEWqgamma}
\delta_{EW}^{q\gamma}(\nu\ell^+\gamma) = +0.011 \,, \;\;
\delta_{EW}^{q\gamma}(\ell^-\bar\nu\gamma) = +0.010 \,.
\end{equation}
Considered in this way, these two contributions essentially cancel and, taken together, do not represent a substantial
further correction to the rate.  Indeed, the biggest impact of the inclusion of the electroweak
corrections results from the coupling factor change, $\alpha(G_\mu) \to \alpha(0)$.  Electroweak
corrections to the $O(\alpha_s)$ channels represented by Eqs.~(\ref{qgprocess})
and ~(\ref{gammagprocess}) can only be defined in the presence of a jet and their effect
on the inclusive rate cannot be directly inferred from the calculations we have performed in
this work. Results for these channels will be presented in the following section.

\subsection{Differential distributions}

We now provide a set of predictions suitable for a possible future
measurement of this process at 13 TeV.  For this we adopt a slightly
different set of cuts, as detailed in Table~\ref{CMSdiffprojectioncuts}.
\begin{table}
\begin{center}
\begin{tabular}{|l|l|}
\hline
$|\eta^\gamma| < 2.5$   & $p^\gamma_T >30$ GeV \\
$|\eta^\ell | < 2.5$       & $p^\ell_T >30$ GeV \\
$\Delta R^{l\gamma} > 0.7$ & $\Etmiss >40$ GeV \\
\hline
\end{tabular}
\caption{Fiducial cuts imposed for the calculation of differential distributions at 13 TeV.}
\label{CMSdiffprojectioncuts}
\end{center}
\end{table}
\renewcommand\baselinestretch{1}
We use the same lepton and photon isolation requirements as in the previous
section, but for this case we identify jets with,
\begin{equation}
\label{jetdefndiff}
p_T(\mbox{jet})>30~\mbox{GeV} \,, \qquad
|\eta(\mbox{jet})|<2.5 \,.
\end{equation}
All plots in this section sum the contributions of $e^+\nu \gamma$ and $e^-\bar{\nu} \gamma$.
The theoretical result for muons will be identical. 

\begin{figure}
\begin{center}
\includegraphics[angle=90,width=0.7\textwidth]{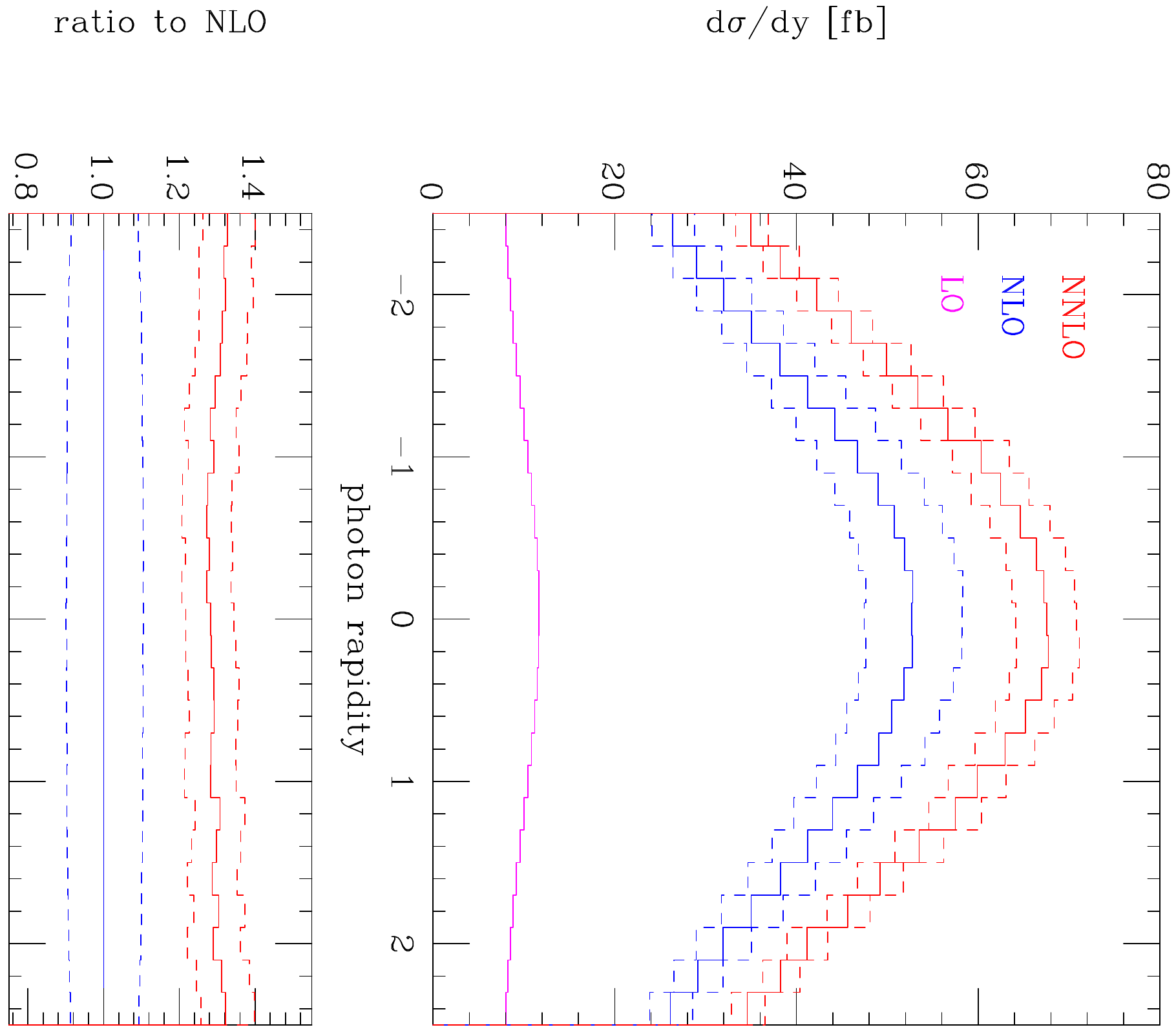}
\renewcommand\baselinestretch{1.2}
\caption{Distribution of the photon rapidity at 13 TeV.
The NNLO and NLO histograms display the error estimates, obtained via a 9-point scale variation,
and the lower panel displays the ratio of the NLO and NNLO predictions to the central NLO result.}
\label{etagamma}
\end{center}
\end{figure}

\begin{figure}
\begin{center}
\includegraphics[angle=90,width=0.7\textwidth]{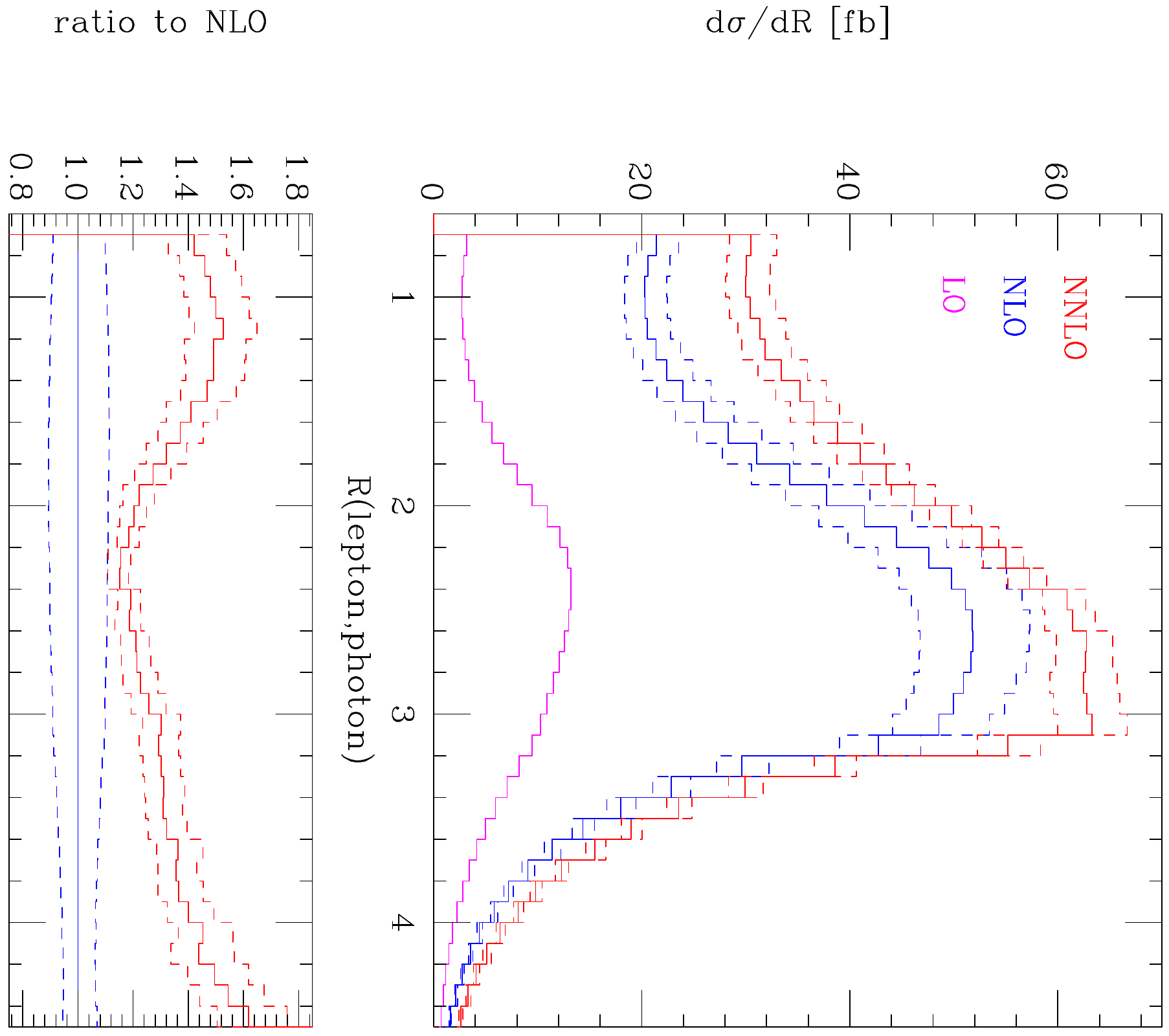}
\renewcommand\baselinestretch{1.2}
\caption{$\Delta R(\ell,\gamma)$ distribution at 13 TeV.
The NNLO and NLO histograms display the error estimates, obtained via a 9-point scale variation,
and the lower panel displays the ratio of the NLO and NNLO predictions to the central NLO result.}
\label{deltaregamma}
\end{center}
\end{figure}

We first show predictions for some basic quantities: the photon rapidity (Figure~\ref{etagamma}) and the
angular separation between the photon and the lepton, $\Delta R^{\ell\gamma}$
(Figure~\ref{deltaregamma}).  The enormous corrections to the total cross section are reflected as a practically-uniform
shift in the photon rapidity distribution, while the shape of the $\Delta R^{\ell\gamma}$ distribution is modified at NNLO.
The scale uncertainties, also shown in the figures, do not overlap between NLO
and NNLO. This is to be expected because of the effectively leading order nature of a large part of the
$O(\alpha_s)$ contribution.

\begin{figure}
\begin{center}
\includegraphics[angle=90,width=0.7\textwidth]{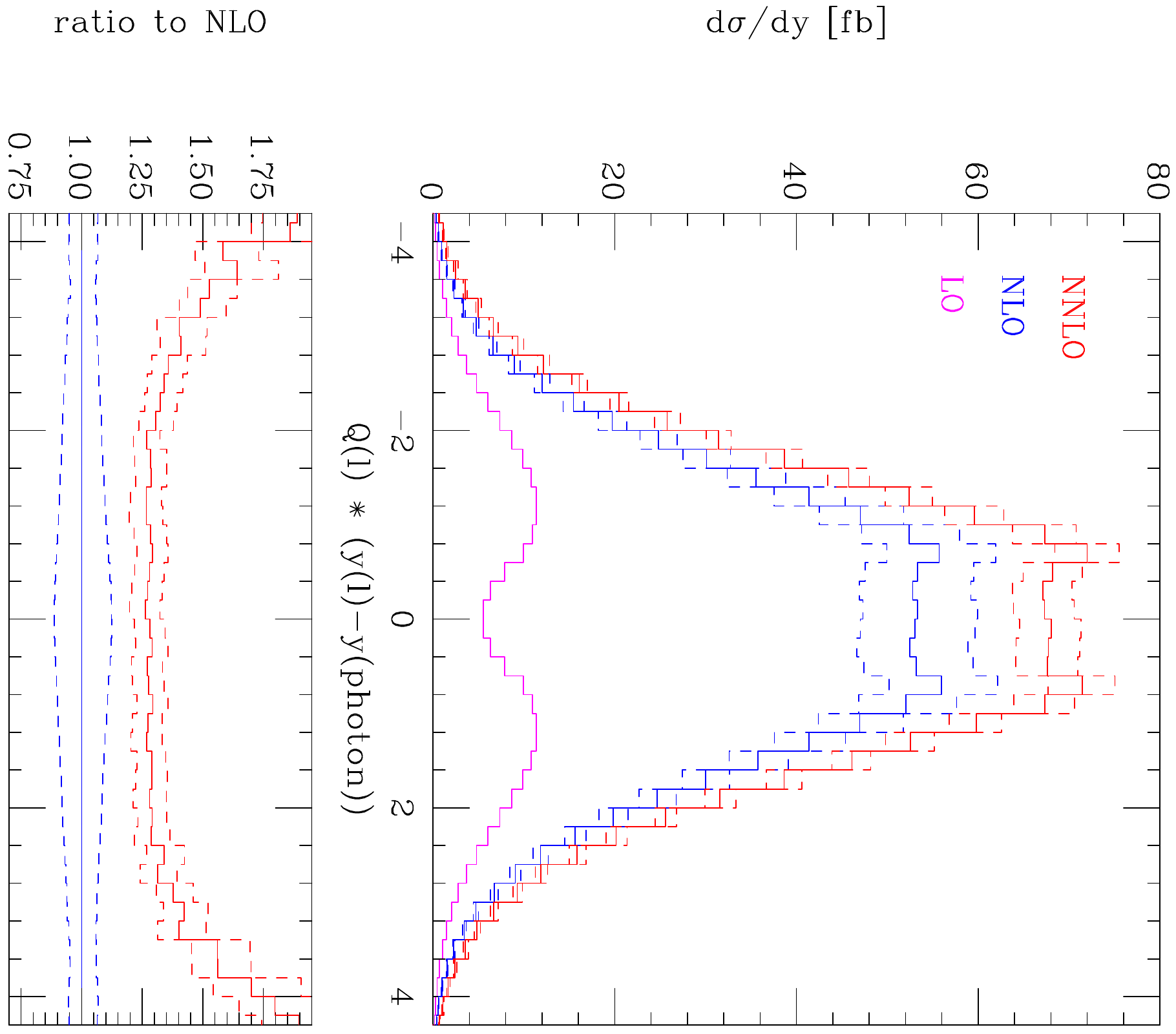}
\renewcommand\baselinestretch{1.2}
\caption{Signed rapidity difference between the lepton and the photon at 13 TeV.
The NNLO and NLO histograms display the error estimates, obtained via a 9-point scale variation,
and the lower panel displays the ratio of the NLO and NNLO predictions to the central NLO result.}
\label{ydiff}
\end{center}
\end{figure}

\begin{figure}
\begin{center}
\includegraphics[angle=90,width=0.7\textwidth]{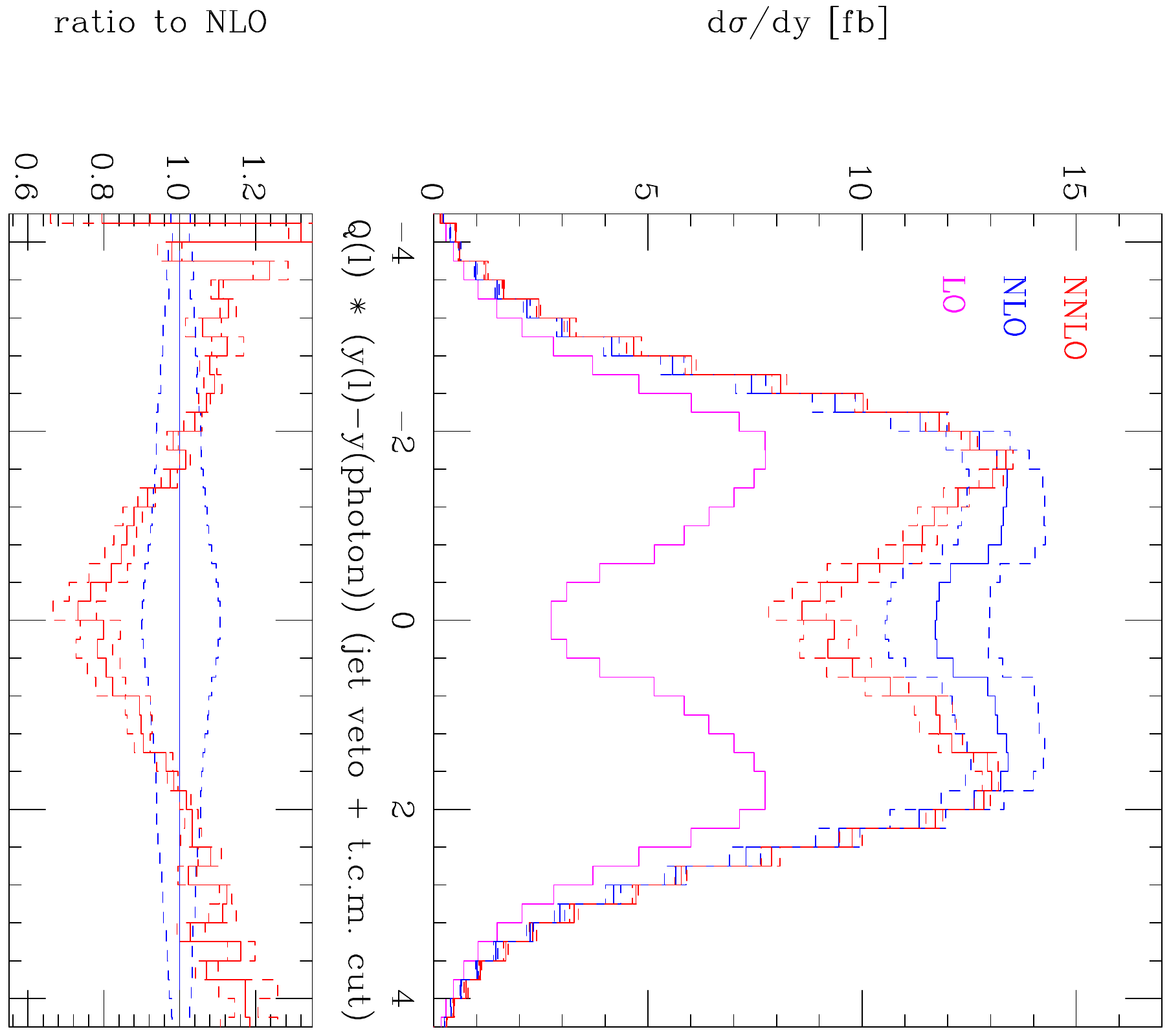}
\renewcommand\baselinestretch{1.2}
\caption{Signed rapidity difference between the lepton and the photon at 13 TeV,
after additional cuts to veto jets and require $M_T(\ell\gamma\Etmiss) > 150$~GeV.
The NNLO and NLO histograms display the error estimates, obtained via a 9-point scale variation,
and the lower panel displays the ratio of the NLO and NNLO predictions to the central NLO result.}
\label{vetoydiff}
\end{center}
\end{figure}

As already discussed in our presentation of results at 7 TeV, the 
signature of the radiation zero for the $W\gamma$ process is the signed rapidity difference
between the lepton and the photon.  As shown in Figure~\ref{ydiff}, the
effect of the radiation zero -- a depletion in the central region at LO -- is almost completely
eliminated at NLO.  As already indicated, it can be partially restored by applying a jet veto (i.e. so that any event containing a jet as
defined in Eq.~(\ref{jetdefndiff}) is removed) and applying a cut on the transverse cluster mass (c.f. Eq.~\ref{eq:TCM}).
Our prediction at 13~TeV, after applying a jet veto and demanding that $M_T(\ell\gamma\Etmiss) > 150$~GeV, 
is shown in Figure~\ref{vetoydiff}.  In this case, in contrast to 7 TeV, the extent of the dip that reflects
the radiation zero is changed from NLO to NNLO.  This reflects the difficulty in capturing the effects of a jet
veto in fixed-order perturbation theory, particularly in the presence of higher-order corrections that are
significantly larger at 13~TeV than 7~TeV.

\begin{figure}
\begin{center}
\includegraphics[angle=90,width=0.7\textwidth]{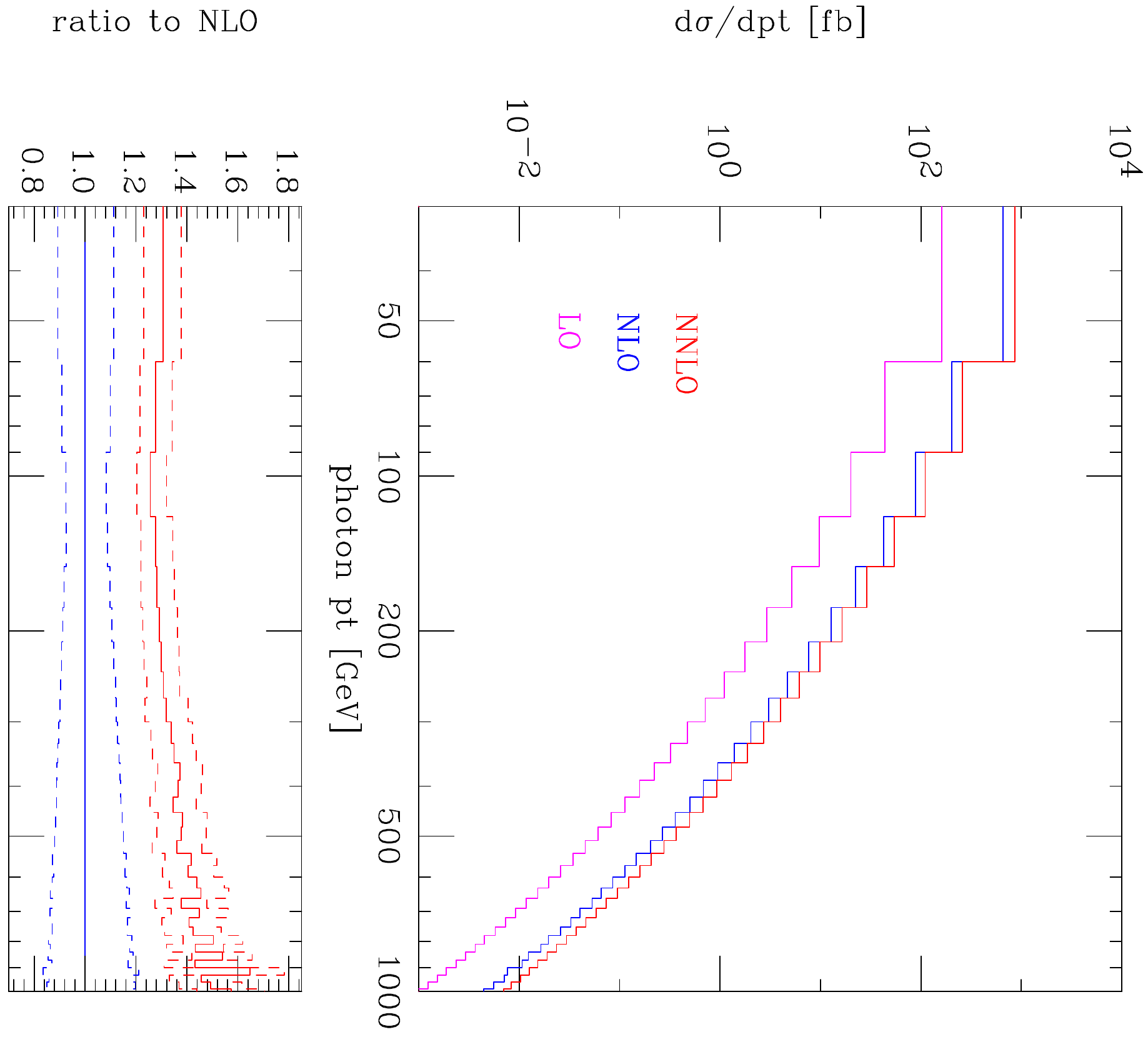}
\caption{$p_T^\gamma$ distribution at 13 TeV.
The lower panel displays the ratio of the NLO and NNLO predictions, including
error estimates obtained via a 9-point scale variation, to the central NLO result.}
\label{ptgamma}
\end{center}
\end{figure}

\begin{figure}
\begin{center}
\includegraphics[angle=90,width=0.7\textwidth]{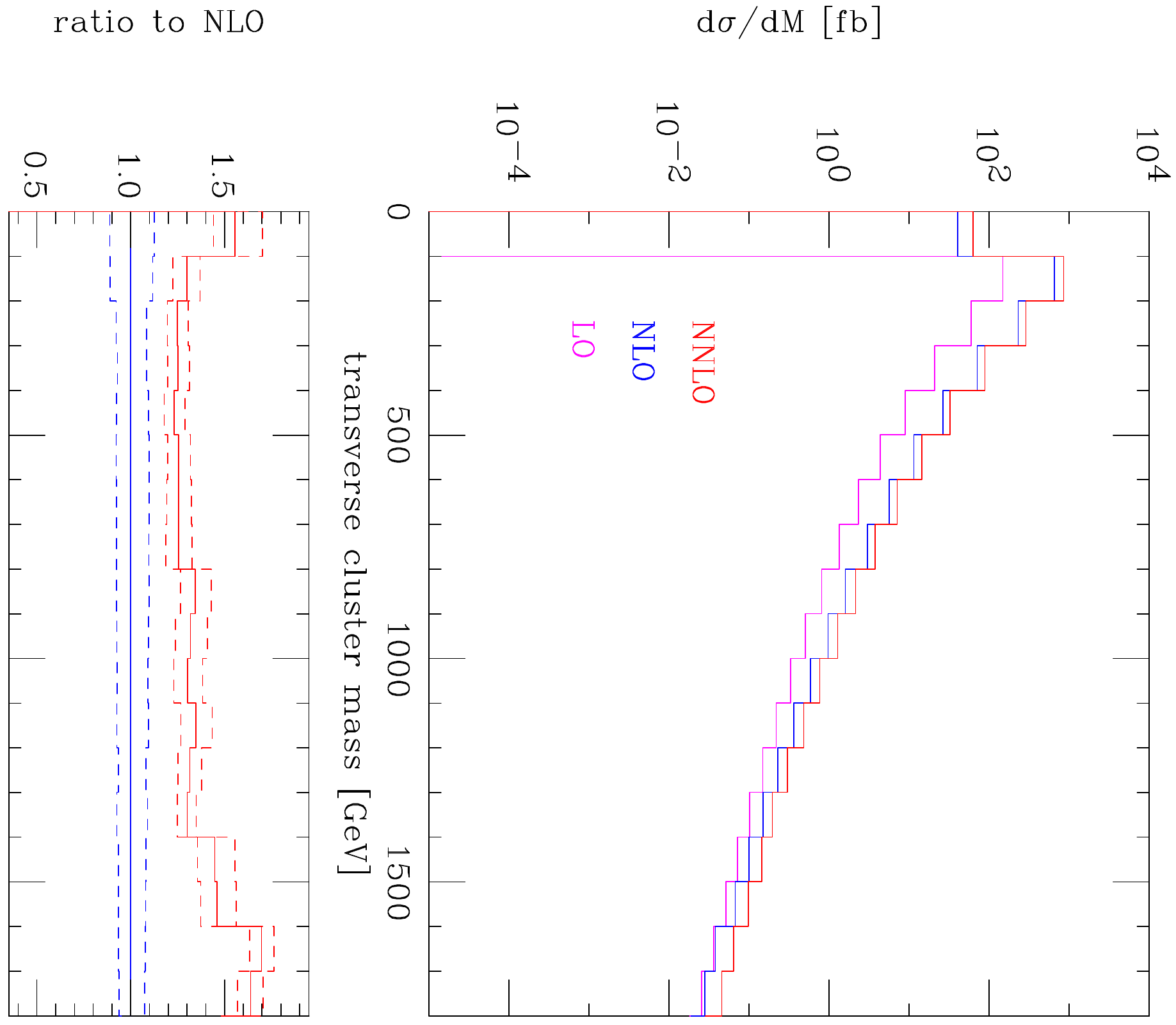}
\caption{Cluster transverse mass of the ($\ell$--$\gamma$--$\nu$) system,
defined in Eq.~(\ref{eq:TCM}), at 13 TeV.
The lower panel displays the ratio of the NLO and NNLO predictions, including
error estimates obtained via a 9-point scale variation, to the central NLO result.}
\label{tcmass}
\end{center}
\end{figure}

Lastly, we turn to two distributions that directly probe a wide range of energy scales:
the photon $p_T$ (Figure~\ref{ptgamma}) and the transverse cluster mass
distribution (Figure~\ref{tcmass}).  Although we have seen that the net effect of
electroweak corrections to the rate in the fiducial volume is small, these distributions are
particularly sensitive to their effects.  

\subsection{Numerical results for electroweak corrections}
\label{Numerical_Electroweak}

We illustrate the size of electroweak corrections that can be expected under this set of cuts by considering
their effect on the $p_T^\gamma$ distribution.  We plot
this distribution out to values of $1$~TeV since this corresponds,
under these cuts, to about $200$ $\ell^\pm \nu \gamma$ events in the full
3~ab${}^{-1}$ HL-LHC data set.

\begin{figure}[t]
\begin{center}
\includegraphics[angle=90,width=0.7\textwidth]{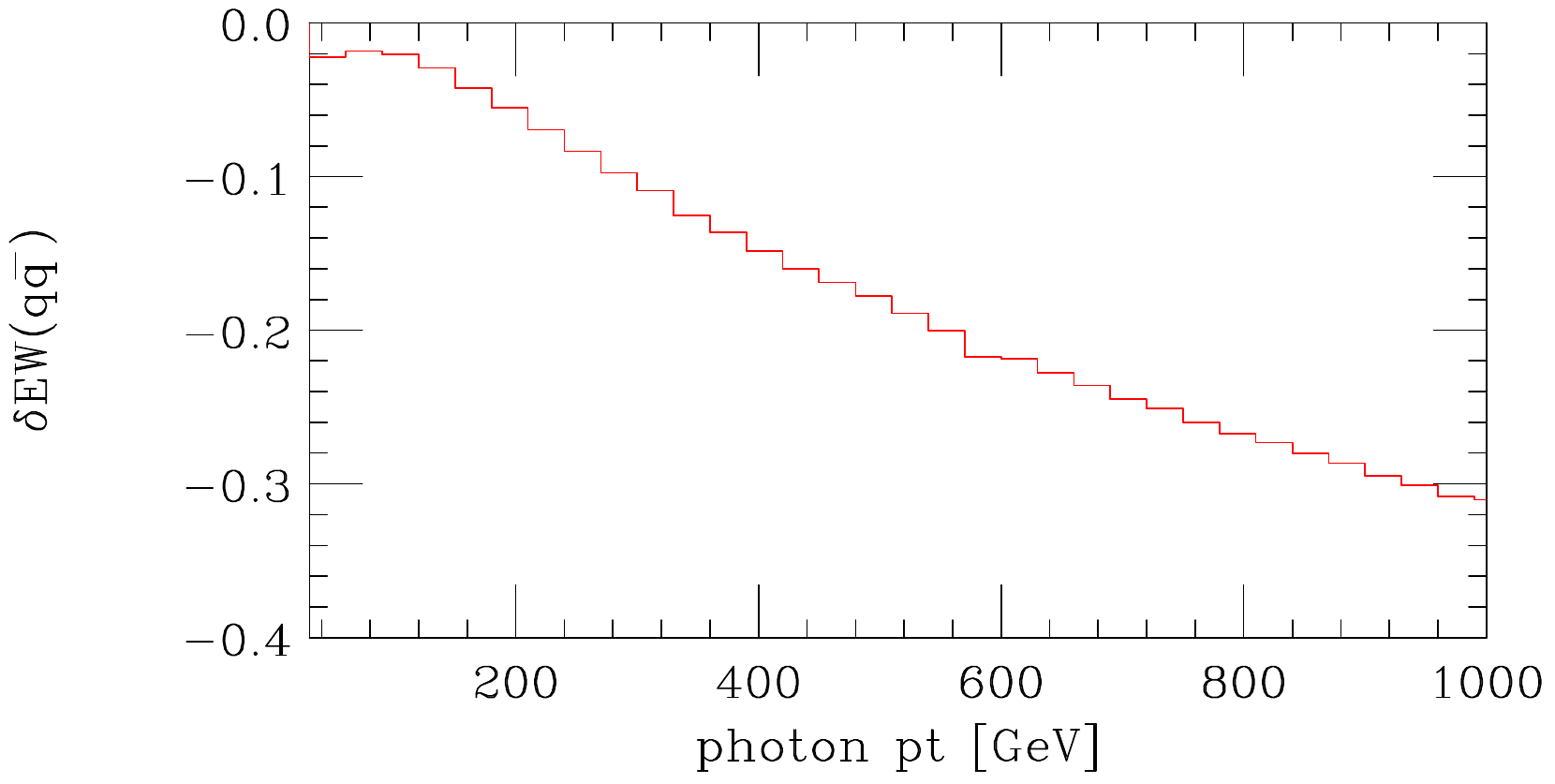}
\renewcommand\baselinestretch{1.2}
\caption{Relative electroweak corrections to the $q\bar q$-initiated process in Eq.~(\ref{udbarprocess}).}
\label{fig:ptgamma_relew_qqbar}
\end{center}
\end{figure}
\begin{figure}[t]
\begin{center}
\includegraphics[angle=90,width=0.7\textwidth]{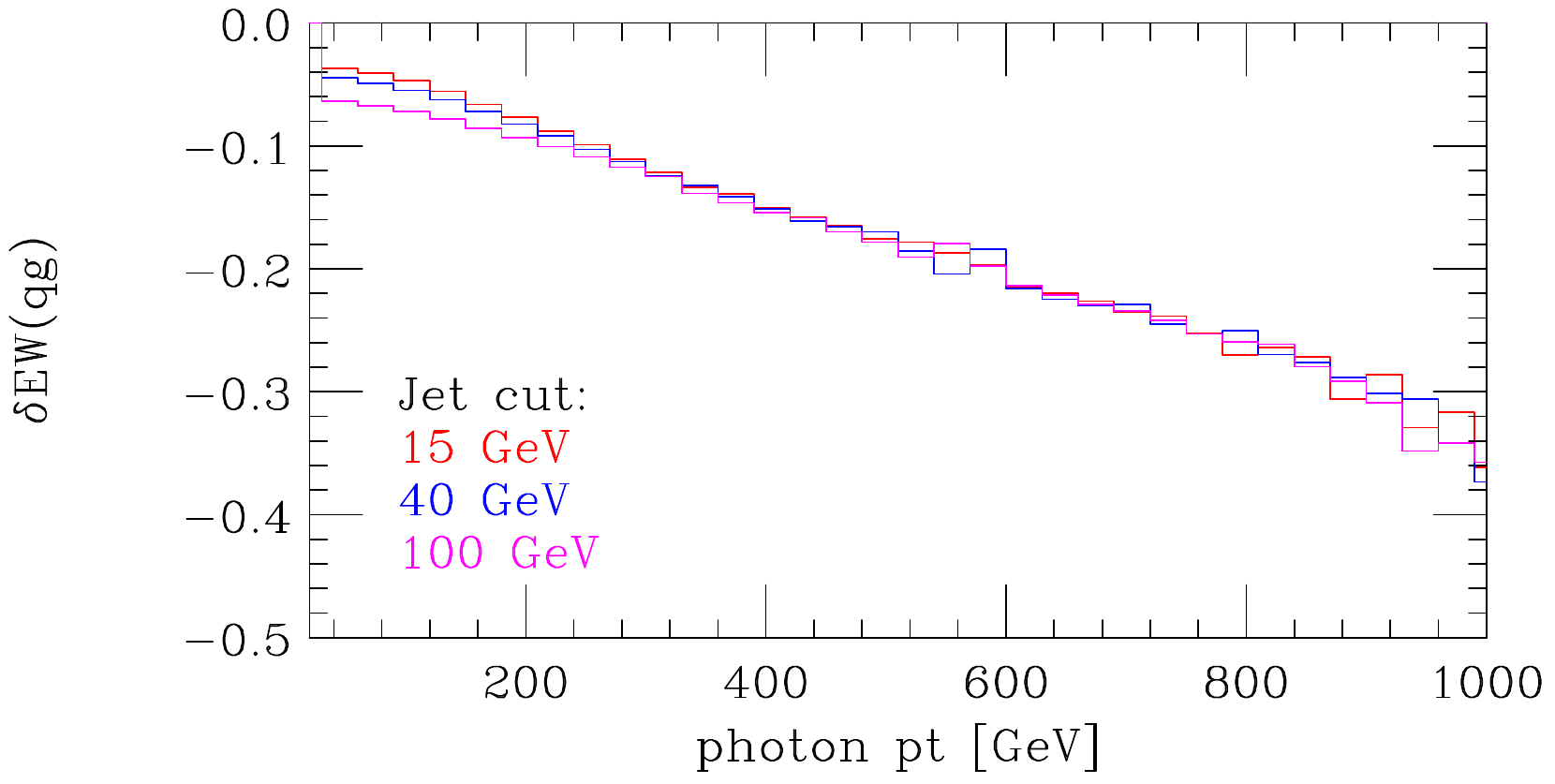}
\caption{Relative electroweak corrections to the $O(\alpha_s)$ process in Eq.~(\ref{qgprocess}), for three
values of the minimum jet $p_T$.}
\label{fig:ptgamma_relew_qqbarjet}
\end{center}
\end{figure}

Corrections from the process in Eq.~(\ref{udbarprocess}) are shown in Fig.~\ref{fig:ptgamma_relew_qqbar}, as a factor
relative to the LO process (c.f. Eq.~(\ref{eq:EWdelta})).  Overall the effect on the total rate is,
\begin{equation}
\label{eq:dEWqqbardiff}
\delta_{EW}^{q\bar q} = -0.013 \,, \;\;
\end{equation}
but the corrections to the $p_T^\gamma$ distribution are much more significant in the tail. 
The size of the corrections is very similar to that already observed, albeit under slightly
different cuts, in Ref.~\cite{Denner:2014bna}.
Relative corrections from the process in Eq.~(\ref{qgprocess}) are shown in Fig.~\ref{fig:ptgamma_relew_qqbarjet},
for three different choices of the jet $p_T$ threshold -- $15$, $40$ and $100$~GeV --  normalized to the 
$O(\alpha_s)$ (leading order) result for the $W\gamma$-jet process.  Although the size of the
corrections differs in the first few bins of this distribution, for $p_T^\gamma > 200$~GeV all three
curves are similar.  This indicates that, although the effect on the inclusive $W\gamma$ rate is hard to
estimate, corrections from this channel are important at large $p_T^\gamma$ and should be taken into account.
The similarity of the electroweak corrections between the $O(1)$ (Fig.~\ref{fig:ptgamma_relew_qqbar})
and $O(\alpha_s)$ (Fig.~\ref{fig:ptgamma_relew_qqbarjet}) channels suggests that a
multiplicative approach to incorporating the effects of electroweak corrections into QCD-corrected predictions
correctly captures these effects, particularly at high energies.  An estimate of the mixed QCD-EW corrections,
that are not correctly captured in such a scheme, can be inferred from the difference between
Figs.~\ref{fig:ptgamma_relew_qqbar} and~\ref{fig:ptgamma_relew_qqbarjet}. A definitive statement cannot be made, of
course, until a proper calculation of the mixed QCD-EW corrections is performed.

\begin{figure}[t]
\begin{center}
\includegraphics[width=0.7\textwidth]{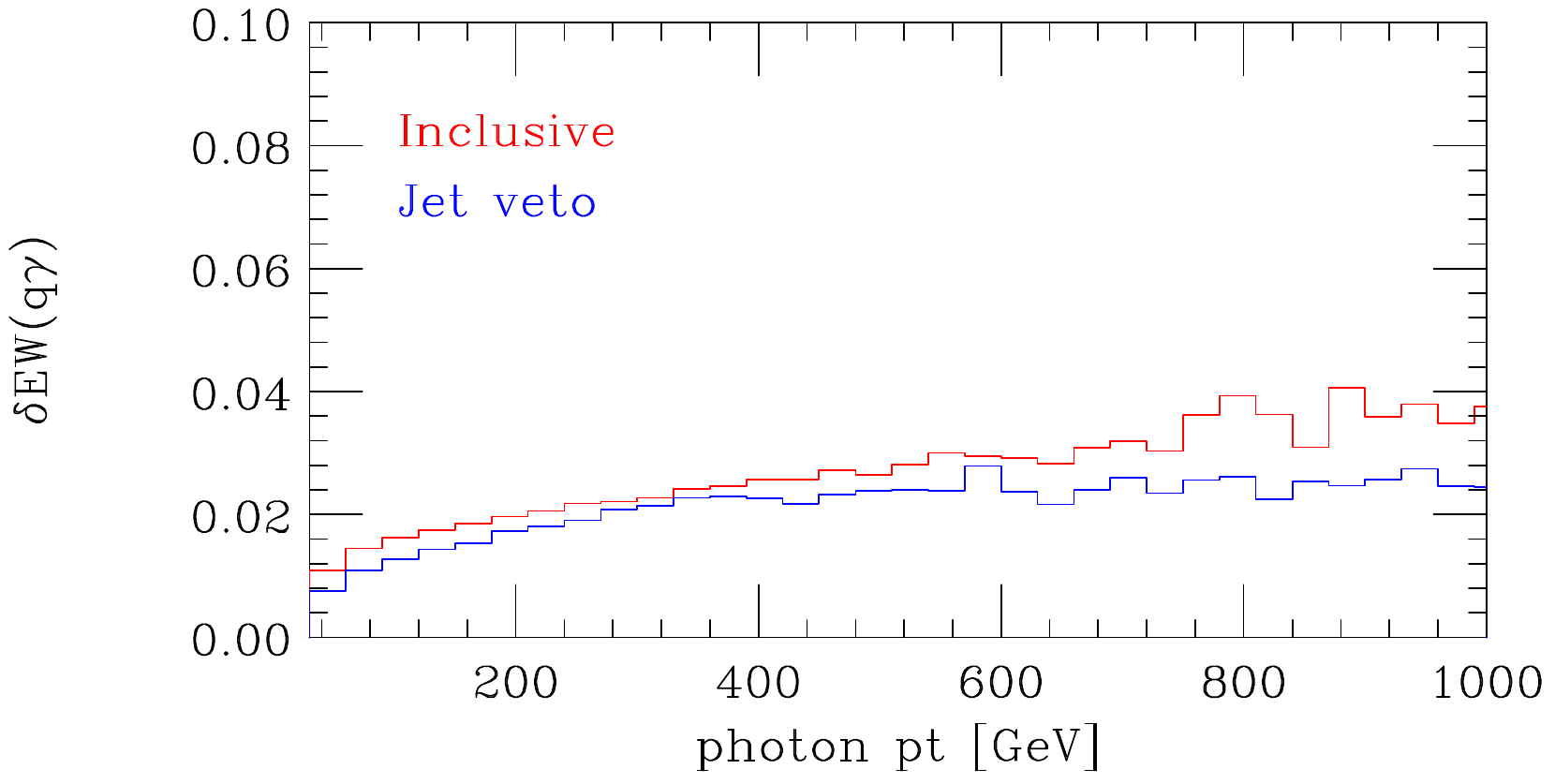}
\caption{Relative electroweak corrections resulting from the $q\gamma$-initiated process in Eq.~(\ref{gammauprocess}).}
\label{fig:ptgamma_relew_gq}
\end{center}
\end{figure}

Results for $\delta^{q\gamma}_{EW}$, resulting from the $q \gamma \to W \gamma q$-channel
in Eq.~(\ref{gammauprocess}),
are shown in Fig.~\ref{fig:ptgamma_relew_gq}.  Under these cuts the effect on the 
full rate is,
\begin{equation}
\label{eq:dEWqgammadiff}
\delta_{EW}^{q\gamma} = +0.013 \,, \;\;
\end{equation}
although in the distribution this manifests as a $4\%$ enhancement for $p_T^\gamma = 1$~TeV.
As expected the application of a jet veto, with jets defined according to Eq.~(\ref{jetdefndiff}),
somewhat reduces the size of the corrections, especially at large $p_T^\gamma$, and 
the size of the electroweak corrections to the overall rate becomes,
\begin{equation}
\label{eq:dEWqgammadiffveto}
\delta_{EW}^{q\gamma}({\rm jet~veto}) = +0.009 \,. \;\;
\end{equation}
Although this channel results in an enhancement of the cross-section we note that, compared to previous
calculations of these effects~\cite{Denner:2014bna}, the size of the
corrections is much reduced.  This is partially due to the fact that here we have normalized to NNLO
predictions;  replacing the denominator in Eq.~(\ref{eq:EWdeltaqgamma}) with the LO result would increase
$\delta_{EW}^{q\gamma}$ by about a factor of $4$.  The remaining large difference with
respect to the results of Ref.~\cite{Denner:2014bna} simply reflects the improved
determination of the photon distribution in the LUX pdf
set~\cite{Manohar:2016nzj,Manohar:2017eqh} compared to
NNPDF2.3QED~\cite{Ball:2013hta}, the pdf set used in
Ref.~\cite{Denner:2014bna}.  The photon pdf in the LUX determination
is significantly smaller at large $x$ than the central result in
NNPDF2.3QED, although the two are compatible within the (large)
uncertainties of the latter.
We find that, after factorization of singularities into the pdfs, corrections from the process in
Eq.~(\ref{gammagprocess}) are negligible, below the per-mille level, across all of the kinematic range.
This is due to the fact that this channel does not open a new and significant kinematic configuration, unlike
the processes in Eqs.~(\ref{qgprocess}) and~(\ref{gammauprocess}).  This lends further credence to the
suggestion that corrections from photon-induced channels should be applied additively, i.e. according to
Eq.~(\ref{eq:EWdeltaqgamma}).
 
Finally, we note that application of a jet veto will reduce the effect of all these corrections,
as demonstrated explicitly in Fig.~\ref{fig:ptgamma_relew_gq} for the $q\gamma$-initiated contributions.  This merely
indicates that, especially for the case of electroweak corrections, the effect of higher orders depends
delicately on the experimental measurement for which the theoretical prediction is being made.

\pagebreak
 
\section{Conclusions}
We have presented NNLO results for the processes $pp\to e^+\nu_e \gamma$ and 
$pp\to e^-\bar{\nu}_e \gamma$ calculated using a jettiness slicing scheme. This allows us to
construct the cross section for an electron of either charge, accompanied by a photon, 
\beq
\sigma = \sigma(pp\to e^+\nu_e \gamma) + \sigma(pp\to e^-\bar{\nu}_e \gamma)
\eeq
which is the quantity usually quoted by experiment. 
Our analytic formulae are valid for the process $pp\to e^+\nu_e \gamma$. The extension to 
the opposite charge process $pp\to e^-\bar{\nu}_e \gamma$ can be performed by exchange, 
and, since the leptons are considered to be massless, the extension to $\mu^+$ and $\mu^-$ is
immediate. 

Our numerical results show full agreement with the results of the
MATRIX collaboration, performed using a different slicing method. We
have included the effects of CKM rotation, which are found to be
numerically small.  Generically we can see that the NNLO effects are large and (in the main) positive, 
(e.g. at $\sqrt{s}=13$~TeV they are about $20\%$), and that the scale dependence at NLO
gives little hint that such a correction is to be expected. However the destructive interference
in the lowest order contribution does lead one to suspect that the NNLO calculation for $W\gamma$  
might behave more like an NLO correction than an NNLO correction, a suspicion that is borne out
by our detailed calculations. 

Our results indicate that the NNLO effects can play a substantial role
in the description of the radiation zero at $\sqrt{s}=13$~TeV. The
clearest signature of the radiation zero requires a veto on jet
activity.  It is well-known that such vetoes can generate large
logarithms, suggesting that a more accurate description of the
distributions might benefit from a resummation of large logarithms, as
was performed for the $Z\gamma$ process in Ref.~\cite{Becher:2020ugp}.
   
We have also considered electroweak effects of various sources. As far as the total cross section
is concerned we find that the decrease in the cross section due to the replacement of one power
of $\alpha$ changes the cross section by a factor $\alpha(0)/\alpha(M_Z^2)\approx 0.965$. The incoming photon
$\gamma q$ process corrects the lowest order process by 0.9--1.3\%, depending on the
photon $p_T$ cut and the presence (or not) of a jet veto,
c.f. Eqs.~(\ref{eq:dEWqgamma}),~(\ref{eq:dEWqgammadiff}) and~(\ref{eq:dEWqgammadiffveto}).
In contrast the $\gamma g$ process gives
a negligible effect. Virtual and real photon emission corrections to the total cross section
can only be evaluated for the $q\bar{q}$ process and are negative and give about $-1.3\%$,
c.f. Eqs.~(\ref{eq:dEWqqbar}) and~(\ref{eq:dEWqqbardiff}).
As is well known the virtual and real photon emission corrections become large at high $p_T$, as much as
$-30\%$ and $+4\%$ respectively, at photon $p_T = 1$~TeV.

We are of the opinion that, compared to the $Z\gamma$ process,  the $W\gamma$ process
has not as yet received the attention from experimenters it deserves. 
When decays to final state leptons are taken
into account the $Z\gamma$ and $W\gamma$ processes have about the same cross section.   
Of course the $Z\gamma$ process has a final state without missing energy, and it is also 
of interest because of its role in the search for the rare Higgs boson decay, $H \to Z\gamma$. 
However the importance of the triple weak boson coupling in the $W\gamma$ process should not be 
underestimated.

\section*{Acknowledgements}
We acknowledge useful discussions with Heribertus Hartanto, Zolt{\'a}n
Kunszt, Tobias Neumann and Ciaran Williams. We thank Daniel Ma\^itre
for the use of his code to generate spinor helicity ans\"atze. This
document was prepared using the resources of the Fermi National
Accelerator Laboratory (Fermilab), a U.S. Department of Energy, Office
of Science, HEP User Facility. Fermilab is managed by Fermi Research
Alliance, LLC (FRA), acting under Contract No. DE-AC02-07CH11359.  The
numerical calculations reported in this paper were performed using the
Wilson High-Performance Computing Facility at Fermilab and the
Vikram-100 High Performance Computing Cluster at Physical Research
Laboratory.

\pagebreak
\appendix
\section{Spinor algebra}
\label{spinorsection}
All results are presented using the
standard notation for the kinematic invariants of the process,
\begin{equation}
s_{ij} = (p_i+p_j)^2 \, ,
s_{ijk} = (p_i+p_j+p_k)^2 \, ,
s_{ijkl} = (p_i+p_j+p_k+p_l)^2 \,.
\end{equation}
and the Gram determinant,
\begin{equation} \label{Delta3eqn}
\DeltaTM{i}{j}{k}{l} =(s_{ijkl}-s_{ij}-s_{kl})^2-4 s_{ij} s_{kl}  \, .
\end{equation}
We express the amplitudes in terms of spinor products defined as,
\begin{equation}
\label{Spinor_products1}
\spa i.j=\bar{u}_-(p_i) u_+(p_j), \;\;\;
\spb i.j=\bar{u}_+(p_i) u_-(p_j), \;\;\;
\spa i.j \spb j.i = 2 p_i \cdot p_j,\;\;\;
\end{equation}
and we further define the spinor sandwiches for light-like momenta $j$ and $k$,
\begin{eqnarray}
\label{Spinor_products2}
\spab{i}.{(j+k)}.{l} &=& \spa{i}.{j} \spb{j}.{l} +\spa{i}.{k} \spb{k}.{l}, \nonumber \\
\spba{i}.{(j+k)}.{l} &=& \spb{i}.{j} \spa{j}.{l} +\spb{i}.{k} \spa{k}.{l}.
\end{eqnarray}
In the Weyl representation the spinor solutions of the massless Dirac equation are,
\begin{equation} \label{eq:explicitspinor}
u_+(p) = |p \rangle =
  \left[ \begin{matrix} \sqrt{p^+} \cr 
   \sqrt{p^-} e^{i\varphi_p} \cr
                  0 \cr 
   0 \cr  \end{matrix}\right] , \hskip3mm
u_-(p) = |p] =
  \left[ \begin{matrix} 0 \cr 
                  0 \cr
                \sqrt{p^-} e^{-i\varphi_p} \cr 
                                 -\sqrt{p^+} \cr  \end{matrix}\right] , 
\end{equation}
where 
\begin{equation} \label{eq:phasekdef}
e^{\pm i\varphi_p}\ \equiv\ 
  \frac{ p^1 \pm ip^2 }{ \sqrt{(p^1)^2+(p^2)^2} }
\ =\  \frac{ p^1 \pm ip^2 }{ \sqrt{p^+p^-} }\ ,
\qquad p^\pm\ =\ p^0 \pm p^3.  
\end{equation}
In this representation the Dirac conjugate spinors are,
\begin{equation}
\label{eq:explicitspinorconjg}
\bar{u}_{+}(p) = [p| \equiv  u_{+}^\dagger(p) \gamma^0 =
  \left[ 0, 0, \begin{matrix} \sqrt{p^+} , 
   \sqrt{p^-} e^{-i\varphi_p}  \cr  \end{matrix}\right] ,
\end{equation}
\begin{equation}
\bar{u}_{-}(p) = \langle p| \equiv  u_{-}^\dagger(p) \gamma^0 =
  \left[ \sqrt{p^-} e^{i\varphi_p}, -\sqrt{p^+}, 0,0 \right] .
\end{equation}

\section{Integral functions in the amplitudes}

Due to the linear vanishing of $\ln(r)$ as $r\rightarrow 1$ it is convenient to introduce the $\Ll_0$ and $\Ll_1$ functions \cite{Bern:1993mq} in order to  make explicit the absence of certain singularities
\begin{equation}
  \Ll_0(r) = \frac{\ln(r)}{1-r} \; , \quad \quad \Ll_1(r) = \frac{\Ll_0(r)+1}{1-r} = \frac{\ln(r)}{(1-r)^2} + \frac{1}{1-r}\, .
\end{equation}
In particular, $r \rightarrow \{0, 1, \infty\}$ are the three physically relevant limits, which can be respectively written as the following Maclaurin series for $\Ll_0$
\begin{eqnarray}
\lim_{r\rightarrow0} \; \Ll_0(r) & \approx & \vphantom{\frac11} \ln(r) + r \, \ln(r) + \dots \; , \\
\lim_{x\rightarrow0} \; \Ll_0(1+x) & \approx & -1 + \frac{x}{2} - \frac{x^2}{3} + \dots \; , \\
\lim_{y\rightarrow0} \; \Ll_0(1/y) & \approx & \vphantom{\frac11} - y \, \ln(1/y) - y^2 \, \ln(1/y) + \dots \; ,
\end{eqnarray}
and for $\Ll_1$
\begin{eqnarray}
  \lim_{r\rightarrow0} \; \Ll_1(r) & \approx & \vphantom{\frac11} (1 + \ln(r)) + r \, (1 + 2 \ln(r)) + \dots \; , \\
  \lim_{x\rightarrow0} \; \Ll_1(1+x) & \approx & -\frac{1}{2} + \frac{x}{3} - \frac{x^2}{4} + \dots \; , \\
  \lim_{y\rightarrow0} \; \Ll_1(1/y) & \approx & \vphantom{\frac11} -y  + y^2 (-1+\ln(1/y)) + \dots \; .
\end{eqnarray}
Both $\Ll_0$ and $\Ll_1$ display a logarithmic divergence for $r\rightarrow 0$, are regular for $r\rightarrow 1$ and vanish linearly for $r\rightarrow \infty$. Let us consider the commonly seen case where $r = s_{ijk}/s_{ij}$. Then, the three limits $r\rightarrow \{0, 1, \infty\}$ correspond respectively to $\{s_{ijk}, \langle k|i+j|k], s_{ij}\} \rightarrow 0$, due to the following relation
\begin{equation}
  s_{ijk} = s_{ij} + s_{ik} + s_{jk} = s_{ij} + \langle k|i+j|k] \; .
\end{equation}
Sums of Mandelstam variables of the form $\langle k|i+j|k]$ appear as poles in scalar bubble integral coefficients. However, these poles are spurious when considering complete one-loop amplitudes. We can use the $\Ll_0$ and $\Ll_1$ functions to explicitly remove them. If $\langle k|i+j|k]$ appears as a simple pole we can write
\begin{equation}
\frac{\Ll_0(\frac{-s_{ijk}}{-s_{ij}})}{s_{ij}} = - \ln(\frac{-s_{ijk}}{-s_{ij}}) \frac{1}{\langle k|i+j|k]} \, ,
\end{equation}
which is regular in $\langle k|i+j|k]\rightarrow 0$ and logarithmically divergent for $\{s_{ijk}, s_{ij}\} \rightarrow 0$. If $\langle k|i+j|k]$ appears as a double pole, then there will be a corresponding rational piece where it appears as a simple pole. This allows to write the following expression
\begin{equation}
\frac{\Ll_1(\frac{-s_{ijk}}{-s_{ij}})}{s_{ij}} = \ln(\frac{-s_{ijk}}{-s_{ij}}) \frac{s_{ij}}{\langle k|i+j|k]^2} - \frac{1}{\langle k|i+j|k]} \, ,
\end{equation}
which is regular in $\{ \langle k|i+j|k], s_{ij}\} \rightarrow 0$ and logarithmically divergent for $s_{ijk} \rightarrow 0$. Alternatively the following combination may also appear
\begin{equation}
\frac{\Ll_1(\frac{-s_{ij}}{-s_{ijk}})}{s_{ijk}} = \ln(\frac{-s_{ij}}{-s_{ijk}}) \frac{s_{ijk}}{\langle k|i+j|k]^2} + \frac{1}{\langle k|i+j|k]} \, ,
\end{equation}
which is regular for $\{ \langle k|i+j|k], s_{ijk}\} \rightarrow 0$ and logarithmically divergent for $s_{ij} \rightarrow 0$.

The following transcendental functions are also needed
\begin{eqnarray}
  \Ls_{-1}(r_1,r_2) &=&
      \Li_2(1-r_1) + \Li_2(1-r_2) + \ln r_1\,\ln r_2 - {\pi^2\over6}\,,\cr
  \Ls_0(r_1,r_2) &=&  {1\over (1-r_1-r_2)}\, \Ls_{-1}(r_1,r_2)\,,\cr
  \Ls_1(r_1,r_2) &=& {1\over (1-r_1-r_2)}\, 
  \LB \Ls_0(r_1,r_2) + \Ll_0(r_1)+\Ll_0(r_2)\RB\,,
\end{eqnarray}
\begin{eqnarray}
  &&\Ls^{2{\rm m}h}_{-1}(s,t;m_1^2,m_2^2) =
    -\Li_2\left(1-{m_1^2\over t}\right)
    -\Li_2\left(1-{m_2^2\over t}\right)
    -{1\over2}\ln^2\left({-s\over-t}\right) \nonumber \\
    &+&{1\over2}\ln\left({-s\over-m_1^2}\right)
              \ln\left({-s\over-m_2^2}\right)
    + \biggl[ {1\over2} (s-m_1^2-m_2^2) + {m_1^2m_2^2\over t} \biggr]
        I_3^{3{\rm m}}(s,m_1^2,m_2^2) \, ,
\end{eqnarray}

\begin{eqnarray}
\Lsnew^{2{\rm m}h}_{-1}(s,t;m_1^2,m_2^2) &=&
    -\Li_2\left(1-{m_1^2\over t}\right)
    -\Li_2\left(1-{m_2^2\over t}\right)
    -{1\over2}\ln^2\left({-s\over-t}\right)\nonumber \\
    &+&{1\over2}\ln\left({-s\over-m_1^2}\right)
              \ln\left({-s\over-m_2^2}\right)\,,
\end{eqnarray}
\begin{eqnarray}
  \Ls^{2{\rm m}e}_{-1}(s,t;m_1^2,m_3^2) &=& 
    -\Li_2\left(1-{m_1^2\over s}\right)                      
    -\Li_2\left(1-{m_1^2\over t}\right)                      
    -\Li_2\left(1-{m_3^2\over s}\right)                      
    -\Li_2\left(1-{m_3^2\over t}\right) \cr                     
&&\quad
    +\Li_2\left(1-{m_1^2m_3^2\over st}\right)
    -{1\over2}\ln^2\left({-s\over-t}\right) \; .
\end{eqnarray}
where the dilogarithm is
\beq
\Li_2(x) = - \int_0^x dy \, {\ln(1-y) \over y}\,.
\eeq
and $I_3^{3{\rm m}}$ is a three-mass scalar triangle integral, defined
according to Appendix II of Ref.~\cite{Bern:1997sc},
\begin{equation}
I_3^{3{\rm m}}(s_{12},s_{34},s_{56}) =
\int_0^1 da_1 \, \int_0^1 da_2 \, \int_0^1 da_3 \, 
\frac{\delta(1-a_1-a_2-a_3)}{[-s_{12} a_1 a_2 - s_{34} a_2 a_3 - s_{56} a_3 a_1+i\ep]} .
\end{equation}
Note that this has the opposite sign to commonly-used definitions of
scalar integrals, such as in QCDLoop~\cite{Ellis:2007qk}.

\section{Six-parton process at one-loop order}
\label{Oneloopappendix}
\subsection{Radiation for $u$ and $d$ quarks}
The one-loop corrections to the process
\begin{equation}
0 \rightarrow \bar u (p_1) + d(p_2) + \nu_e (p_3) + e^+ (p_4)  + g (p_5) + g(p_6)
\end{equation}
have been presented in Ref.~\cite{Bern:1997sc}.
Although certain terms that we need can be derived from these results, that is not true for all terms,
so we present the full analytic terms here.
A computer readable representation of the results in this
appendix accompanies the arXiv version of this article.
\subsubsection{Leading colour, $5_{\gamma}^+, 6_g^+$}
\begin{eqnarray}
 && A^u_\lc  (5_{\gamma}^+, 6_g^+ ) = 
      \frac{\spa2.3^2 \spb4.3}{\spa1.5 \spa2.6 \spa5.6}
          \, \LsminOne{-s_{15}}{-s_{156}}{-s_{16}}
 \nonumber \\ &&
      +\frac{\spa1.2 \spa2.3^2 \spb3.4}{\spa1.5 \spa1.6 \spa2.5 \spa2.6}
          \, \left(\LsminOneTME{s_{126}}{s_{156}}{s_{16}}{s_{34}}
          +\LsminOne{-s_{16}}{-s_{126}}{-s_{26}}\right)
 \nonumber \\ &&
      +  \frac{\spa1.2   \spb4.5}{2 \spa1.5 \spa1.6 \spa2.6}
       \left[ 3 \, \spa2.3\,\Lzero{ - s_{126}}{ - s_{34}}
      + \spa3.5 \spab2.1+6.5
       \,\frac{\Lone{ - s_{126}}{ - s_{34}}}{s_{34}}  \right]
 \nonumber \\ &&
      + \frac{1}{2} \frac{\spa1.2 \spb1.6 }{\spa1.5 \spa1.6 \spa2.5} \Bigg(
	 \spa2.3  (3 \spa1.3 \spb3.4 - 2 \spa1.5 \spb4.5)
	  -\spa1.3 \spab2.1+5.4 \Bigg)
       \,\frac{\Lzero{ - s_{126}}{ - s_{26}}}{s_{26}} 
 \nonumber \\ &&
      - \frac{\spa1.2^2 \spa1.3 \spb1.6 \spb1.4}{2 \spa1.5 \spa1.6 \spa2.5}
       \,\frac{\Lone{ - s_{126}}{ - s_{26}}}{s_{26}}
      + \frac{\spa1.2^2 \spa2.3 \spb1.4}{2 \spa1.5 \spa1.6 \spa2.5 \spa2.6}
 \nonumber \\ &&
      + \frac{\spa1.2 \spa2.3  \spb4.5}{\spa1.5 \spa1.6 \spa2.6}
      + \frac{\spa2.3 \spb5.6 \spab1.2+3.4}{2 \spa1.5 \spa1.6  s_{234}}
 \nonumber \\ &&
      + \Vpole_\lc(s_{126}) \, A^u_\tree  ( 1_{\bar u}^+,2_{d}^-,3_{\nu}^-,4_{\ell}^+, 5_{\gamma}^+, 6_g^+ )
\end{eqnarray}
 
In this formula the term containing the poles in $\epsilon$ is given by,
\begin{equation}
V_\lc(s_x) = -\frac{1}{\epsilon^2} \left[
 \left( \frac{\mu^2}{-s{_{16}}} \right)^\epsilon
+\left( \frac{\mu^2}{-s{_{26}}} \right)^\epsilon \right]
-\frac{3}{2\epsilon}
 \left( \frac{\mu^2}{-s_x} \right)^\epsilon 
- 3 \, ,
\end{equation}
and the corresponding leading-order subamplitude has been given in Eq.(\ref{eq:LOpp-Qu}).

\begin{eqnarray}
 && A^d_\lc (5_{\gamma}^+, 6_g^+ ) = \nonumber \\ && 
      - \frac{\spa1.2^3 \spa3.5^2 \spb4.3}{\spa1.5^3 \spa1.6 \spa2.5 \spa2.6}
         \, \LsminOneTME{s_{126}}{s_{256}}{s_{26}}{s_{34}}
 \nonumber \\ &&
      +\frac{\left(\spa2.3 \spa5.6 (\spa2.5 \spa3.6+\spa2.6 \spa3.5)
      -(\spa2.5 \spa3.6)^2\right) \spb4.3}
      {\spa1.6 \spa2.5 \spa5.6^3}
         \, \LsminOne{-s_{26}}{-s_{256}}{-s_{25}}
 \nonumber \\ &&
      - \frac{\spa1.2^3 \spb1.4 \spab3.2+6.1}{2 \spa1.6 \spa1.5 \spa2.6 \spa2.5}
         \, \frac{\Lone{ - s_{126}}{ - s_{26}}}{s_{26}} 
      + \frac{\spa1.2 \spa2.5 \spb4.5^2 \spa4.3}{2 \spa1.6 \spa1.5 \spa2.6}
         \, \frac{\Lone{ - s_{126}}{ - s_{34}}}{s_{34}} 
 \nonumber \\ &&
      + \frac{\spa1.2^3 \spb1.4^2 \spa4.3}{2 \spa1.6 \spa1.5 \spa2.6 \spa2.5}
         \, \frac{\Lone{ - s_{134}}{ - s_{34}}}{s_{34}} 
      - \frac{\spa2.6 \spa3.5^2 \spb5.6^2 \spb4.3}{2 \spa1.5 \spa5.6}
         \, \frac{\Lone{ - s_{26}}{ - s_{256}}}{s_{256}^2}
 \nonumber \\ &&
      - \frac{\spa2.5 \spa3.6^2 \spb5.6^2 \spb4.3}{2 \spa1.6 \spa5.6}
         \, \frac{\Lone{ - s_{25}}{ - s_{256}}}{s_{256}^2} 
      -  \frac{\spa1.2^2 \spb1.6 \spb4.3}{\spa1.5 \spa1.6 \spa2.5} \left(
           \frac{\spa1.3 \spa3.5}{\spa1.5}
         + \frac{\spab3.1+5.4}{2 \spb3.4}
         \right) \, \frac{\Lzero{ - s_{126}}{ - s_{26}}}{s_{26}} 
 \nonumber \\ &&
      +  \frac{\spa1.2 \spb4.5 (\spa1.2 \spa3.5 - \spa2.3 \spa1.5)}{\spa1.5^2 \spa1.6 \spa2.6}        
         \, \Lzero{ - s_{126}}{ - s_{34}}  
      - \frac{\spa1.2^3 \spa3.5 \spb1.4}{ \spa1.6 \spa1.5^2 \spa2.6 \spa2.5}
         \, \Lzero{ - s_{134}}{ - s_{34}} 
 \nonumber \\ &&
      +  \frac{\spa3.5^2 \spb5.6 \spb4.3 \left( \spa1.2\spa5.6 + \spa1.5\spa2.6 \right)}{\spa1.5^2 \spa5.6^2} 
         \, \frac{\Lzero{ - s_{256}}{ - s_{26}}}{s_{26}} 
      +  \frac{ \spa3.6 \spb5.6 \spb4.3 \left(\spa2.6 \spa3.5 - \spa2.3 \spa5.6 \right)}{\spa1.6 \spa5.6^2}
         \, \frac{\Lzero{ - s_{256}}{ - s_{25}}}{s_{25}} 
 \nonumber \\ &&
      - \frac{\spa1.2 \spa2.3 \spb4.6}{2 \spa1.6 \spa1.5 \spa2.5}
	\,\Lnrat{ - s_{126}}{ - s_{26}} 
      +  \frac{\spa1.2 \spa2.3 \spb4.3}{\spa1.5 \spa1.6 \spa2.6} \left(
           \frac{3 \spa2.3}{2 \spa2.5} 
         - \frac{\spa1.3}{\spa1.5}
         \right) \, \Lnrat{ - s_{256}}{ - s_{126}}  
 \nonumber \\ &&
      + \frac{\spa1.3 \spb5.6 \spab2.1+3.4}{2 \spa1.6 \spa1.5 s_{256}}
      - \frac{\spa1.2 \spa2.3 \spb4.5}{2 \spa1.6 \spa1.5 \spa2.6} 
\end{eqnarray}

\subsubsection{Subeading colour, $5_{\gamma}^+, 6_g^+$}
\begin{eqnarray}
 && A^u_\slc ( 5_{\gamma}^+, 6_g^+ ) = \frac{\spa2.5^2 \spa3.6^2 \spb3.4}
             {\spa1.5 \spa2.6 \spa5.6^3}
         \, \LsminOneTME{s_{125}}{s_{126}}{s_{12}}{s_{34}}
\nonumber \\ &&
         -\frac{\spa2.3^2 \spb3.4}{\spa1.6 \spa2.5 \spa5.6}
         \, \left(\LsminOneTME{s_{126}}{s_{156}}{s_{16}}{s_{34}}
         + \LsminOne{-s_{16}}{-s_{126}}{-s_{12}}\right)
 \nonumber \\ &&
       +\frac{\spa2.3^2 \spb3.4}{\spa1.5 \spa2.6 \spa5.6}
         \, \left(\LsminOneTME{s_{125}}{s_{156}}{s_{15}}{s_{34}}
         + \LsminOne{-s_{15}}{-s_{125}}{-s_{12}} \right)
  \nonumber \\ &&
      +\frac{\spa1.2^2 \spa3.6 (\spa1.2 \spb4.3 \spa3.6-\spa2.6 \spa1.5 \spb5.4)}
            {\spa1.5 \spa1.6^3 \spa2.5 \spa2.6}
         \, \LsminOne{-s_{12}}{-s_{126}}{-s_{26}}
 \nonumber \\ &&
      + \frac{\spa2.6 \spb4.6^2 \spa4.3}{\spa1.5 \spa5.6}
        \, \frac{\Lone{ - s_{125}}{ - s_{34}}}{s_{34}}
      - \frac{\spa1.2 \spa2.6 \spa3.6 \spb1.6 \spb4.6}{\spa1.6 \spa2.5 \spa5.6}
        \, \frac{\Lone{ - s_{126}}{ - s_{12}}}{s_{12}}
 \nonumber \\ &&
      + \frac{\spa1.2^2 \spa1.3 \spb1.6 \spb1.4}{2 \spa1.5 \spa1.6 \spa2.5}
        \, \frac{\Lone{ - s_{126}}{ - s_{26}}}{s_{26}}
      - \frac{\spa1.2 \spa2.5 \spb4.5^2 \spa4.3}{2 \spa1.5 \spa1.6 \spa2.6}
        \, \frac{\Lone{ - s_{126}}{ - s_{34}}}{s_{34}}
 \nonumber \\ &&
      - \frac{\spa1.2 \spa3.5^2 \spb1.5 \spb4.3}{\spa1.5 \spa5.6^2}
        \, \frac{\Lzero{ - s_{125}}{ - s_{12}}}{s_{12}}
      - \frac{\spa2.5 \spa3.6 \spb4.6}{\spa1.5 \spa5.6^2}
        \, \Lzero{ - s_{125}}{ - s_{34}}
 \nonumber \\ &&
      - \frac{\spa1.2 \spb1.6}{\spa1.6 \spa5.6}   \left(
           \frac{\spa1.3 \spa2.6^2 \spb2.4}{\spa1.6 \spa2.5}
         + \frac{\spa1.3 \spa2.6 \spb1.4}{\spa2.5}
         - \frac{\spa3.6 \spab2.1+3.4}{\spa2.5} 
         + \frac{\spa3.6^2 \spb3.4}{\spa5.6}
         \right)
	\, \frac{\Lzero{ - s_{126}}{ - s_{12}}}{s_{12}}
 \nonumber \\ &&
      - \frac{\spa1.2 \spb1.6}{\spa1.6 \spa2.5}   \left(
           \frac{\spa1.2 \spa1.3 \spa2.6 \spb2.4}{\spa1.5 \spa1.6 }
         + \frac{ \spa1.3 \spa2.3 \spb3.4}{2 \spa1.5}
         - \frac{3 \spa1.3 \spab2.1+5.4}{2 \spa1.5}
         - \spa2.3 \spb4.5
         \right)
	\, \frac{\Lzero{ - s_{126}}{ - s_{26}}}{s_{26}}
 \nonumber \\ &&
      + \frac{\spb4.5}{\spa1.5 \spa1.6} \left(
           \frac{\spa1.6 \spa2.5 \spa3.5}{\spa5.6^2}
         - \frac{\spa1.2 \spa2.3}{\spa2.6}
         \right)
	\, \Lzero{ - s_{126}}{ - s_{34}}
      + \frac{\spa2.3 \spa3.5 \spb4.3}{\spa1.5 \spa5.6^2}
        \, \Lnrat{ - s_{126}}{ - s_{125}}
 \nonumber \\ &&
      - \frac{\spa2.3 \spab1.2+3.4}{2 \spa5.6 s_{234}}   \left(
           \frac{\spb1.5}{\spa1.6} 
         - \frac{\spb1.6}{\spa1.5} 
         \right)
      - \frac{\spa2.3 \spab2.1+3.4}{2 \spa1.6 \spa2.5 \spa5.6}
 \nonumber \\ &&
      + \left( \Vpole_\slc + \frac{1}{2} \right) \, A^u_\tree  ( 1_{\bar u}^+,2_{d}^-,3_{\nu}^-,4_{\ell}^+, 5_{\gamma}^+, 6_g^+ )
\end{eqnarray}
 
The contribution that includes the poles in $\epsilon$ is,
\begin{equation}
V_\slc = \frac{1}{\epsilon^2} 
 \left( \frac{\mu^2}{-s{_{12}}} \right)^\epsilon
+\frac{3}{2\epsilon}
 \left( \frac{\mu^2}{-s{_{126}}} \right)^\epsilon 
+ 3 \, .
\end{equation}

\begin{eqnarray}
 && A^d_\slc ( 5_{\gamma}^+, 6_g^+ ) = 
        \frac{\spa2.6^2 \spa3.5^2 \spb4.3}
        {\spa1.6 \spa2.5 \spa5.6^3}
         \, \LsminOneTME{s_{126}}{s_{125}}{s_{12}}{s_{34}}
\nonumber \\ && -\frac{\spa1.2^2 \spa3.5^2 \spb4.3}{\spa1.5^2 \spa5.6}\left[
       \frac{1}{\spa1.5 \spa2.6}
         \, \LsminOneTME{s_{126}}{s_{256}}{s_{26}}{s_{34}}
      - \frac{1}{\spa1.6 \spa2.5 }
         \, \LsminOne{-s_{12}}{-s_{125}}{-s_{25}}\right]
 \nonumber \\ && +\frac{\spa1.2^2 \spa3.6}{\spa1.6^3 \spa2.5 \spa5.6}\left[
      \spa3.6 \spb4.3
         \, \LsminOneTME{s_{125}}{s_{256}}{s_{25}}{s_{34}}
      +\spab6.1+2.4 \, \LsminOne{-s_{12}}{-s_{126}}{-s_{26}}\right]
 \nonumber \\ &&
      - \frac{\spa2.5 \spa3.6^2 \spb5.6^2 \spb4.3}{2 \spa5.6 \spa1.6}
         \, \frac{\Lone{ - s_{25}}{ - s_{134}}}{s_{134}^2}
      + \frac{\spa2.6 \spa3.5^2 \spb5.6^2 \spb4.3}{2 \spa5.6 \spa1.5}
         \, \frac{\Lone{ - s_{26}}{ - s_{134}}}{s_{134}^2} 
      - \frac{\spa2.6^2 \spb4.6^2 \spa4.3}{\spa2.5 \spa5.6 \spa1.6}
         \, \frac{\Lone{ - s_{125}}{ - s_{34}}}{s_{34}}
 \nonumber \\ &&
      - \frac{\spa1.2^2 \spa1.3 \spb1.6 \spb1.4}{2 \spa2.5 \spa1.5 \spa1.6}
         \, \frac{\Lone{ - s_{126}}{ - s_{26}}}{s_{26}}
      - \frac{\spa2.5 \spa1.2 \spb4.5^2 \spa4.3}{2 \spa2.6 \spa1.5 \spa1.6}
         \, \frac{\Lone{ - s_{126}}{ - s_{34}}}{s_{34}}
 \nonumber \\ &&
      - \frac{\spa2.6^2 \spb4.6 \spab3.1+2.6}{\spa2.5 \spa5.6 \spa1.6}
         \, \frac{\Lone{ - s_{126}}{ - s_{12}}}{s_{12}}
      - \frac{\spa1.2^3 \spb1.4^2 \spa4.3}{2 \spa2.5 \spa2.6 \spa1.5 \spa1.6}
         \, \frac{\Lone{ - s_{134}}{ - s_{34}}}{s_{34}}
 \nonumber \\ &&
      + \frac{\spa1.2 \spa1.3^2 \spb1.5 \spb4.3}{\spa1.5 \spa1.6^2}
         \, \frac{\Lzero{ - s_{125}}{ - s_{25}}}{s_{25}}
      + \frac{\spa2.6 \spa3.6 \spb4.6 (\spa2.5 \spa1.6 - \spa1.2 \spa5.6)}{\spa1.6^2 \spa2.5 \spa5.6^2}
	 \, \Lzero{ - s_{125}}{ - s_{34}}
 \nonumber \\ &&
      + \frac{\spa3.5^2 \spa1.2 \spb1.5 \spb4.3}{\spa5.6^2 \spa1.5}
         \, \frac{\Lzero{ - s_{125}}{ - s_{12}}}{s_{12}}
      + \spb4.5 \left(
           \frac{\spa2.5 \spa3.5 \spa1.2}{\spa2.6 \spa5.6 \spa1.5^2}
         - \frac{\spa2.6 \spa3.5}{\spa5.6^2 \spa1.6}
         + \frac{\spa2.3 \spa1.2}{\spa2.6 \spa1.5 \spa1.6}
        \right)
	\, \Lzero{ - s_{126}}{ - s_{34}}
 \nonumber \\ &&
      +  \frac{\spa1.2 \spb1.6 \spb4.3}{\spa1.5 \spa1.6 \spa2.5} \left( 
           \frac{\spa2.6 \spa1.3 \spb4.6}{2 \spb3.4}
         + \frac{\spa3.5 \spa1.2 \spa1.3}{\spa1.5}
         - \frac{\spa3.6 \spa1.2 \spab1.2+6.4}{\spa1.6 \spb3.4}
         \right)
	 \, \frac{\Lzero{ - s_{126}}{ - s_{26}}}{s_{26}}
 \nonumber \\ &&
      + \frac{\spa2.6 \spa3.6 \spa1.2 \spb4.6}{ \spa1.6 \spa5.6} \left(
           \frac{\spb1.2}{\spa5.6}
         + \frac{\spab1.2+6.1}{\spa2.5 \spa1.6}
        \right)
	\, \frac{\Lzero{ - s_{126}}{ - s_{12}}}{s_{12}} 
      - \frac{\spa3.6^2 \spa1.2 \spb5.6 \spb4.3}{\spa5.6 \spa1.6^2}
        \, \frac{\Lzero{ - s_{134}}{ - s_{25}}}{s_{25}}
 \nonumber \\ &&
      - \frac{\spa3.5^2 \spa1.2 \spb5.6 \spb4.3}{ \spa5.6 \spa1.5^2}
        \, \frac{\Lzero{ - s_{134}}{ - s_{26}}}{s_{26}}
      + \frac{\spa1.2^2  \spb1.4}{\spa1.5 \spa1.6 \spa2.6} \left(
           \frac{\spa3.6 \spa1.2}{\spa2.5 \spa1.6}
         + \frac{\spa1.3}{\spa1.5}
        \right)
	\, \Lzero{ - s_{134}}{ - s_{34}}  
 \nonumber \\ &&
      + \frac{(\spa3.6 \spa1.2 \spa 5.6-\spa2.6 \spa1.6 \spa3.5)}{\spa2.5  \spa1.6^2 \spa5.6^2}
      \, \left( \spa1.2 \spb1.4 \, \Lnrat{ - s_{126}}{ - s_{12}}
         +\spa2.3 \spb4.3 \, \Lnrat{ - s_{126}}{ - s_{125}}\right)
 \nonumber \\ &&
      + \frac{\spa1.2 \spa2.3 \spb4.3}{\spa1.5 \spa1.6} \left(
           \frac{3\spa2.3}{2 \spa2.5 \spa2.6} 
         - \frac{\spa1.3}{\spa2.5 \spa1.6}
         - \frac{\spa1.3}{\spa2.6 \spa1.5}
        \right)
	\, \Lnrat{ - s_{126}}{ - s_{134}}
 \nonumber \\ &&
      + \frac{\spb5.6 \spb4.3}{\spa1.6 s_{134}}   \left(
           \frac{\spa2.6 \spa1.3 \spb1.4}{\spa5.6 \spb3.4}
         - \frac{\spa2.3 \spa3.6}{\spa5.6}
         + \frac{\spa1.3 \spab2.1+3.4}{2 \spa1.5 \spb3.4}
         \right)
      - \frac{\spa2.3 \spa1.2 \spab2.1+5.4}{2 \spa2.5 \spa2.6 \spa1.5 \spa1.6}
\end{eqnarray}
 
\subsubsection{Leading colour, $5_{\gamma}^-, 6_g^+$}
\begin{eqnarray}
 && A^d_\lc  ( 5_{\gamma}^-, 6_g^+ ) = s_{34} \Bigg[
 \nonumber \\ &&
       \Big(\frac{\spb1.4^2 \spa2.5^2}{s_{256} \spa2.6 \spb3.4 \spab6.2+5.1}
      +\frac{\spab3.2+5.6^2}{\spb2.5 s_{256} \spa3.4 \spab1.2+6.5}
      \Big)
         \, \LsminOneTMH{s_{16}}{s_{256}}{s_{25}}{s_{34}}
 \nonumber \\ &&
      -\frac{\spa1.2 \spab2.1+6.4^2}
      {\spa1.6 \spb3.4 \spa2.6 \spab1.2+6.5 \spab2.1+6.5}
         \, \LsminOneTMH{s_{25}}{s_{126}}{s_{16}}{s_{34}}
 \nonumber \\ &&
      +\frac{\spab3.2+5.6^2}{\spb2.5 s_{256} \spa3.4 \spab1.2+6.5}
         \, \LsminOne{-s_{25}}{-s_{256}}{-s_{26}}
 \nonumber \\ &&
      - \frac{\spab2.1+6.4 (\spa2.6 \spb6.1 \spa1.3-\spa1.2 \spb1.4 \spa3.4)}
        {\spab1.2+6.5 \spb1.5 \spa1.6 \spa2.6 s_{34}}
         \, \LsminOne{-s_{16}}{-s_{126}}{-s_{26}}
 \nonumber \\ &&
      + \Big( \frac{2 \spa2.5^2 \spab3.2+5.6 \spb1.4}{\spa6.2 s_{256}^2}
      +\frac{\spa1.2}{\spa6.2 s_{126} s_{256}}
       \Big(\frac{\spab5.2+6.1 \spab3.2+5.6 \spab2.1+6.4}{\spab1.2+6.5}
	-\spb1.6 \spa2.5 \spa5.3 \spb1.4 \Big) \Big)
 \nonumber \\ &&
         \times \iTm{s_{16}}{s_{25}}{s_{34}}
       + \frac{3}{2} \frac{\spab3.1+4.6^2}{ \spb2.5 \spa3.4 \spab1.2+6.5 s_{134}}
         \, \Big( \Lnrat{ - s_{256}}{ - s_{26}} - \Lnrat{ - s_{34}}{ - s_{126}} \Big)
 \nonumber \\ &&
       + \frac{3 \spa1.3 \spb1.6 \spab2.1+6.4
          + \spa1.2 \spb1.4 (\spa3.5 \spb5.6-\spa3.4 \spb4.6)}{2 \spa1.6 \spb1.5 \spa3.4 \spb3.4 \spab1.2+6.5} 
	 \, \Lnrat{ - s_{26}}{ - s_{126}}      
 \nonumber \\ &&
       - \frac{2 \spa1.3 \spb1.6 \spab3.1+4.6}{\spa3.4 \spb2.5 \spab1.2+6.5 s_{134}}
         \, \Lzero{ - s_{34}}{ - s_{134}} 
       - \frac{3 \spa3.5 \spa1.2 \spab2.3+5.4}{2 \spa3.4 \spb3.4 \spa1.6 \spa2.6 \spab1.2+6.5}
         \, \Lzero{ - s_{126}}{ - s_{34}} 
 \nonumber \\ &&
       + \frac{\spa1.2^2 \spb1.4}{2 \spa1.6 \spa2.6 \spb3.4 \spab1.2+6.5}  \Big(
           \frac{\spb1.4}{\spb1.5}
          -\frac{\spa3.5}{\spa3.4}
          \Big)
	  \, \Lzero{ - s_{126}}{ - s_{26}} 
 \nonumber \\ &&
       - \frac{\spa3.5 \spb5.6 \spab3.2+5.6}{\spa2.6 \spa3.4 \spb2.5 \spb2.6 \spab1.2+6.5}
         \, \Lzero{ - s_{256}}{ - s_{26}} 
       - \frac{1}{2} \frac{\spa1.3^2 \spb1.6^2}{\spb2.5 \spa3.4 \spab1.2+6.5 s_{134}} 
         \, \Lone{ - s_{34}}{ - s_{134}}
 \nonumber \\ &&
       + \frac{1}{2} \frac{\spa3.5 \spa1.2 \spa2.5 \spb4.5}{\spa1.6 \spa2.6 \spab1.2+6.5 s_{126}} 
         \, \Lone{ - s_{34}}{ - s_{126}}
       + \frac{1}{2} \frac{\spa3.5^2 \spb5.6^2}{\spb2.5 \spa3.4 \spab1.2+6.5 s_{256}}
         \, \Lone{ - s_{26}}{ - s_{256}}
 \nonumber \\ &&
       + \frac{1}{2 \spab1.2+6.5} \Big(
         \frac{\spab3.1+4.6^2}{ \spb2.5 \spa3.4 s_{134}}
       +  \frac{\spa1.2^2 \spb1.4^2}{ \spb3.4 \spb1.5 \spa1.6 \spa2.6}
       +  \frac{\spb4.6^2}{ \spb3.4 \spb2.5}
       -  \frac{\spa3.5 \spa1.2 \spa2.3}{ \spa3.4 \spa1.6 \spa2.6}
         \Big)
\Bigg]
 \nonumber \\ &&
      + \Vpole_\lc(s_{126}) \, A^d_\tree  ( 1_{\bar u}^+,2_{d}^-,3_{\nu}^-,4_{\ell}^+, 5_{\gamma}^-, 6_g^+ )
\end{eqnarray}
 where the corresponding tree-level amplitude has been given in Eq.(\ref{eq:LOmp-Qd}).
An alternative form may also be useful for simplification:
\begin{eqnarray}
&& A^d_\tree  (  5_{\gamma}^-, 6_g^+ ) =
 \nonumber \\ && \frac{1}{\spab1.2+6.5} \Bigg(
  \frac{\spa1.2 \spb1.4 \spab2.1+6.4 \spa4.3}{\spb1.5 \spa1.6 \spa2.6}
- \frac{\spa1.3 \spb1.6 \spab2.1+6.4}{\spb1.5 \spa1.6}
- \frac{\spab3.1+4.6^2 \spb4.3}{\spb2.5 s_{134}}
\Bigg)
\end{eqnarray}

\begin{eqnarray}
 && A^u_\lc  (5_{\gamma}^-, 6_g^+ ) = 
 \nonumber \\ &&
      \frac{1}{s_{156}} \Big(
       \frac{\spa3.4 \spab6.1+5.4^2 \spab5.1+6.2^2}{\spa1.6 \spab6.1+5.2^3}
      +\frac{\spa2.3^2 \spb1.6^2 \spb3.4}{\spab2.1+6.5 \spb1.5}
       \Big) 
        \, \LsminOneTMH{s_{26}}{s_{156}}{s_{15}}{s_{34}}
 \nonumber \\ &&
      - \frac{\spa1.2^3 \spa3.4 \spb4.5^2 s_{126}^2}
             {\spa1.6 \spa6.2 \spab1.2+6.5^3 \spab2.1+6.5}
        \, \LsminOneTMH{s_{15}}{s_{126}}{s_{26}}{s_{34}}
 \nonumber \\ &&
      + \frac{\spb1.6^2 \spa2.3^2 \spb3.4}{\spb1.5 s_{156} \spab2.1+6.5}
        \, \LsminOne{-s_{15}}{-s_{156}}{-s_{16}}
 \nonumber \\ &&
      - \frac{\spb1.6 \spab2.1+6.4 \spa2.3}{\spb1.5 \spa1.6 \spab2.1+6.5}
        \, \LsminOne{-s_{16}}{-s_{126}}{-s_{26}}
\nonumber \\ &&
  - \left( T_\text{sum} - T_\text{SL} \right) \, \iTm{s_{15}}{s_{26}}{s_{34}} 
  + L(5_{\gamma}^-,6_g^+)
\nonumber \\ &&
  + \Vpole_\lc(s_{126}) \, A^u_\tree  ( 1_{\bar u}^+,2_{d}^-,3_{\nu}^-,4_{\ell}^+, 5_{\gamma}^-, 6_g^+ )
\end{eqnarray}
 where the triangle coefficient is expressed in terms of,
\begin{eqnarray}
 && T_\text{sum} = s_{34} \Big[
 \nonumber \\ &&
 2\frac{\spa1.5 \spb1.6 \spa2.3 \spab5.1+6.4}{ \spa1.6 s_{234}^2}
-\frac{\spa1.2 \spb1.5 \spa2.3 \spab2.1+6.4 \spab5.1+6.2}{ \spa1.6 \spab2.1+6.5 \spab6.1+2.5 s_{234}}
 \nonumber \\ &&
-2\frac{\spa1.5 \spb1.6 \spa2.3 \spb2.4 \spa5.6}{ \spa1.6 \spab6.1+5.2 s_{234}}
+\frac{\spa1.2 \spab3.2+5.1 \spab5.1+6.4}{ \spa1.6 \spab6.1+2.5 s_{234}}
 \nonumber \\ &&
-\frac{\spa1.2 \spb2.6 \spa3.6 \spb4.5 \spab5.2+6.1}{ \spab1.2+6.5 \spab6.1+5.2 \spab6.1+2.5}
+\frac{2 \spa2.6 \spb2.6 \spab1.5.6 \spab3.1+5.4 \spab5.2+6.1}{\spab1.2+6.5 \spab6.1+5.2 \, \DeltaTM1526}
 \nonumber \\ &&
+\frac{\spa1.2 \spb2.6 \spab3.1+5.4 \spab5.2+6.1 (s_{12}+s_{25}+s_{16}+s_{56})}{\spab1.2+6.5 \spab6.1+5.2 \, \DeltaTM1526}
 \nonumber \\ &&
-\frac{\spa1.2 \spb1.6 \spb2.4 \spa3.6 \spa5.6}{ \spa1.6 \spab6.1+5.2 \spab6.1+2.5}
+\frac{\spa1.2 \spb1.4 \spa3.5}{ \spa1.6 \spab6.1+2.5}
\Big]
\end{eqnarray}
 and,
\begin{eqnarray}
 && T_\text{SL} = s_{34} \Big[
 \nonumber \\ &&
+\frac{\spa1.3 \spb5.6 \spab5.2+6.1 \spab1.3+5.4 (s_{134}-s_{345}) (s_{12}-s_{56})}{\spab1.2+6.5^2 \spab6.1+5.2 \, \DeltaTM1526}
 \nonumber \\ &&
+\frac{\spa3.6 \spb4.5 \spab5.1+2.6 (s_{34}-s_{56})}{2 \spab1.2+6.5 \spab6.1+5.2 \spab6.1+2.5}
+\frac{\spa1.3 \spb1.5 \spb3.4 \spa3.6 \spab5.1+2.6}{ \spab1.2+6.5 \spab6.1+5.2 \spab6.1+2.5}
 \nonumber \\ &&
-\frac{3 \spa1.5 \spb1.5 \spa1.6 \spb2.6 \spab2.1+5.6 \spab3.1+5.4 \spab5.2+6.1 (s_{12}+s_{16}+s_{25}+s_{56})}{\spab1.2+6.5 \spab6.1+5.2 \, \DeltaTM1526^2}
 \nonumber \\ &&
-\frac{6 \spa1.5^2 \spb1.5 \spb2.5 \spa2.6 \spb2.6 \spab2.1+5.6 \spab3.1+5.4 \spab5.2+6.1}{\spab1.2+6.5 \spab6.1+5.2 \, \DeltaTM1526^2}
 \nonumber \\ &&
+\frac{\spa1.3 \spb4.6 \spa5.6 \spb5.6 \spab5.2+6.1 (s_{134}-s_{345} -s_{24})}{\spab1.2+6.5 \spab6.1+5.2 \, \DeltaTM1526}
 \nonumber \\ &&
+\frac{2 \spb1.2 \spa1.3 \spb5.6 \spab5.2+6.1}{\spab1.2+6.5 \spab6.1+5.2 \, \DeltaTM1526}
 \nonumber \\ &&
 \, (-\spa1.2 \spb1.4 \spa1.5-2 \spa1.2 \spb2.4 \spa2.5-\spa1.2 \spb3.4 \spa3.5-2 \spa1.3 \spa2.5 \spb3.4+2 \spa1.5 \spa2.5 \spb4.5)
 \nonumber \\ &&
+\frac{\spa1.3 \spa5.6 \spb5.6 \spab5.2+6.1}{2 \spab1.2+6.5 \spab6.1+5.2 \, \DeltaTM1526}
 \, (-4 \spa1.2 \spb1.4 \spb2.6 - \spa2.3 \spb2.4 \spb3.6-3 \spa2.4 \spb2.4 \spb4.6)
 \nonumber \\ &&
+\frac{4 \spa1.2 \spb1.2 \spa1.3 \spa1.6 \spb1.6 \spb4.6 \spab5.2+6.1}{ \spab1.2+6.5 \spab6.1+5.2 \, \DeltaTM1526}
 \nonumber \\ &&
-\frac{\spa1.2 \spa3.5 (\spb1.2 \spb4.6 + \spb1.4 \spb2.6) (s_{134}-s_{345}) (s_{125}-s_{156})}{2 \spab1.2+6.5 \spab6.1+5.2 \, \DeltaTM1526} 
 \nonumber \\ &&
+\frac{\spa1.2 \spa3.5 (-3 \spb1.2 \spb4.6 +\spb1.4 \spb2.6)}{2 \spab1.2+6.5 \spab6.1+5.2}
+\frac{\spa2.3 \spb4.5 \spab5.1+2.6}{\spab1.2+6.5 \spab6.1+2.5}
 \nonumber \\ &&
+\frac{\spb4.6 \spab5.2+6.1 (4 \spa1.2 (\spa3.5 \spb5.6 + \spa1.3 \spb1.6) - \spa1.3 \spa2.5 \spb5.6)}{\spab1.2+6.5 \, \DeltaTM1526}
 \nonumber \\ &&
-\frac{3 \spa1.2 \spb1.6 \spa3.5 \spb4.6 (s_{134}-s_{345})}{\spab1.2+6.5 \, \DeltaTM1526}
-\frac{\spb1.6 \spa2.5 \spa3.5 \spb4.6}{2 \, \DeltaTM1526}
\Big]
 \nonumber \\ &&
+\Big\{ 1 \leftrightarrow 2, 3 \leftrightarrow 4, 5 \leftrightarrow 6, \langle\,\rangle\leftrightarrow[\,] \Big\}
\end{eqnarray}
 The term containing logarithmic functions is,
\begin{eqnarray}
 && L(5_{\gamma}^-, 6_g^+ ) = s_{34} \Bigg[
 \nonumber \\ &&
       + \frac{1}{\DeltaTM1526}  \frac{ \spab2.1+5.6  \spab5.2+6.1 \spab3.1+5.4}{2 \spa3.4 \spb3.4 \spab1.2+6.5 \spab6.1+5.2} \Big(
\frac{\spa1.6 \, \deltaof}{\spa2.6}
          - \frac{\spb2.5 \, \deltats}{\spb1.5}
          \Big)
 \nonumber \\ &&
       - \frac{\spb1.4 \spb2.6 \spa1.2^2 \spa1.3}{2 \spb3.4 \spa1.6 \spa3.4 \spab1.2+6.5^2}
         \, \Lzero{ - s_{126}}{ - s_{26}}
       - \frac{1}{2} \frac{\spb2.4^2 \spa2.5^2 s_{234}}{\spa1.6 s_{34}^2 \spb3.4 \spab6.1+5.2}
         \, \Lone{ - s_{234}}{ - s_{34}}
 \nonumber \\ &&
       + \frac{\spa5.6 \spb4.6}{\spa1.6 \spb1.5 \spb3.4 \spab6.1+5.2} \Big(
           \frac{\spa5.6 \spb1.5 \spb4.2}{\spab6.1+5.2}
          -2 \frac{\spab5.1+6.4}{\spa1.5}
          \Big)
	  \, \Lzero{ - s_{156}}{ - s_{15}}
 \nonumber \\ &&
       + \frac{\spb2.4}{\spa1.6 \spb3.4 \spab6.1+5.2 s_{34}} \Big(
            (\spab5.2+3.4 + \spb1.4 \spa1.5) \spa2.5
          + \frac{\spa5.6 \spab2.1+5.4 \spab5.1+6.2}{\spab6.1+5.2}
          \Big)
	  \, \Lzero{ - s_{234}}{ - s_{34}}
 \nonumber \\ &&
       +  \frac{\spa1.2 \spa3.5}{\spa1.6 \spa2.6 \spa3.4 \spab1.2+6.5} \Big(
            \frac{\spb4.5 \spa1.2 \spa3.4}{\spab1.2+6.5}
          - \frac{\spab2.1+6.4}{\spb3.4}
          \Big)
	  \, \Lzero{ - s_{345}}{ - s_{34}}
 \nonumber \\ &&
       - \frac{1}{2} \frac{\spb4.6^2 \spa5.6^2 s_{156}}{\spb3.4 \spa1.6 s_{15}^2 \spab6.1+5.2}
         \, \Lone{ - s_{156}}{ - s_{15}}
       + \frac{1}{2} \frac{\spb3.4 \spa1.2 \spa3.5^2 \spab2.3+4.5}{\spa1.6 \spa2.6 s_{34}^2 \spab1.2+6.5}
         \, \Lone{ - s_{345}}{ - s_{34}}
 \nonumber \\ &&
       + \Lnrat{ - s_{15}}{ - s_{26}} \, \frac{1}{2 \, \DeltaTM1526} \frac{1}{\spab1.2+6.5 \spab6.1+5.2}  \Big(
         -12 \spb1.2 \spa1.2 \spb1.6 \spa1.5 \spab3.1+5.4
 \nonumber \\ &&
         -10 \spb1.2 \spa1.2 \spb2.6 \spa2.5 \spab3.1+5.4
          -3 \frac{\spa1.3 \spb1.2 \spa3.5 \spab2.1+5.6 (s_{15}-s_{26}-s_{34})}{\spa3.4}
 \nonumber \\ &&
          +3 \frac{\spa1.3 \spb2.6 \spa2.3 \spab5.2+6.1 (s_{15}-s_{26})       }{\spa3.4}
          -  \frac{\spb2.4 \spa1.5 \spab2.1+5.6 \spb1.4 (s_{15}-s_{26})       }{\spb3.4}
 \nonumber \\ &&
          +  \frac{\spb2.4 \spa1.5 \spab2.1+5.6 \spb4.5 \spab5.2+6.1     }{\spb3.4}
          -3 \frac{\spb2.4 \spb4.6 \spa1.2 \spab5.2+6.1 (s_{15}-s_{26})  }{\spb3.4}
 \nonumber \\ &&
          -2 \frac{\spb4.5 \spb4.6 \spa1.5 \spab5.2+6.1 (s_{15}-s_{26})  }{\spb3.4}
          +8 \spab3.1.4 \spab5.1.2    \spab2.1+5.6 
          +6 \spab3.1.4 \spab5.6.2    \spab2.1+5.6 
 \nonumber \\ &&
          +6 \spab3.5.4 \spab5.1.2    \spab2.1+5.6 
          +2 \spab3.5.4 \spab5.6.2    \spab2.1+5.6 
          +  \spb1.4 \spb2.5 \spa1.5 \spa3.5 \spab2.1+5.6
 \nonumber \\ &&
          -2 \spb1.2 \spb4.6 \spa3.5 \spa2.5 \spab1.2+6.5
          -8 \spb1.6 \spb4.6 \spa3.5 \spa5.6 \spab1.2+6.5
          +2 \spb4.6 \spa1.3 \spa1.6 \spb1.6 \spab5.2+6.1 
 \nonumber \\ &&
          +  \spab5.2+6.1 \Big(
           2 \spb4.6 \spa1.3 \spab1.2+5.1    
          +2 \spb4.6 \spab1.3+4.2 \spa2.3   
          +6 \spb4.6 \spa1.6 \spab3.2+5.6   
 \nonumber \\ &&
          -4 \spb2.4 \spab1.3+4.6 \spa2.3   
          +  \spa1.2 \spb2.6 \spab3.1+5.4  \Big)
          +  \frac{\spb2.6 \spa1.3 \spa3.6 \spab2.1+5.6 \spab5.2+6.1}{\spa3.4}
 \nonumber \\ &&
          -2 \frac{\spb2.6 \spa5.6 \spab3.1+5.4 (s_{12}-s_{56}) (s_{125}-s_{156})}{\spab6.1+5.2}
          -2 \frac{\spb5.6 \spa1.5 \spab3.1+5.4 (s_{12}-s_{56}) (s_{134}-s_{345})}{\spab1.2+6.5}
          \Big)
 \nonumber \\ &&
       + \frac{1}{\spab6.1+5.2} \Big(
            \frac{\spa1.5 \spb1.4 \spab1.2+6.4}{\spb3.4 \spa1.6 \spab1.2+6.5} 
          -\frac{3}{2} \frac{\spb1.6 \spa1.3 \spa3.5}{\spa3.4 \spab1.2+6.5} 
          +\frac{3}{2} \frac{\spab5.1+6.4^2}{\spb3.4 \spa1.6 s_{234}}
          \Big)
	  \, \Lnrat{ - s_{15}}{ - s_{26}}
 \nonumber \\ &&
       + \Lnrat{ - s_{15}}{ - s_{34}} \, \frac{1}{2 \, \DeltaTM1526} \frac{1}{\spab1.2+6.5 \spab6.1+5.2} \Big(
 \nonumber \\ &&
          -3 \spa1.2 \spa3.5 \spb1.2 \spb2.4 \spab2.1+5.6
          -  \spa1.5 \spb1.4 \spab3.4+5.2    \spab2.1+5.6
 \nonumber \\ &&
          -  \spa1.5 \spa2.5 \spb1.5 \spb2.6 \spab3.1+5.4
          +  \spa1.5 \spa1.6 \spb1.6^2       \spab3.1+5.4
          +3 \spa1.2 \spa3.5 \spb1.6 \spb4.6 \spab6.1+5.2
 \nonumber \\ &&
      + \spab5.2+6.1 \Big(
           5 \spa1.2 \spa2.3 \spb2.4 \spb2.6
          +5 \spa1.5 \spa3.6 \spb4.6 \spb5.6 
          +4 \spa1.5 \spa2.3 \spb2.4 \spb5.6 
          +  \spa1.6 \spa3.4 \spb4.6^2       \Big)
 \nonumber \\ &&
          +  \frac{\spa1.2 \spa3.6 \spb4.6 \spab5.2+6.1 (s_{15}-s_{34})}{\spa2.6}
          -2 \frac{\spa1.2 \spa2.3 \spb2.4 \spab5.2+6.1 (s_{15}-s_{34})}{\spa2.6}
 \nonumber \\ &&
          +  \frac{\spa1.2 \spab3.1+5.4    \spab5.2+6.1 \spab6.1+5.6    }{\spa2.6}
          +3 \frac{\spa1.2 \spa2.5 \spb1.2 \spab3.1+5.4 (s_{15}-s_{34}) }{\spa2.6}
 \nonumber \\ &&
          +2 \frac{\spa5.6 \spab2.1+5.6    \spab3.1+5.4 (s_{15}-s_{34}) }{\spa2.6}
          +2 \frac{\spab3.1+5.4 \spab5.3+4.2 (s_{12}-s_{56}) (s_{125}-s_{156})}{\spab6.1+5.2}
          \Big)
 \nonumber \\ &&
       + \frac{1}{2 \, \DeltaTM1526}  \frac{1}{\spab1.2+6.5 \spab6.1+5.2} \bigg(
           \frac{\spb1.2}{\spb1.5}\Big(
 \nonumber \\ &&
           3 \spb1.4 \spa3.2 \spa1.5 \spb5.6 (s_{26}-s_{34})
           - \spab2.1+5.6 \spab3.1+5.4 (s_{256}-s_{34}) \Big)
 \nonumber \\ &&
       + 2 \spa3.5 \spab2.1+5.6 \Big(
           \spb1.2 \spb2.4 \spa1.2 
          -3 \spb1.2 \spb3.4 \spa1.3  \Big)
          -3 \spb1.2 \spb4.6 \spa1.2 \spa3.5 (s_{16}+s_{26}-s_{34})
 \nonumber \\ &&
          -4 \spb1.2 \spb1.4 \spa1.3 \spa1.5 \spab2.1+5.6
          -  \spb2.4 \spb2.5 \spa2.5 \spa3.5 \spab2.1+5.6
          -8 \spb1.2 \spa1.2 \spa2.5 \spb2.6 \spab3.1+5.4
 \nonumber \\ &&
          +3 \spb2.4 \spa1.2 \spab3.4+5.6    \spab5.2+6.1
          -3 \spb2.4 \spa1.3 \spa2.5 \spb5.6 \spab5.2+6.1
          +3 \spa2.5 \spb2.6 \spab3.1+5.4 s_{156}
 \nonumber \\ &&
          +\spa1.5 \spb1.6 \spab3.1+5.4 (3 s_{234}-3 s_{34}+3 s_{12} + s_{12}+s_{26})
          +\spb2.6 \spb2.5 \spa2.5^2 \spab3.1+5.4
 \nonumber \\ &&
          + \spb2.6 \spb3.4 \spa1.3 \spa2.3 \spab5.2+6.1
          - \spb4.6 \spab2.1+5.6 \spa3.5 \spab6.1+5.2
 \nonumber \\ &&
          -3 \spb2.5 \spb4.6 \spa1.2 \spa3.5 \spa5.6 \spb1.6
          -2 \frac{\spab1.3+4.6 \spab3.1+5.4}{\spab1.2+6.5} (s_{12}-s_{56}) (s_{134}-s_{345})
          \bigg)
	  \, \Lnrat{ - s_{34}}{ - s_{26}}
 \nonumber \\ &&
       + \frac{3 \spa1.5 \spab2.1+5.6 \spab3.1+5.4 \spab5.2+6.1}{\DeltaTM1526^2 \spab1.2+6.5 \spab6.1+5.2} \Big(
            2 \spa1.6 \spb1.2 \spb5.6 
          + \spb2.5 (s_{12}+s_{25}+s_{56}-s_{16})
          \Big)
	  \, \Lnrat{ - s_{15}}{ - s_{34}} 
 \nonumber \\ &&
       + \frac{3 \spb2.6 \spab2.1+5.6 \spab3.1+5.4 \spab5.2+6.1}{\DeltaTM1526^2 \spab1.2+6.5 \spab6.1+5.2}  \Big(
            2 \spa1.2 \spa5.6 \spb2.5
          + \spa1.6 (s_{12}+s_{16}+s_{56}-s_{25})
          \Big)
	  \, \Lnrat{ - s_{34}}{ - s_{26}}
 \nonumber \\ &&
       + \frac{\spb4.6}{\spab6.1+5.2} \Big(
           \frac{\spb2.4 \spa5.6^2}{\spb3.4 \spa1.6 \spab6.1+5.2}
          +\frac{1}{2} \frac{\spa3.5}{\spab1.2+6.5}
          \Big)
	  \, \Lnrat{ - s_{34}}{ - s_{26}}
 \nonumber \\ &&
       + \frac{\spa1.2}{\spa2.6 \spb3.4 \spab1.2+6.5 \spab6.1+5.2}  \Big(
           \frac{\spb4.5 \spab5.1+3.4 (s_{12}-s_{56})}{\spab1.2+6.5}
          -\spb1.4 \big (\spab5.6.4 + \frac{\spb2.4 \spa2.5}{2} \big)
          \Big)
	  \, \Lnrat{ - s_{126}}{ - s_{15}}
 \nonumber \\ &&
       + \frac{\spa1.2 \spb1.4}{2 \spa1.6 \spb3.4 \spab1.2+6.5} \Big(
          3 \frac{\spb1.2 \spab1.2+3.4}{\spb1.5 \spab6.1+5.2}
          + \frac{\spb2.4 \spa1.5}{\spab6.1+5.2}
          + \frac{\spb2.6 \spa1.2 \spb4.5}{\spb1.5 \spab1.2+6.5}
          + \frac{\spb2.6 \spa1.2 \spa1.3}{\spa3.4 \spab1.2+6.5}
 \nonumber \\ &&
          + \frac{\spb4.6}{\spb1.5}
          - \frac{\spab3.1+2.6}{\spb1.5 \spa3.4}
          - 3 \frac{\spa2.3 \spb1.6 \spab5.2+3.4 \spab1.2+6.5}{\spa1.2 \spa3.4 \spb1.4 \spb1.5 s_{234}}
          \Big)
	  \, \Lnrat{ - s_{126}}{ - s_{26}}
 \nonumber \\ &&
       +  \frac{\spa1.2}{\spa2.6 \spab1.2+6.5 \spab6.1+5.2} \Big(
          -\frac{3}{2} \frac{\spb1.2 \spb1.4 \spa2.3}{\spb1.5}
          + \frac{\spb1.4 \spa1.3 (s_{12}-s_{56})}{\spab1.2+6.5}
          - \frac{\spb4.5 \spa3.5 \spa5.6}{\spa1.6}
          \Big)
	  \, \Lnrat{ - s_{126}}{ - s_{34}}
 \nonumber \\ &&
       - \frac{3}{2} \frac{\spab5.1+6.4^2}{\spb3.4 \spa1.6 \spab6.1+5.2 s_{234}}
         \, \Lnrat{ - s_{156}}{ - s_{34}}
 \nonumber \\ && 
       + \frac{1}{2 \spa1.6 \spb3.4} \Big(
       - \frac{\spab5.1+6.4 \spb4.6 \spa5.6}{\spa1.5 \spb1.5 \spab6.1+5.2}
       - \frac{\spb1.4 \spa1.3 \spa2.5 \spab2.1+6.5}{\spb1.5 \spa2.6 \spa3.4 \spab1.2+6.5}
       + \frac{\spb1.6 \spa2.3 \spa3.6 \spa5.6 \spb3.4}{\spb1.5 \spa2.6 \spa3.4 \spab6.1+5.2}
 \nonumber \\ && 
       + \frac{\spb1.6 \spa2.3 \spab5.1+6.4}{\spb1.5 \spa3.4 s_{234}}
       - \frac{\spb2.4 \spa2.5^2 \spab6.1+5.4}{\spa2.6 s_{34} \spab6.1+5.2}
       - \frac{\spa1.6 \spb4.5 \spa3.6 \spab2.1+5.6 \spab5.2+6.1}{\spb1.5 \spa2.6 \spa3.4 \spab1.2+6.5 \spab6.1+5.2}
 \nonumber \\ && 
       + \frac{\spb3.4 \spa1.2 \spa2.3 \spa3.5}{\spa2.6 \spa3.4 \spab1.2+6.5}
       \Big)
\Bigg]
\end{eqnarray}
 The tree-level amplitude has been given in Eq.(\ref{eq:LOmp-Qu}).

\subsubsection{Subleading colour, $5_{\gamma}^-, 6_g^+$}

\begin{eqnarray}
 && A^d_\slc  ( 5_{\gamma}^-, 6_g^+ ) = 
 \frac{\spab2.1+6.4 \spab3.2+5.1}{ \spa1.6 \spb1.5 \spab6.1+2.5}
          \, \LsminOne{ - s_{12}}{ - s_{126}}{ - s_{16}} 
 \nonumber \\ &&          
       + \frac{\spab3.2+5.1^2 \spb4.3}{\spb2.5 \spab6.1+2.5 \spab6.2+5.1}
          \Big( \LsminOne{ - s_{12}}{ - s_{125}}{ - s_{25}}
              + \LsminOneTMHT{s_{16}}{s_{125}}{s_{25}}{s_{34}} \Big)
 \nonumber \\ &&
       + \frac{\spa1.2^2 \spab6.1+2.4 (
            \spa1.3 \spa2.6 \spb1.2
          - \spa1.6 \spa3.4 \spb1.4
          )}{\spa1.6^3 \spa2.6 \spb1.5 \spab1.2+6.5}  
	  \, \LsminOne{ - s_{12}}{ - s_{126}}{ - s_{26}}       
 \nonumber \\ &&
       - \frac{\spa2.5^2 \spa4.3 \spb1.4^2}{ \spa2.6 \spab6.2+5.1 s_{256}}
          \, \LsminOne{ - s_{25}}{ - s_{256}}{ - s_{56}}
       - \frac{\spb5.6^2 \spab3.5+6.2^2 \spb4.3}{\spb2.5^3 \spab1.2+6.5 s_{256}}
          \, \LsminOne{ - s_{26}}{ - s_{256}}{ - s_{56}} 
 \nonumber \\ &&
       -  \frac{1}{s_{256}}  \Big(
            \frac{\spa2.5^2 \spa5.6 \spa4.3 \spb1.4^2}{\spa2.6 \spa5.6 \spab6.2+5.1}
          + \frac{\spb5.6 \spab1.2+5.6^2 \spab3.5+6.2^2 \spb4.3}{ \spb2.5 \spb5.6 \spab1.2+6.5 \spab1.5+6.2^2}
          \Big)
	  \, \LsminOneTMHT{s_{12}}{s_{256}}{s_{56}}{s_{34}}
 \nonumber \\ &&
       - \frac{\spab2.1+6.4^2 \spa4.3}{ \spa1.6 \spab2.1+6.5 \spab6.1+2.5}
          \, \LsminOneTMHT{s_{25}}{s_{126}}{s_{16}}{s_{34}}
 \nonumber \\ &&
       + \frac{\spa3.6^2 \spb1.5^2 \spb4.3 s_{125}^2}{\spb2.5 \spab6.1+2.5^3\spab6.2+5.1}
          \, \LsminOneTMHT{s_{56}}{s_{125}}{s_{12}}{s_{34}}
 \nonumber \\ &&
       + \frac{\spa4.3 \spab2.1+6.5^2 \spab6.1+2.4^2}{ \spa2.6 \spab1.2+6.5 \spab6.1+2.5^3}
          \, \LsminOneTMHT{s_{56}}{s_{126}}{s_{12}}{s_{34}}
\nonumber \\ &&
       + \Big(
	    \frac{ \spb1.6 \spb4.3 (\spa2.3^2 s_{16} s_{25}-\spaa2.1+6.2+5.3^2)}{\spab2.1+6.5}
          - \frac{\spa2.5 \spa4.3 (\spb1.4^2 s_{16} s_{25}-\spbb1.2+5.1+6.4^2)}{\spab6.2+5.1}
          \Big)
\nonumber \\ &&
         \times \frac{\iTm{s_{16}}{s_{25}}{s_{34}}}{2 \spa1.6 \spb2.5 \spab2.6+5.1}
+ \Big( T_\text{antisym} - T_\text{sym} \Big) \, \iTm{s_{12}}{s_{34}}{s_{56}} 
 \nonumber \\ &&
       - \frac{\spab5.1+2.6}{\spab1.3+4.2 \spab6.1+2.5} \Big(
            \frac{(s_{15}-s_{26}) \spa3.6 \spb4.5}{\spab6.1+2.5}
          + \spa3.6 \spb4.6
 \nonumber \\ &&
          + \frac{(s_{12}-s_{34}-s_{56}) s_{234} \spab3.1+2.4}{\DeltaTM1234}
          + \frac{2 s_{34} s_{56} \spab3.1+2.4}{\DeltaTM1234}
          \Big)
	  \, \Lnrat{ - s_{12}}{ - s_{56}} 
 \nonumber \\ &&
       + \frac{\spa4.3 \spb4.5 \spab2.1+6.4 s_{345}}{\spab1.2+6.5 \spab6.1+2.5^2}
          \, \Lnrat{ - s_{12}}{ - s_{126}}  
       + \frac{\spa1.3 \spb2.6 \spab3.1+4.6 \spb4.3}{\spab1.2+6.5 \spab1.3+4.2 \spb2.5}
          \, \Lnrat{ - s_{26}}{ - s_{12}} 
 \nonumber \\ &&
       + \frac{\spb2.6 \spab3.1+4.2 \spab3.1+4.6 \spb4.3}{\spab1.3+4.2 \spb2.5^2 s_{134}}
          \, \Lnrat{ - s_{26}}{ - s_{56}}
 \nonumber \\ &&
       + \frac{\spab3.1+2.4 s_{34}}{\spab1.3+4.2 \spab6.1+2.5}  \Big(
            \frac{\spb2.6}{\spb2.5}
          + \frac{\spab5.1+2.6 (s_{234}-s_{134})}{\DeltaTM1234}
          \Big)
	  \, \Lnrat{ - s_{34}}{ - s_{12}}
 \nonumber \\ &&
       + \frac{\spa1.3 \spb2.6 \spb4.6 s_{34}}{\spab1.2+6.5 \spab1.3+4.2 \spb2.5} 
          \, \Lnrat{ - s_{34}}{ - s_{256}}
 \nonumber \\ &&
       + \frac{\spa3.6}{\spab6.1+2.5^2} \Big(
            \frac{\spa1.2 \spbb4.1+2.2+6.1}{\spa1.6}
          - \frac{\spb1.5 \spab3.1+2.6 \spb4.3}{\spb2.5}
          \Big)
	  \, \Lnrat{ - s_{126}}{ - s_{12}}
 \nonumber \\ &&
       + \frac{1}{\spa1.6}  \Big(
            \frac{\spa1.3 \spa2.6 \spb1.6 \spb4.6-\spa1.2 \spa2.3 \spb1.4 \spb2.6}{2 \spab1.2+6.5 \spb1.5}
          - \frac{\spa1.2 \spa1.3 \spb1.6 \spab6.1+2.4}{\spab1.2+6.5 \spb1.5 \spa1.6}
 \nonumber \\ &&
          + \frac{\spa1.2 \spa3.4 \spb1.4 \spb4.6}{\spab1.2+6.5 \spb1.5}
          - \frac{3 \spab3.1+4.6^2 \spa1.6 \spb4.3}{2 \spab1.2+6.5 \spb2.5 s_{134}}
          \Big)
	  \, \Lnrat{ - s_{126}}{ - s_{26}} 
 \nonumber \\ &&
       + \frac{\spb4.3}{\spab6.1+2.5^2} \Big(
            \frac{\spa1.3 \spb1.5 \spab3.2+6.1}{\spb2.5}
          - \frac{\spa3.6 \spb1.6 \spb4.5 \spa3.4}{\spb2.5}
          + \spa2.3 \spab3.2+6.1
          \Big)
	  \, \Lnrat{ - s_{126}}{ - s_{125}}
 \nonumber \\ &&
       - \frac{3 \spab3.1+4.6^2 \spb4.3}{2 \spab1.2+6.5 \spb2.5 s_{134}}
          \, \Lnrat{ - s_{256}}{ - s_{34}}
       -  \frac{\spa1.2^2 \spb1.4 (\spab3.4.1 - \spab3.5.1)}{2 \spab1.2+6.5 \spa1.6 \spa2.6 \spb1.5} 
	  \, \Lzero{ - s_{126}}{ - s_{26}}
 \nonumber \\ &&
       - \spa2.6  \Big(
          - \frac{\spa1.2 \spa3.6 \spb1.2 \spb4.6}{\spab6.1+2.5^2 \spa1.6}
          - \frac{\spa1.3 \spa2.6 \spb1.6 \spb2.4}{\spab6.1+2.5 \spb1.5 \spa1.6^2}
          + \frac{(\spa3.6 \spb4.6 - \spa1.3 \spb1.4) \spb1.6}{\spab6.1+2.5 \spb1.5 \spa1.6}
          \Big)
	  \, \Lzero{ - s_{126}}{ - s_{12}}
 \nonumber \\ &&
       + \Big(
	    \frac{3 \spa1.2 \spa2.3 \spa3.5 \spb4.3}{2 \spab1.2+6.5  \spa1.6 \spa2.6}
	  - \frac{\spa1.2 \spa3.5 \spb4.5 \spab2.5+6.2}{\spab1.2+6.5 \spb2.5 \spa1.6 \spa2.6}
 \nonumber \\ &&
	  - \frac{\spa4.3 \spb4.5^2 \spab5.2+6.1}{\spab6.1+2.5^2 \spb2.5}
	  - \frac{\spa3.5 \spb4.5 \spab6.1+2.6}{\spab6.1+2.5 \spb2.5 \spa1.6}
	  \Big)
	 \, \Lzero{ - s_{126}}{ - s_{34}}
 \nonumber \\ &&
       - \frac{\spa3.5 \spb1.5 \spab3.1+2.5 \spb4.3}{\spab6.1+2.5^2 \spb2.5}
          \, \Lzero{ - s_{125}}{ - s_{12}}
       - \frac{\spa3.6 \spb1.6 \spb4.5 s_{34}}{\spab6.1+2.5^2 \spb2.5} 
          \, \Lzero{ - s_{125}}{ - s_{34}}
 \nonumber \\ &&
       + \frac{\spa1.3 \spb1.6}{\spab1.2+6.5} \Big(
            \frac{\spa1.3 \spb2.6 \spb4.3}{\spab1.3+4.2 \spb2.5}
          - \frac{3 \spa1.3 \spb1.6}{2 \spb2.5 \spa3.4}
          + 2 \frac{\spb4.6}{\spb2.5}
          \Big)
	  \, \Lzero{ - s_{134}}{ - s_{34}}
 \nonumber \\ &&
       + \frac{\spa3.5 \spb5.6^2 \spb3.4}{2 \spb2.5^2 \spab1.2+6.5 s_{26}}  \Big(
            2 \spab3.1+4.2 - \spab3.5.2
          \Big)
	  \, \Lzero{ - s_{256}}{ - s_{26}} 
 \nonumber \\ &&
       - \frac{\spa2.3 \spb2.6^2 \spab3.1+4.2 \spb4.3}{\spab1.3+4.2 \spb2.5^2 \spb5.6 \spa5.6}
          \, \Lzero{ - s_{256}}{ - s_{56}}
       + \frac{\spa2.6 \spa3.6 \spb1.6 \spb4.6}{\spab6.1+2.5 \spb1.5 \spa1.6}
          \, \Lone{ - s_{126}}{ - s_{12}}
 \nonumber \\ &&
       + \frac{\spa1.2 \spa2.5 \spa3.5 \spb4.5}{2 \spab1.2+6.5 \spa1.6 \spa2.6}
          \, \Lone{ - s_{126}}{ - s_{34}}
       - \frac{\spa3.6 \spb1.6 \spb4.6}{\spab6.1+2.5 \spb2.5}
          \, \Lone{ - s_{125}}{ - s_{34}}
 \nonumber \\ &&
       - \frac{\spa1.3^2 \spb1.6^2}{2 \spab1.2+6.5 \spb2.5 \spa3.4}
          \, \Lone{ - s_{134}}{ - s_{34}}
       - \frac{ \spa3.5^2 \spb5.6^2 \spb4.3}{2 \spab1.2+6.5 \spb2.6 \spb2.5 \spa2.6}
          \, \Lone{ - s_{256}}{ - s_{26}}
 \nonumber \\ &&
          - \frac{\spa2.6 \spa4.3 \spb1.6 \spb4.6 \spb4.5}{\spab1.2+6.5 \spab6.1+2.5 \spb1.5}
          - \frac{\spa1.2 \spa2.3 \spa4.3 \spb1.4 \spb3.4}{\spab1.2+6.5 \spb1.5 \spa1.6 \spa2.6}
          + \frac{\spa1.2 \spa2.5 \spb1.4 \spb4.5 \spa4.3}{2 \spab1.2+6.5 \spb1.5 \spa1.6 \spa2.6}
 \nonumber \\ &&
          + \frac{3 \spa1.3 \spa2.6 \spb1.6 \spb4.6}{2 \spab1.2+6.5 \spb1.5 \spa1.6}
          + \frac{\spa1.3 \spa2.3 \spb1.6 \spb3.4}{2 \spab1.2+6.5 \spb1.5 \spa1.6}
          + \frac{\spa1.3 \spa2.6 \spb1.6 \spb4.6}{2 \spab1.2+6.5 \spb2.5 \spa2.6}
 \nonumber \\ &&
          - \frac{\spa2.3^2 \spb2.6 \spb3.4}{2 \spab1.2+6.5 \spb2.5 \spa2.6}
          - \frac{\spab3.1+4.6^2 \spb4.3}{2 \spab1.2+6.5 \spb2.5 s_{134}}
 \nonumber \\ &&
          - \frac{\spa3.6 \spb1.6 \spb4.6}{2 \spb1.5 \spb2.5 \spa1.6}
          - \frac{\spa4.3 \spb1.4 \spb4.6}{2 \spb1.5 \spb2.5 \spa1.6}
          + \frac{\spa2.3^2 \spb1.2 \spb4.3}{2 \spb1.5 \spb2.5 \spa1.6 \spa2.6}
 \nonumber \\ &&
      + \Vpole_\slc \, A^d_\tree  ( 1_{\bar u}^+,2_{d}^-,3_{\nu}^-,4_{\ell}^+, 5_{\gamma}^-, 6_g^+ )
\end{eqnarray}
 where the quantities that define one of the triangle coefficients are,
\begin{eqnarray}
&& T_{sym} = \frac{1}{2 \spab6.1+2.5} \Big[
 \nonumber \\ &&          
      -\frac{\spa1.3 \spb2.4 \, \DeltaTM1234 (s_{15}-s_{26})^2}
       {4 \spab1.3+4.2^2 \spab6.1+2.5}
      -\frac{\spab1.2+5.1 \, \DeltaTM1234
       \, (\spab3.2.4 - 2 \spab3.6.4 - 3 \spab3.1.4 )}
      {2 \spab1.3+4.2 \spab6.1+2.5}
 \nonumber \\ &&          
      +\frac{2 \spa1.3 \spb1.6 \spb3.4 \spa3.5
       (s_{345}-s_{346})- \spab3.1+2.4 \spab5.1+2.6 s_{125}
      +\spa3.5 \spb4.6 \, \DeltaTM1234}
      {2 \spab1.3+4.2}
 \nonumber \\ &&          
      -\frac{2 \spab5.1+2.6 \spa1.3 \spb3.4 \spab3.2+5.1}{\spab1.3+4.2}
      +\frac{3 \spb1.4 \spa2.3 \, \DeltaTM1234}{4 \spab6.1+2.5}
\Big]
 \nonumber \\ &&          
+\Big\{ 1 \leftrightarrow 2, 3 \leftrightarrow 4, 5 \leftrightarrow 6, \langle\,\rangle\leftrightarrow[\,] \Big\}
\end{eqnarray}
 and,
\begin{eqnarray}
&& T_\text{antisym} = \frac{1}{2 \spab1.3+4.2 \spab6.1+2.5} \Big[
      \frac{-\spa1.3 \spb1.5 \spb3.4 \spa3.6 \, \DeltaTM1234
       \, (\spab1.3+4.1 + \spab2.3+4.2)}
      {\spab6.1+2.5^2}
  \nonumber \\ &&	   
      +\frac{\spb2.4 \spa3.4 \spab1.2+3.4
       \, \DeltaTM1234 \, (\spab1.2+5.1-\spab2.1+6.2)}
      {2 \spab1.3+4.2 \spab6.1+2.5}
  \nonumber \\ &&	   
       +\frac{\spa1.2 s_{34} \spb2.4 \, (s_{345}-s_{346})
\, (\spa3.6 \spb6.1 - \spa3.5 \spb5.1)}
       {\spab6.1+2.5}
       -\frac{\spa3.4 \spb2.4 \, \DeltaTM1234
	\, (\spa1.2 \spb1.4+2 \spa2.3 \spb3.4)}
       {2 \spab6.1+2.5}
  \nonumber \\ &&	   
       +\frac{\spb2.4 \spa2.6 \spa3.4 \spb4.5
	\, \spab5.1+2.6 (\spab1.3+4.1+\spab2.3+4.2-2 s_{125})}
       {\spab6.1+2.5}
  \nonumber \\ &&	   
       -\frac{\spb2.4 \spa3.4
	\, \spab1.2+3.4 \spab5.1+2.6 \, (s_{156}-s_{256})}
       {2 \spab1.3+4.2}
       -\frac{\spab3.1+2.4
	\, \spab5.1+2.6 s_{123} \, (s_{156}-s_{256}) s_{34}}
       {\DeltaTM1234}
  \nonumber \\ &&	   
       -\spb2.4 \spa2.5 \spa3.4 \spb4.6 \, (s_{345}-s_{346})
       +\spb1.4 \spa3.4
	\, \spab5.1+2.6 (\spa1.2 \spb2.4-3 \spa1.3 \spb3.4)
\Big]
 \nonumber \\ &&          
-\Big\{ 1 \leftrightarrow 2, 3 \leftrightarrow 4, 5 \leftrightarrow 6, \langle\,\rangle\leftrightarrow[\,] \Big\}
\end{eqnarray}
 
The remaining amplitude is expressed in terms of quantities that have already
been defined for other amplitudes,
\begin{eqnarray}
 && A^u_\slc  ( 5_{\gamma}^-, 6_g^+ )=
+ \frac{\spb1.2^2 \spab3.1+2.5^2 \spb3.4}{ \spb1.5 \spb2.5^2 \spab6.1+2.5 \spab6.1+5.2}
         \, \LsminOne{ - s_{15}}{ - s_{125}}{ - s_{12}}
 \nonumber \\ &&
       - \frac{\spa1.2 \spa3.6 s_{12} \spab6.1+2.4}{ \spa1.6^3 \spb1.5 \spab6.1+2.5}
         \, \LsminOne{ - s_{12}}{ - s_{126}}{ - s_{26}}
       + \frac{\spa2.3 \spb1.6 \spab2.1+6.4}{ \spa1.6 \spb1.5 \spab2.1+6.5}
         \, \LsminOne{ - s_{12}}{ - s_{126}}{ - s_{16}} 
 \nonumber \\ &&
       - \frac{\spa5.6^2 \spab1.5+6.4^2 \spa3.4}{ \spa1.6^3 \spab6.1+5.2 s_{156}}
         \, \LsminOne{ - s_{15}}{ - s_{156}}{ - s_{56}} 
       - \frac{\spa2.3^2 \spb1.6^2 \spb3.4}{ \spb1.5 \spab2.1+6.5 s_{156}}
         \, \LsminOne{ - s_{16}}{ - s_{156}}{ - s_{56}}
 \nonumber \\ &&
       +  \frac{1}{s_{156}} \Big(
            \frac{\spa1.5 \spab1.5+6.4^2 \spab5.1+6.2^2 \spa3.4}{ \spa1.6 \spa5.6 \spab1.5+6.2^3}
          - \frac{\spa2.3^2 \spb1.6^3 \spb3.4}{ \spb1.5 \spb5.6 \spab2.5+6.1}
 \nonumber \\ &&
          + \frac{\spa2.3^2 \spb1.6^2 \spab2.1+5.6 \spb3.4}{ \spb5.6 \spab2.1+6.5 \spab2.5+6.1}
          - \frac{\spab1.5+6.4^2 \spab5.1+6.2^3 \spa3.4}{ \spa5.6 \spab1.5+6.2^3 \spab6.1+5.2}
          \Big)
	  \, \LsminOneTMHT{s_{12}}{s_{156}}{s_{56}}{s_{34}}
 \nonumber \\ &&
       + \frac{\spab3.1+2.5^2 \spab6.2+5.1^2 \spb3.4}{ \spb1.5 \spab6.1+2.5^3 \spab6.1+5.2}
         \, \LsminOneTMHT{s_{56}}{s_{125}}{s_{12}}{s_{34} }
 \nonumber \\ &&
       + \frac{\spa2.6^2 \spa3.4 \spb4.5^2 s_{126}^2}{ \spa1.6 \spab2.1+6.5 \spab6.1+2.5^3}
         \, \LsminOneTMHT{s_{56}}{s_{126}}{s_{12}}{s_{34}}
 \nonumber \\ &&
       - \frac{\spa1.2^2 \spa3.4 \spb4.5^2 s_{126}^2}{ \spa2.6 \spab1.2+6.5^3 \spab6.1+2.5}
         \, \LsminOneTMH{s_{15}}{s_{126}}{s_{26}}{s_{34}}
 \nonumber \\ &&
       + \frac{\spa3.6^2 \spb1.2^2 s_{125}^2 \spb3.4}{ \spb1.5 \spab6.1+2.5 \spab6.1+5.2^3}
         \, \LsminOneTMH{s_{26}}{s_{125}}{s_{15}}{s_{34}}
\nonumber \\ &&
- \Big( T_\text{antisym} + T_\text{sym} \Big) \, \iTm{s_{12}}{s_{34}}{s_{56}} 
\nonumber \\ &&
- T_\text{SL} \, \iTm{s_{15}}{s_{26}}{s_{34}}
 \nonumber \\ &&
          + \frac{\spa1.3 \spb2.6 \spb4.3 \spab3.1+4.6}{\spb2.5 \spab1.2+6.5 \spab1.3+4.2}
	    \, \Lnrat{ - s_{12}}{ - s_{15}}
          + \frac{\spa1.3 \spa2.3 \spb4.3 s_{345}}{\spa1.6 \spab1.2+6.5 \spab6.1+2.5}
	    \, \Lnrat{ - s_{15}}{ - s_{34}} 
\nonumber \\ &&
          - \Big( \frac{\spa1.2 \spa3.6 \spb4.6 s_{345}}{\spa1.6 \spab1.2+6.5 \spab6.1+2.5}
          + \frac{\spa1.2 \spa3.5 \spb4.6 }{\spa1.6 \spab1.2+6.5}
          - \frac{\spa1.3 \spa2.6 \spb1.6 \spb4.6 }{\spb1.5 \spa1.6 \spab1.2+6.5}
 \nonumber \\ &&
          + \frac{\spa3.4 \spb2.6 \spb3.4 \spab3.1+2.4}{\spb2.5 \spab1.3+4.2 \spab6.1+2.5} 
          + \frac{\spa2.6 \spa3.5 \spb4.6}{\spa1.6 \spab6.1+2.5}
          + \frac{\spa3.4 \spb4.6 \spab2.1+6.5 \spab6.1+2.4}{\spab1.2+6.5 \spab6.1+2.5^2} 
	  \Big) \, \Lnrat{ - s_{12}}{ - s_{34}}
\nonumber \\ &&
          - \frac{\spa1.3 \spa2.6 \spb4.6 s_{345}}{\spa1.6 \spab1.2+6.5 \spab6.1+2.5}
	    \, \Lnrat{ - s_{34}}{ - s_{126}} 
          - \frac{\spa1.3 \spb1.4 s_{345} \spab2.1+6.5 }{\spb1.5 \spa1.6 \spab1.2+6.5 \spab6.1+2.5}
	    \, \Lnrat{ - s_{12}}{ - s_{126}} 
 \nonumber \\ &&
          - 2 \, \Lnrat{ - s_{34}}{ - s_{15}} \frac{\spa1.2 \spb2.6 \spab3.1+5.4 \spab5.2+6.1 s_{34}}{\DeltaTM1526 \spab1.2+6.5 \spab6.1+5.2}
 \nonumber \\ &&
          - \Lnrat{ - s_{15}}{ - s_{26}} \frac{\spab1.2+5.6 \spab3.1+5.4 \spab5.2+6.1 s_{34}}{\DeltaTM1526 \spab1.2+6.5 \spab6.1+5.2}
 \nonumber \\ &&
          + \Lnrat{ - s_{15}}{ - s_{26}} \frac{(s_{15}-s_{26}) \spab1.2+5.6 \spab3.1+5.4 \spab5.2+6.1}{\DeltaTM1526 \spab1.2+6.5 \spab6.1+5.2}
 \nonumber \\ &&
- \Lnrat{ - s_{34}}{ - s_{26}} \frac{s_{34} (s_{34}-s_{15}-s_{26}) \spb5.6 \spab3.1+5.4 \spab5.2+6.1}{\spb1.5 \, \DeltaTM1526 \spab1.2+6.5 \spab6.1+5.2}
 \nonumber \\ &&
          - \Lnrat{ - s_{56}}{ - s_{12}} 
      \frac{\spab3.1+2.4 \spab5.1+2.6 (2 s_{34} s_{56}-(s_{34}+s_{56}-s_{12}) s_{234})}
      {\spab1.3+4.2 \spab6.1+2.5 \, \DeltaTM1234}
 \nonumber \\ &&
          + \Lnrat{ - s_{12}}{ - s_{34}} \frac{s_{34} (s_{234}-s_{134}) \spab3.1+2.4 \spab5.1+2.6}{\spab1.3+4.2 \spab6.1+2.5 \, \DeltaTM1234}
 \nonumber \\ &&
          - \Big( \frac{\spa1.3 \spa5.6 \spb2.4 \spab2.1+5.6}{\spa1.6 \spab1.2+6.5 \spab6.1+5.2}
          - \frac{\spa1.3 \spa2.5 \spb4.6 }{\spa1.6 \spab1.2+6.5}
 \nonumber \\ &&
          + \frac{\spa1.3 \spa2.3 \spb1.2 \spb3.4}{\spb2.5 \spa1.6 \spab6.1+2.5}
          + \frac{\spa3.5 \spa5.6 \spb4.6}{\spa1.6 \spab6.1+5.2}
	  \Big) \, \Lnrat{ - s_{15}}{ - s_{34}}
\nonumber \\ &&
          - \Big( \frac{\spa1.3 \spa2.6 \spb1.2 \spb2.4  \spab2.1+5.6}{ \spb1.5 \spa1.6 \spab1.2+6.5 \spab6.1+5.2}
          + \frac{\spa3.6 \spa5.6 \spb1.6 \spb4.6 }{\spb1.5 \spa1.6 \spab6.1+5.2}
          - \frac{\spa3.6 \spa5.6 \spb1.2 \spb2.4 \spab2.1+5.6}{\spb1.5 \spa1.6 \spab6.1+5.2^2} 
            \Big) \, \Lnrat{ - s_{26}}{ - s_{34}}
 \nonumber \\ &&
          + \frac{\spa3.6 \spab5.1+2.6 (\spab1.5+6.1 \spb4.5+\spab6.1+2.4 \spb5.6)}{\spab1.3+4.2 \spab6.1+2.5^2}
	    \, \Lnrat{ - s_{12}}{ - s_{56}}
 \nonumber \\ &&
          - \frac{\spa1.2 \spa2.6 \spa1.3 \spb1.6 \spb2.4 }{\spb1.5 \spa1.6^2 \spab1.2+6.5}
	    \, \Lnrat{ - s_{12}}{ - s_{26}}
          - \frac{\spa1.5 \spa3.4 \spab1.2+3.4 \spab5.1+6.4}{\spa1.6^2 \spab1.3+4.2 s_{234}}
	    \, \Lnrat{ - s_{15}}{ - s_{56}}
 \nonumber \\ &&
          - \frac{\spa1.5 \spa3.4 \spa5.6 \spb2.4 \spb4.6 }{\spa1.6 \spab1.3+4.2 \spab6.1+5.2}
	    \, \Lnrat{ - s_{156}}{ - s_{34}}
          - \frac{\spa3.6 \spb1.5 \spb3.4 \spab3.1+2.6 }{\spab6.1+2.5^2 \spb2.5}
	    \, \Lnrat{ - s_{125}}{ - s_{12}}
 \nonumber \\ &&
          + \frac{\spb3.4}{\spb2.5} \Big( -\frac{\spa3.6 \spb1.2^2 \spab3.1+5.6}{\spb1.5 \spab6.1+5.2^2} 
          - \frac{\spb1.2 \spb2.4 \spab3.2+4.1 \spa3.4}{\spb1.5 \spab6.1+5.2^2} 
          + \frac{\spb4.5 \spab3.2+6.1 \spa3.4}{\spab6.1+2.5^2} 
	  \Big) \, \Lnrat{ - s_{34}}{ - s_{125}}
\nonumber \\ &&
          - \Big( \frac{\spb4.5 s_{345} \spab2.1+6.5 \spab5.1+2.4}{\spb3.4 \spab1.2+6.5 \spab6.1+2.5^2} 
          + \frac{\spa2.5 \spb4.6 \spb4.5 s_{345}}{\spb3.4 \spab1.2+6.5 \spab6.1+2.5}
	    \Big) \, \Lzero{ - s_{126}}{ - s_{34}}
 \nonumber \\ &&
          - \Big( \frac{\spb2.4 \spab2.1+5.4 \spab5.1+6.2^2}{\spb3.4 \spab1.3+4.2 \spab6.1+5.2^2 }
          - \frac{\spa2.5 \spb2.4 \spb4.6 \spab5.1+6.2}{\spb3.4 \spab1.3+4.2 \spab6.1+5.2} 
	    \Big) \, \Lzero{ - s_{156}}{ - s_{34}}
 \nonumber \\ &&
          - \frac{\spa1.5^2 \spa3.4 \spb1.4 \spab1.2+3.4}{\spb5.6 \spa1.6^2 \spa5.6 \spab1.3+4.2}
	    \, \Lzero{ - s_{156}}{ - s_{56}}
 \nonumber \\ &&
          + s_{34} \Big( \frac{\spa3.6^2 \spb1.6 \spb3.4 }{\spab6.1+2.5^2 \spab6.1+5.2}
          - \frac{\spa3.6 \spb1.6 \spb2.4 \spab6.2+5.1}{\spb1.5 \spab6.1+2.5 \spab6.1+5.2^2}
	    \Big) \, \Lzero{ - s_{125}}{ - s_{34}}
 \nonumber \\ &&
          - \Big( \frac{\spa1.3 \spa2.6 \spb1.6 s_{345} \spab6.1+2.4}{\spb1.2 \spb1.5 \spa1.2 \spa1.6^2 \spab6.1+2.5}
          - \frac{\spa2.6 \spa3.6 \spb4.6 s_{345}}{\spab6.1+2.5^2 \spa1.6}
	    \Big) \, \Lzero{ - s_{126}}{ - s_{12}}
 \nonumber \\ &&
          - \frac{\spa3.5 \spb1.5 \spb3.4 \spab3.1+2.5}{\spab6.1+2.5^2 \spb2.5}
	    \, \Lzero{ - s_{125}}{ - s_{12}}
          - \frac{\spa2.3 \spb1.2^2 \spb3.4 \spab3.1+5.2}{\spb1.5 \spb2.5 \spab6.1+5.2^2}
	    \, \Lzero{ - s_{125}}{ - s_{15}}
 \nonumber \\ &&
          - \Big( \frac{\spa3.4 \spa5.6 \spb1.2 \spb4.6 \spab6.1+5.4}{\spb1.5 \spa1.6 \spab6.1+5.2^2}
          - \frac{\spa3.4 \spa5.6 \spb4.6 \spab6.1+5.4}{\spb1.5 \spa1.6^2 \spab6.1+5.2}
	   \Big) \, \Lzero{ - s_{156}}{ - s_{15}} 
\nonumber \\ &&
          + \frac{\spa3.6 \spb1.6 \spb4.6 \spab6.2+5.1}{\spb1.5 \spab6.1+2.5 \spab6.1+5.2 }
	    \, \Lone{ - s_{125}}{ - s_{34}}
          - \frac{\spa3.4 \spa5.6^2 \spb4.6^2 }{\spb1.5 \spa1.6 \spa1.5 \spab6.1+5.2}
	    \, \Lone{ - s_{156}}{ - s_{15}}
 \nonumber \\ &&
          + \frac{\spa2.6 \spa3.6 \spb1.6 \spb4.6 s_{345}}{\spb1.2 \spb1.5 \spa1.2 \spa1.6 \spab6.1+2.5}
	    \, \Lone{ - s_{126}}{ - s_{12}}
          - \frac{\spa3.4 \spa5.6 \spb1.4 \spb4.6 }{\spb1.5 \spa1.6 \spab6.1+5.2}
 \nonumber \\ &&
  - L(5_{\gamma}^-,6_g^+)
+ \Vpole_\slc \, A^u_\tree  ( 1_{\bar u}^+,2_{d}^-,3_{\nu}^-,4_{\ell}^+, 5_{\gamma}^-, 6_g^+ )
\end{eqnarray}

\subsection{Amplitudes for radiation from the $W$-boson and positron}
These pieces are characterized as being proportional to the difference of the quark charges.
\subsubsection{Decomposition of leading colour amplitude}

It is convenient to decompose the leading colour amplitude into two contributions,
\begin{eqnarray}
A^\dk_\lc  ( 1_{\bar u}^+,2_{d}^-,3_{\nu}^-, 4_{\ell}^+, 5_{\gamma}^{h_5}, 6_g^{h_6} )
&=& A^\lrad_\lc  ( 1_{\bar u}^+,2_{d}^-,3_{\nu}^-, 4_{\ell}^+, 5_{\gamma}^{h_5}, 6_g^{h_6} ) \nonumber \\
&+& A^\Wrad_\lc  ( 1_{\bar u}^+,2_{d}^-,3_{\nu}^-, 4_{\ell}^+, 5_{\gamma}^{h_5}, 6_g^{h_6} )
\end{eqnarray}
because the contribution $A^\Wrad_\lc$ exhibits a simple rule for flipping the helicities of the photon
and gluon,
\begin{equation}
\label{Wradflip}
  A^\Wrad_\lc  ( 1_{\bar u}^+,2_{d}^-,3_{\nu}^-, 4_{\ell}^+, 5_{\gamma}^{-h_5}, 6_g^{-h_6} ) \\
= - A^\Wrad_\lc  ( 2_{\bar u}^+,1_{d}^-,4_{\nu}^-, 3_{\ell}^+, 5_{\gamma}^{h_5}, 6_g^{h_6} )
 \, \bigl\{ \langle \rangle \leftrightarrow [] \bigr\}
\end{equation}

\subsubsection{Leading colour, radiation from positron}
There are four independent contributions for $A^\lrad_\lc$:

\begin{eqnarray}
 && A^\lrad_\lc  ( 5_{\gamma}^+, 6_g^+ ) = 
 \frac{1}{\spa2.5 \spa1.6 \spa2.6} \Bigg[ \nonumber \\ &&
      \frac{\spa1.2 \spa2.3}{\spa4.5}
      \left(\frac{\spa2.6 \spb1.6 s_{45}}{2 s_{345}}-\spab2.4+5.1 \right)
       \, \Lzero{-s_{26}}{-s_{345}}
 \nonumber \\ &&
      +\frac{\spa1.2^2 \spab2.4+5.1 \spab3.2+6.1}{2 \spa4.5 s_{345}} 
       \, \Lone{-s_{26}}{-s_{345}}
      +\frac{\spa2.3^2 \spab2.1+6.3}{\spa4.5}
       \, \LsminOne{-s_{16}}{-s_{345}}{-s_{26}}
        \Bigg]
 \nonumber \\ &&
      + \Vpole_\lc(s_{26}) \, A^\lrad_\tree  ( 1_{\bar u}^+,2_{d}^-,3_{\nu}^-,4_{\ell}^+, 5_{\gamma}^+, 6_g^+ )
\end{eqnarray}
 \begin{eqnarray}
 && A^\lrad_\lc  ( 5_{\gamma}^+, 6_g^- ) =
\frac{1}{\spa2.5 \spb1.6 \spb2.6} \Bigg[  \nonumber \\ &&
        \frac{\spa2.3 \spb1.2}{2 \spa4.5}
        \, (2 \spab2.4+5.1 s_{345} - \spa2.6 \spb1.6 s_{45})
        \, \frac{\Lzero{-s_{16}}{-s_{345}}}{s_{345}}
 \nonumber \\ &&
       -\frac{(\spa2.3 \spb1.2)^2 \spab2.4+5.3}{2 \spa4.5}
        \, \frac{\Lone{-s_{16}}{-s_{345}}}{s_{345}}
       +\frac{\spab3.2+6.1 \spab2.4+5.1}{\spa4.5}
        \, \LsminOne{-s_{26}}{-s_{345}}{-s_{16}}
	\Bigg]
 \nonumber \\ &&
      + \Vpole_\lc(s_{16}) \, A^\lrad_\tree  ( 1_{\bar u}^+,2_{d}^-,3_{\nu}^-,4_{\ell}^+, 5_{\gamma}^+, 6_g^- )
\end{eqnarray}
 \begin{eqnarray}
 && A^\lrad_\lc  ( 5_{\gamma}^-, 6_g^+ ) =
-\frac{1}{\spb1.5 \spa2.6 \spa1.6} \, \Bigg[  \nonumber \\ &&
       \spa1.2 \spb1.4 \Big(\frac{\spa2.3 \spb1.4 s_{345}}{\spb4.5}
      -\frac{1}{2} \spa2.6 \spa3.5 \spb1.6 \Big)
      \, \frac{\Lzero{-s_{26}}{-s_{345}}}{s_{345}}
 \nonumber \\ &&
      -\frac{(\spa1.2 \spb1.4)^2 \spab3.2+6.1}{2 \spb4.5}
       \, \frac{\Lone{-s_{26}}{-s_{345}}}{s_{345}}
      -\frac{\spa2.3 \spb1.4 \spab2.1+6.4}{\spb4.5}
       \, \LsminOne{-s_{16}}{-s_{345}}{-s_{26}}
        \Bigg]
 \nonumber \\ &&
      + \Vpole_\lc(s_{26}) \, A^\lrad_\tree  ( 1_{\bar u}^+,2_{d}^-,3_{\nu}^-,4_{\ell}^+, 5_{\gamma}^-, 6_g^+ )
\end{eqnarray}
 \begin{eqnarray}
 && A^\lrad_\lc  ( 1_{\bar u}^+,2_{d}^-,3_{\nu}^-, 4_{\ell}^+, 5_{\gamma}^-, 6_g^- ) =
\frac{1}{\spb1.5 \spb1.6 \spb2.6} \Bigg[  \nonumber \\ &&
       \frac{\spa2.3 \spb1.2 \spb1.4^2}{\spb4.5}
        \, \Lzero{-s_{16}}{-s_{345}}
      -\frac{1}{2} \spa2.6 \spa3.5 \spb1.2 \spb1.4 \spb1.6
       \, \frac{\Lzero{-s_{16}}{-s_{345}}}{s_{345}}
 \nonumber \\ &&
      +\frac{\spab2.1+6.4 \spa2.3 \spb1.2^2 \spb1.4}{2 \spb4.5}
       \, \frac{\Lone{-s_{16}}{-s_{345}}}{s_{345}}
      +\frac{\spab3.2+6.1 \spb1.4^2}{\spb4.5}
       \, \LsminOne{-s_{26}}{-s_{345}}{-s_{16}}
      \Bigg]
 \nonumber \\ &&
      + \Vpole_\lc(s_{16}) \, A^\lrad_\tree  ( 1_{\bar u}^+,2_{d}^-,3_{\nu}^-,4_{\ell}^+, 5_{\gamma}^-, 6_g^- )
\end{eqnarray}
 
\subsubsection{Leading colour, radiation from $W$-boson}
Only two extra pieces for $A^\Wrad_\lc$, the other two obtained by symmetry, Eq.~(\ref{Wradflip}):

\begin{eqnarray}
 && A^\Wrad_\lc ( 5_{\gamma}^+, 6_g^+ ) =
 - \frac{1}{\spa2.5 \spa1.6 \spa2.6} \Bigg[
 \nonumber \\ &&
      \spa2.1 \spa2.3 \, \Lzero{-s_{26}}{-s_{345}}
       \left(\spb1.5 \spb4.3 \spa2.3
      +\frac{\spb1.6 \spb4.5 \spa2.6}{2}  (1+s_{34}/s_{345})\right)
 \nonumber \\ &&
      -\spb4.3 \spab2.3+4.5 \spa2.3^2
       \, \LsminOne{-s_{16}}{-s_{345}}{-s_{26}}
 \nonumber \\ &&
      +\frac{1}{2} \spa2.1^2 \spab3.2+6.1 (\spb1.5 \spb4.3 \spa2.3+\spb4.5 \spb1.6 \spa2.6)
       \, \frac{\Lone{-s_{26}}{-s_{345}}}{s_{345}}
       \Bigg]
 \nonumber \\ &&
      + \Vpole_\lc(s_{26}) \, A^\Wrad_\tree  ( 1_{\bar u}^+,2_{d}^-,3_{\nu}^-,4_{\ell}^+, 5_{\gamma}^+, 6_g^+ )
\end{eqnarray}
 \begin{eqnarray}
 && A^\Wrad_\lc  ( 5_{\gamma}^+, 6_g^- ) =
\frac{1}{\spa2.5 \spb1.6 \spb2.6} \Bigg[  \nonumber \\ &&
        \frac{1}{2} \spb1.2 \spa2.3 \Big(2\spa2.3 \spb3.4 \spb1.5 s_{345}-\spa2.6 \spb1.6 \spb4.5 (s_{345}+s_{34}) \Big)
        \, \frac{\Lzero{-s_{16}}{-s_{345}}}{s_{345}}
 \nonumber \\ &&
       +\spab3.2+6.1 (\spb1.4 \spab2.3+4.5+\spa2.5 \spb1.5 \spb4.5) \, \LsminOne{-s_{26}}{-s_{345}}{-s_{16}}
 \nonumber \\ &&
       +\frac{1}{2} (\spa2.3 \spb1.2)^2 \spb3.4 \spab2.1+6.5
        \, \frac{\Lone{-s_{16}}{-s_{345}}}{s_{345}} 
  \Bigg]
 \nonumber \\ &&
      + \Vpole_\lc(s_{16}) \, A^\Wrad_\tree  ( 1_{\bar u}^+,2_{d}^-,3_{\nu}^-,4_{\ell}^+, 5_{\gamma}^+, 6_g^- )
\end{eqnarray}
 
\subsubsection{Decomposition of subleading colour amplitude}
It is convenient to decompose into two contributions,
\begin{eqnarray}
A^\dk_\slc  ( 1_{\bar u}^+,2_{d}^-,3_{\nu}^-, 4_{\ell}^+, 5_{\gamma}^{h_5}, 6_g^{h_6} )
&=& A^\lrad_\slc  ( 1_{\bar u}^+,2_{d}^-,3_{\nu}^-, 4_{\ell}^+, 5_{\gamma}^{h_5}, 6_g^{h_6} ) \nonumber \\
&+& A^\Wrad_\slc  ( 1_{\bar u}^+,2_{d}^-,3_{\nu}^-, 4_{\ell}^+, 5_{\gamma}^{h_5}, 6_g^{h_6} )
\end{eqnarray}
because the contribution $A^\Wrad_\slc$ exhibits a simple rule for flipping the helicities of the photon
and gluon,
\begin{equation}
  A^\Wrad_\slc  ( 1_{\bar u}^+,2_{d}^-,3_{\nu}^-, 4_{\ell}^+, 5_{\gamma}^{-h_5}, 6_g^{-h_6} ) \nonumber \\
= - A^\Wrad_\slc  ( 2_{\bar u}^+,1_{d}^-,4_{\nu}^-, 3_{\ell}^+, 5_{\gamma}^{h_5}, 6_g^{h_6} )
 \, \bigl\{ \langle \rangle \leftrightarrow [] \bigr\}
\end{equation}

\subsubsection{Subeading colour, radiation from positron}
The results for all four helicities for the contribution $A^\lrad_\slc$ are:
\begin{eqnarray}
 && A^\lrad_\slc  ( 5_{\gamma}^+, 6_g^+ ) =
\frac{1}{\spa1.6 \spa2.5 \spa4.5} \Bigg[  \nonumber \\ &&
      -\frac{\spa1.3 \spb1.6^2 \spaa1.2+6.4+5.2}{2 \spb2.6}
       \, \frac{\Lone{-s_{345}}{-s_{26}}}{s_{26}}
 \nonumber \\ &&
      + \spb1.6 \bigg( \spa2.3 \spa1.3 \spab2.4+5.3
      +\frac{\spa1.2 \spa3.6 \spaa1.2+6.4+5.2}{\spa1.6}
      \bigg)
       \, \frac{\Lzero{-s_{345}}{-s_{26}}}{s_{26}}
 \nonumber \\ &&
      +\frac{\spb1.6 \spa2.6 \spab2.4+5.6 \spa3.6 s_{126}}
            {\spa2.6 \spb1.2}
       \, \frac{\Lone{-s_{345}}{-s_{12}}}{s_{12}}
\nonumber \\ &&
      +\frac{\spb1.6 (\spa1.6 \spa2.3 s_{45}-\spa3.6 \spaa1.2+6.4+5.2)}{\spa1.6 \spb1.2}
       \, \Lzero{-s_{345}}{-s_{12}}
 \nonumber \\ &&
      +\frac{\spa2.3^2 \spab2.4+5.3}{\spa2.6}
       \, \LsminOne{-s_{12}}{-s_{345}}{-s_{16}}
 \nonumber \\ &&
      -\frac{\spa1.2^2 \spa3.6 \spaa2.4+5.1+2.6}{\spa1.6^2 \spa2.6}
       \, \LsminOne{-s_{12}}{-s_{345}}{-s_{26}}
 \nonumber \\ &&
      +\frac{1}{2 \spb2.6 \spb1.2} \bigg(
       2 \spa2.3 \spb1.6 \spb2.6 s_{45}
      -\spab3.1+2.6 \Big( \spb1.6 \spab2.4+5.2 + \spb2.6 \spab2.4+5.1
      \Big) \bigg)
      \Bigg]
 \nonumber \\ &&
      + \left( \Vpole_\slc + \frac{1}{2}\right) \, A^\lrad_\tree  ( 1_{\bar u}^+,2_{d}^-,3_{\nu}^-,4_{\ell}^+, 5_{\gamma}^+, 6_g^+ )
\end{eqnarray}
 \begin{eqnarray}
 && A^\lrad_\slc  ( 5_{\gamma}^+, 6_g^- ) =
\frac{1}{\spa2.5 \spa4.5 \spb2.6} \Bigg[ \nonumber \\ &&
     -\frac{(s_{25}+s_{24}) \spa2.6^2 \spab3.1+6.2 }{2 \spa1.6}
      \, \frac{\Lone{-s_{345}}{-s_{16}}}{s_{16}}
 \nonumber \\ &&
     + \frac{\spa2.6 \Big(\spb1.2 \spb2.6 \spb4.5 \spa2.3 \spa4.5
     - \big(\spab3.2+6.1 \spb2.6 - \spab3.1+2.6 \spb1.2) (s_{24}+s_{25} \big)
          \Big)}{\spb2.6}
      \, \frac{\Lzero{-s_{345}}{-s_{16}}}{s_{16}}
 \nonumber \\ &&
     -\frac{\spa2.6 s_{126} \spab2.4+5.6 \spa3.6}{\spa1.2}
      \, \frac{\Lone{-s_{345}}{-s_{12}}}{s_{12}}
     +\frac{\spa2.6 \spab3.4+5.6 \spab2.4+5.2}{\spa1.2 \spb2.6}
      \, \Lzero{-s_{345}}{-s_{12}}
 \nonumber \\ &&
     -\frac{\spab2.4+5.1 \spab3.2+6.1}{\spb1.6}
      \, \LsminOne{-s_{12}}{-s_{345}}{-s_{26}}
 \nonumber \\ &&
     - \frac{\spb1.2 \Big( \spab3.2+6.1 \spab2.1+3.6 \spb2.6 + \spa1.3 \spb1.6^2 s_{45}
            +\spab3.1+2.6 \spb1.6 s_{245}\Big)}{\spb2.6^2 \spb1.6}
      \, \LsminOne{-s_{12}}{-s_{345}}{-s_{16}}
 \nonumber \\ &&
     -\frac{1}{2 \spa1.2 \spa1.6}
      \, \Big((\spa1.6 \spa2.3+\spa1.3 \spa2.6) (\spa1.6 \spab2.4+5.1 + \spa2.6 \spab2.4+5.2)
     +2 \spa2.3 \spa2.6 \spa1.6 s_{45} \Big)
     \Bigg]
 \nonumber \\ &&
     + \left( \Vpole_\slc + \frac{1}{2}\right) \, A^\lrad_\tree  ( 1_{\bar u}^+,2_{d}^-,3_{\nu}^-,4_{\ell}^+, 5_{\gamma}^+, 6_g^- )
\end{eqnarray}
 \begin{eqnarray}
 && A^\lrad_\slc  ( 5_{\gamma}^-, 6_g^+ ) =
\frac{1}{\spa1.6 \spb1.5 \spb4.5} \Bigg[ \nonumber \\ &&
       \spa1.2 \spa3.6 \spb1.6 \spb1.4 \spb4.6
       \, \frac{\Lone{-s_{345}}{-s_{12}}}{s_{12}}
      -\frac{\spab1.2+6.4 \spa1.3 \spb1.4 \spb1.6^2}{2 \spb2.6}
       \, \frac{\Lone{-s_{345}}{-s_{26}}}{s_{26}}
 \nonumber \\ &&
      +\frac{\spb1.4 \spb1.6 \big(
            \spa1.6 \spab3.1+2.4 -\spa3.6 \spab1.2+6.4 \big)}{\spa1.6 \spb1.2}
      \, \Lzero{-s_{345}}{-s_{12}}
 \nonumber \\ &&
      +  \spb1.4 \spb1.6 \Big( \frac{\spa1.2 \spa3.6 \spab1.2+6.4}{\spa1.6}  - \spa1.3 \spab2.1+6.4 \Big)
        \, \frac{\Lzero{-s_{345}}{-s_{26}}}{s_{26}}
 \nonumber \\ &&
      +\frac{\spa1.2^2 \spa3.6 \spb1.4 \spab6.1+2.4}{\spa1.6^2 \spa2.6}
       \, \LsminOne{-s_{12}}{-s_{345}}{-s_{26}}
 \nonumber \\ &&
      -\frac{\spa2.3 \spb1.4 \spab2.1+6.4}{\spa2.6}
       \, \LsminOne{-s_{12}}{-s_{345}}{-s_{16}}
-\frac{\spb1.2 \spb1.4 \spb4.6}{2 \spb1.2 \spb2.6}
       \, \big(\spa1.3 \spb1.6 - \spa2.3 \spb2.6 \big)
       \Bigg]
 \nonumber \\ &&
      + \left( \Vpole_\slc + \frac{1}{2}\right) \, A^\lrad_\tree  ( 1_{\bar u}^+,2_{d}^-,3_{\nu}^-,4_{\ell}^+, 5_{\gamma}^-, 6_g^+ )
\end{eqnarray}
 \begin{eqnarray}
 && A^\lrad_\slc  ( 5_{\gamma}^-, 6_g^- ) =
     \frac{1}{\spb1.5 \spb2.6 \spb4.5} \Bigg[  \nonumber \\ &&
      -\frac{\spa2.6^2 \spb1.4 \spb2.4 \spab3.1+6.2}{2 \spa1.6}
       \, \frac{\Lone{-s_{345}}{-s_{16}}}{s_{16}}
      -\spa2.6 \spa3.6 \spb1.2 \spb1.4 \spb4.6
       \, \frac{\Lone{-s_{345}}{-s_{12}}}{s_{12}}
 \nonumber \\ &&
      -\frac{\spa2.6 \spb1.4}{\spb2.6}
       \, \Big(
       \spab3.1+6.2 \spb1.2 \spb4.6
      +\spab3.2+6.1 \spb2.4 \spb2.6 \Big)
      \, \frac{\Lzero{-s_{345}}{-s_{16}}}{s_{16}}
 \nonumber \\ &&
      -\frac{\spa2.6 \spb1.4}{\spa1.2 \spb2.6}
        \Big( \spab3.1+2.6 \spb2.4 + \spa3.6 \spb2.6 \spb4.6 \Big)
      \, \Lzero{-s_{345}}{-s_{12}}
 \nonumber \\ &&
      -\frac{\spb1.4^2 \spab3.2+6.1}{\spb1.6}
       \, \LsminOne{-s_{12}}{-s_{345}}{-s_{26}}
 \nonumber \\ &&
      +\frac{\spb1.2^2 \spb1.4 \spb4.6 \spab3.1+2.6}{\spb2.6^2 \spb1.6}
       \, \LsminOne{-s_{12}}{-s_{345}}{-s_{16}}
      -\frac{\spa3.6 \spb1.4 \big(
        \spa2.6 \spb2.4 - \spa1.6 \spb1.4
       \big)}
      {2 \spa1.6}
      \Bigg]
 \nonumber \\ &&
      + \left( \Vpole_\slc + \frac{1}{2}\right) \, A^\lrad_\tree  ( 1_{\bar u}^+,2_{d}^-,3_{\nu}^-,4_{\ell}^+, 5_{\gamma}^-, 6_g^- )
\end{eqnarray}
 
\subsubsection{Subeading colour, radiation from $W$-boson}
The symmetry noted above means that we only have to give results for two of the helicities for $A^\Wrad_\slc$: 
\begin{eqnarray}
 && A^\Wrad_\slc  ( 5_{\gamma}^+, 6_g^+ ) = \nonumber \\ &&
       \frac{1}{2 \spa1.6 \spa2.5 \spb2.6 \spb1.2} \Big(
       \spa1.2 \spa3.5 \spb1.6 (\spb1.5 \spb2.6+\spb1.6 \spb2.5) \spb4.5
 \nonumber \\ &&
      -(\spb1.4 \spb2.6+\spb1.6 \spb2.4) \spab2.3+4.5 \spab3.1+2.6
 \nonumber \\ &&
      - \spa2.3 \spb4.5 \spb5.6 \spb2.6 \spab5.3+4.1
      + \spa2.3 \spb4.5 \spb1.6 \spb2.5 \spab5.3+4.6
      \Big)
 \nonumber \\ &&
      +\frac{\spb1.6}{\spa1.6^2 \spa2.5} \bigg(
       \spa1.2^2 \spa3.6 \spb4.5 \spab5.3+4.5
      +(\spa1.6 \spa2.3-\spa1.2 \spa3.6) \spa1.3 \spb3.4 \spab2.3+4.5
       \bigg)
       \, \frac{\Lzero{-s_{26}}{-s_{345}}}{s_{345}}
 \nonumber \\ &&
      -\frac{\spa1.2 \spb1.6}{\spa1.6^2 \spa2.5} \bigg(
       \spa1.3 \spa3.6 \spb3.4 \spab2.3+4.5
      -\frac{1}{2} \left(\spa1.2 \spa3.6+\spa1.3 \spa2.6\right) \spb4.5 \spab5.3+4.5
       \bigg)
       \, \frac{\Lzero{-s_{345}}{-s_{12}}}{s_{12}}
 \nonumber \\ &&
      -\frac{\spa1.2^2 \spa3.6}{\spa1.6^3 \spa2.6 \spa2.5} \bigg(
       \spa2.5 \spab6.3+4.5 \spb4.5
      +\spab2.3+4.5 \spab6.3+5.4
      \bigg)
       \,  \LsminOne{-s_{12}}{-s_{345}}{-s_{26}}
 \nonumber \\ &&
      +\frac{\spa2.3^2 \spb3.4 \spab2.3+4.5}{\spa1.6 \spa2.6 \spa2.5}
       \, \LsminOne{-s_{12}}{-s_{345}}{-s_{16}}
 \nonumber \\ &&
      +\frac{\spa1.2 \spb4.5 \spab5.3+4.5
      -\spa1.3 \spb3.4 \spab2.3+4.5}{2 \spa1.6 \spa2.6 \spa2.5 \spb2.6^2}
       \, \spa1.3 \spb1.6^2 \, \Lone{-s_{345}}{-s_{26}}
 \nonumber \\ &&
      +\frac{\spb1.6}{\spa1.6 \spa2.5 \spa1.2 \spb1.2^2}
       \, \Lone{-s_{345}}{-s_{12}}
       \, \bigg((\spab2.3+4.5 \spa3.6 \spb4.6
      +\spa2.6 \spa3.5 \spb4.5 \spb5.6
 \nonumber \\ &&
      +\frac{1}{2} \spa2.3 \spb4.5 (s_{346}-s_{56})) s_{345}
      -\frac{1}{2} \spa2.3 \spb4.5 s_{12} s_{34}
      \bigg)
 \nonumber \\ &&
      + \left( \Vpole_\slc + \frac{1}{2}\right) \, A^\Wrad_\tree  ( 1_{\bar u}^+,2_{d}^-,3_{\nu}^-,4_{\ell}^+, 5_{\gamma}^+, 6_g^+ )
\end{eqnarray}
 \begin{eqnarray}
 && A^\Wrad_\slc  ( 5_{\gamma}^+, 6_g^- ) = \nonumber \\ &&
      \Lzero{-s_{345}}{-s_{16}} \frac{\spa2.6 \spab3.4+5.2}{\spa1.6 \spa2.5 \spb2.6 \spb2.6}
      \, \bigg(
      \spab2.1+6.2 \spb4.5 
     +\spa2.3 \spb3.4 \spb2.5 \bigg)
 \nonumber \\ &&
     -\Lzero{-s_{345}}{-s_{16}} \frac{\spa2.6}{\spa1.6 \spa2.5 \spb2.6 \spb1.6}
      \, \bigg(-\spb1.2 \spa2.3 \spb4.5 (s_{35}+s_{45})
 \nonumber \\ &&
     -2 \spa2.6 \spb1.6 \spb4.5 \spab3.4+5.2
     +2 \spa2.3 \spb3.4 \spb1.5 \spab3.4+5.2
      \bigg)
 \nonumber \\ &&
     - \Lone{-s_{345}}{-s_{16}} \frac{\spa2.6^2 \spab3.4+5.2}{2 \spa1.6^2 \spb1.6 \spa2.5 \spb2.6}
      \, (\spa2.3 \spb3.4 \spb2.5-\spb4.5 \spab2.3+4+5.2)
 \nonumber \\ &&
     + \frac{\Lone{-s_{345}}{-s_{12}}}{s_{12}}
       \, \frac{\spa2.6 \spa3.6 (\spa2.5 \spb4.5 \spb5.6 +\spab2.3+4.5 \spb4.6) s_{345}}{\spa1.2 \spa2.5 \spb2.6}
 \nonumber \\ &&
     + \Lzero{-s_{345}}{-s_{12}} \frac{\spa2.6}{\spa1.2 \spa2.5 \spb2.6^2}
      \, \Bigg( \spab3.1+6.2 (\spa1.2 \spb1.6 \spb4.5 + \spa2.3 \spb3.4 \spb5.6)
 \nonumber \\ &&
     +\spa2.3 \spb2.6 \spb4.5 ( s_{345} - s_{34})
     \Bigg)
 \nonumber \\ &&
     - \frac{\spab3.2+6.1 (\spa2.3 \spb1.5 \spb3.4-\spa2.6 \spb1.6 \spb4.5)}
            {\spa2.5 \spb2.6 \spb1.6}
      \, \LsminOne{-s_{12}}{-s_{345}}{-s_{26}} 
 \nonumber \\ &&
     - 
      \frac{\spb1.2^2}{\spa2.5 \spb2.6^2}
       \Bigg(
       \frac{\spa1.2 \spb4.5 \spab3.4+5.6}{\spb2.6}
     + \frac{\spa2.3^2 \spb3.4 \spb5.6}{\spb1.6}
     + \frac{\spa2.3 \spa1.3 \spb3.4 \spb5.6}{\spb2.6}
     \Bigg) \, \LsminOne{-s_{12}}{-s_{345}}{-s_{16}}
 \nonumber \\ &&
     +\frac{(\spa1.6 \spa2.3+\spa1.3 \spa2.6) \spab6.1+2.5 \spb3.4 \spa2.3
-\spa2.6 \spb4.5 \big(
      2 \spa1.6 \spa2.3  s_{34}
      +\spa1.2 \spa3.6 s_{126}
     \big)}{2\spb2.6 \spa2.5 \spa1.6 \spa1.2}
 \nonumber \\ &&
     + \left( \Vpole_\slc + \frac{1}{2}\right) \, A^\Wrad_\tree  ( 1_{\bar u}^+,2_{d}^-,3_{\nu}^-,4_{\ell}^+, 5_{\gamma}^+, 6_g^- )
\end{eqnarray}

\section{Seven-parton process at tree level}
\label{Sevenpointtrees}
A computer readable representation of the results in this
appendix accompanies the arXiv version of this article.
\subsection{Gluon radiation}
\label{Sevenpointtrees:gluons}

We can employ the partial fraction relation for the W-boson propagators (Eq.~\ref{eq:Wpropagatorspartialfraction}) to express the entire helicity tree amplitude in terms of just three components
\begin{eqnarray}
  A^{(0)} (5^{h_5}_{\gamma},6^{h_6}_g,7^{h_7}_g) &=&
                                                    Q_{u} \, P(s_{34}) \, A^{u\Wrad}_\tree (5^{h_5}_{\gamma},6^{h_6}_g,7^{h_7}_g) + \nonumber \\
                                                 && Q_{d} \, P(s_{34}) \, A^{d\Wrad}_\tree (5^{h_5}_{\gamma},6^{h_6}_g,7^{h_7}_g) +  \nonumber \\
                                                 && (Q_u-Q_d) P(s_{345}) \, A^{\lrad\Wrad}_\tree (5^{h_5}_{\gamma},6^{h_6}_g,7^{h_7}_g) , 
\end{eqnarray}
where $A^{u\Wrad}_\tree$, $A^{d\Wrad}_\tree$ and $A^{\lrad\Wrad}_\tree$ are given by
\begin{gather}
  A^{u\Wrad}_\tree = A^{u}_\tree + \frac{A^{\Wrad}_\tree}{\spab5.(3+4).5} \; , \quad A^{d\Wrad}_\tree = A^{d}_\tree - \frac{A^{\Wrad}_\tree}{\spab5.(3+4).5} \nonumber \\
  A^{\lrad\Wrad}_\tree = A^{\lrad}_\tree - \frac{A^{\Wrad}_\tree}{\spab5.(3+4).5} .
\end{gather}
The following relation now holds 
\begin{gather}
  A^{u\Wrad}_\tree  ( 1_{\bar u}^+,2_{d}^-,3_{\nu}^-, 4_{\lrad}^+, 5_{\gamma}^{-h_5}, 6_g^{-h_6}, 7_g^{-h_7} ) =  \nonumber \\
  A^{d\Wrad}_\tree  ( 2_{\bar u}^+,1_{d}^-,4_{\nu}^-, 3_{\lrad}^+, 5_{\gamma}^{h_5}, 7_g^{h_7}, 6_g^{h_6} )  \, \bigl\{ \langle \rangle \leftrightarrow [] \bigr\},
\end{gather}
which suggests the following generalisation for $(n-5)$-gluon emission
\begin{gather}
  A^{u\Wrad}_\tree  ( 1_{\bar u}^+,2_{d}^-,3_{\nu}^-, 4_{\lrad}^+, 5_{\gamma}^{-h_5}, 6_g^{-h_6}, \dots, n_g^{-h_n} ) = \nonumber \\
  A^{d\Wrad}_\tree  ( 2_{\bar u}^+,1_{d}^-,4_{\nu}^-, 3_{\lrad}^+, 5_{\gamma}^{h_5}, n_g^{h_n}, \dots , 6_g^{h_6} ) \, \bigl\{ \langle \rangle \leftrightarrow [] \bigr\} . 
\end{gather}
Therefore, $A^{\lrad\Wrad}_\tree$ is required for all helicity configurations while it suffices to provide $A^{u\Wrad}_\tree$ and $A^{d\Wrad}_\tree$ for half of them.

\subsubsection{Tree $5_{\gamma}^-, 6_g^-, 7_g^-$}
\begin{gather}
  A^{u\Wrad}_\tree (5_{\gamma}^{-}, 6_g^{-}, 7_g^{-} ) = \frac{-2[14]^2⟨34⟩⟨5|3+4|2]}{[17][25][26][67]⟨5|3+4|5]} \\
  A^{d\Wrad}_\tree (5_{\gamma}^{-}, 6_g^{-}, 7_g^{-} ) = \frac{2[14]^2⟨34⟩⟨5|3+4|1]}{[15][17][26][67]⟨5|3+4|5]} \\
  A^{e\Wrad}_\tree (5_{\gamma}^{-}, 6_g^{-}, 7_g^{-} ) = \frac{-2[14]^2⟨45⟩s_{345}}{[17][26][35][67]⟨5|3+4|5]}
\end{gather}

\subsubsection{Tree $5_{\gamma}^+, 6_g^+, 7_g^+$}
\begin{equation}
  A^{e\Wrad}_\tree (5_{\gamma}^{+}, 6_g^{+}, 7_g^{+} ) = \frac{2⟨23⟩^2[45]s_{345}}{⟨17⟩⟨26⟩⟨35⟩⟨67⟩⟨5|3+4|5]}
\end{equation}

\subsubsection{Tree $5_{\gamma}^-, 6_g^-, 7_g^+$}
\begin{eqnarray}
  A^{u\Wrad}_\tree (5_{\gamma}^{-}, 6_g^{-}, 7_g^{+} ) &=& \frac{2⟨16⟩⟨34⟩⟨5|3+4|2]⟨6|1+7|4]^2}{⟨17⟩[25]⟨67⟩⟨1|6+7|2]⟨5|3+4|5]s_{167}}+\nonumber\\
                                                       && \frac{-2[27][34]⟨3|1+4|7]^2}{[25][26][67]⟨1|3+4|5]s_{134}}+\nonumber\\
                                                       && \frac{-2⟨15⟩[27]⟨34⟩[4|3+5|2+6|7]^2}{[26][67]⟨1|6+7|2]⟨1|3+4|5]⟨5|3+4|5]s_{267}}\phantom{+}\\
  A^{d\Wrad}_\tree (5_{\gamma}^{-}, 6_g^{-}, 7_g^{+} ) &=& \frac{2⟨16⟩⟨34⟩⟨6|1+7|4]^2⟨5|3+4|1+7|6⟩}{⟨17⟩⟨67⟩⟨1|6+7|2]⟨5|3+4|5]⟨6|1+7|5]s_{167}}+\nonumber\\
                                                       &&\frac{-2[17]^2⟨34⟩⟨6|2+3|4]^2⟨6|1+5|7]}{[15]⟨6|1+7|5][2|3+4|1+5|7]s_{157}s_{234}}+\nonumber\\
                                                       &&\frac{-2[14][27]⟨34⟩⟨5|2+6|7][4|3+5|2+6|7]}{[15][26][67]⟨1|6+7|2]⟨5|3+4|5]s_{267}}+\nonumber\\
                                                       &&\frac{[17][27]⟨34⟩[45]⟨5|2+6|7](2⟨5|2+6|7][24]-2[27][34]⟨35⟩)}{[15][26][67]⟨1|6+7|2]⟨5|3+4|5][2|3+4|1+5|7]}+\nonumber\\
                                                       &&\frac{-2[17]^2[27][34]⟨35⟩⟨3|2+6|7]}{[15][26][67]⟨5|3+4|5][2|3+4|1+5|7]}\phantom{+}\\
  A^{e\Wrad}_\tree (5_{\gamma}^{-}, 6_g^{-}, 7_g^{+} ) &=& \frac{2⟨16⟩⟨45⟩⟨6|1+7|4]^2s_{345}}{⟨17⟩[35]⟨67⟩⟨1|6+7|2]⟨5|3+4|5]s_{167}}+\nonumber\\
                                                       && \frac{-2[27]⟨45⟩[4|3+5|2+6|7]^2}{[26][35][67]⟨1|6+7|2]⟨5|3+4|5]s_{267}}\phantom{+}
\end{eqnarray}

\subsubsection{Tree $5_{\gamma}^+, 6_g^-, 7_g^+$}
\begin{eqnarray}
  A^{e\Wrad}_\tree (5_{\gamma}^{+}, 6_g^{-}, 7_g^{+} ) &=&\frac{2⟨16⟩[45]⟨6|1+7|4+5|3⟩^2}{⟨17⟩⟨35⟩⟨67⟩⟨1|6+7|2]⟨5|3+4|5]s_{167}}+\nonumber\\
                                                       && \frac{-2[27][45]⟨3|2+6|7]^2s_{345}}{[26]⟨35⟩[67]⟨1|6+7|2]⟨5|3+4|5]s_{267}}\phantom{+}
\end{eqnarray}

\subsubsection{Tree $5_{\gamma}^-, 6_g^+, 7_g^-$}
\begin{eqnarray}
  A^{u\Wrad}_\tree (5_{\gamma}^{-}, 6_g^{+}, 7_g^{-} ) &=& \frac{-2[14]^2⟨34⟩⟨7|2+5|6]^3}{[25]⟨7|2+6|5][1|3+4|2+5|6]s_{134}s_{256}}+\nonumber\\
                                                       &&\frac{2[14]^2⟨27⟩^3⟨34⟩⟨7|2+6|3+4|5⟩}{⟨26⟩⟨67⟩⟨2|6+7|1]⟨5|3+4|5]⟨7|2+6|5]s_{267}}+\nonumber\\
                                                       &&\frac{2[16]^3⟨2|3+5|4](⟨34⟩⟨5|2+3|4]-⟨35⟩s_{167})}{[17][25][67]⟨2|6+7|1]⟨5|3+4|5]s_{167}}+\nonumber\\
                                                       &&\frac{2[14][16]^3⟨25⟩⟨3|2+5|6]}{[17][25][67]⟨2|6+7|1][1|3+4|2+5|6]}\phantom{+}\\
  A^{d\Wrad}_\tree (5_{\gamma}^{-}, 6_g^{+}, 7_g^{-} ) &=& \frac{[16]^3⟨2|3+5|4](2[14]⟨25⟩⟨34⟩-2⟨35⟩⟨2|6+7|1])}{[15][17][67]⟨2|6+7|1]⟨5|3+4|5]s_{167}}+\nonumber\\
                                                       &&\frac{-2[14]^2⟨27⟩^3⟨34⟩⟨5|3+4|1]}{[15]⟨26⟩⟨67⟩⟨2|6+7|1]⟨5|3+4|5]s_{267}}+\nonumber\\
                                                       &&\frac{2[16]^3⟨23⟩⟨5|2+3|4]}{[15][17][67]s_{167}s_{234}}\phantom{+}\\
  A^{e\Wrad}_\tree (5_{\gamma}^{-}, 6_g^{+}, 7_g^{-} ) &=& \frac{-2[16]^3⟨45⟩⟨2|3+5|4]^2}{[17][35][67]⟨2|6+7|1]⟨5|3+4|5]s_{167}}+\nonumber\\
                                                       && \frac{2[14]^2⟨27⟩^3⟨45⟩s_{345}}{⟨26⟩[35]⟨67⟩⟨2|6+7|1]⟨5|3+4|5]s_{267}}\phantom{+}
\end{eqnarray}

\subsubsection{Tree $5_{\gamma}^+, 6_g^+, 7_g^-$}
\begin{eqnarray}
  A^{e\Wrad}_\tree (5_{\gamma}^{+}, 6_g^{+}, 7_g^{-} ) &=&\frac{-2[16]^3⟨23⟩^2[45]s_{345}}{[17]⟨35⟩[67]⟨2|6+7|1]⟨5|3+4|5]s_{167}}+\nonumber\\
                                                       && \frac{2⟨27⟩^3[45]⟨3|4+5|1]^2}{⟨26⟩⟨35⟩⟨67⟩⟨2|6+7|1]⟨5|3+4|5]s_{267}}\phantom{+}
\end{eqnarray}

\subsubsection{Tree $5_{\gamma}^-, 6_g^+, 7_g^+$}
\begin{eqnarray}
  A^{u\Wrad}_\tree (5_{\gamma}^{-}, 6_g^{+}, 7_g^{+} ) &=& \frac{2⟨23⟩^2[34]⟨5|3+4|2+6|7⟩}{⟨17⟩⟨26⟩⟨67⟩⟨5|3+4|5]⟨7|2+6|5]}+\nonumber\\
                                                       &&\frac{⟨25⟩⟨35⟩(-4⟨23⟩[34]+2⟨25⟩[45])}{⟨17⟩⟨26⟩⟨67⟩⟨5|3+4|5]}+\nonumber\\
                                                       &&\frac{2[14]⟨25⟩^2⟨3|1+4|5]}{⟨26⟩⟨67⟩⟨7|2+6|5]s_{134}}+\nonumber\\
                                                       &&\frac{2[34]⟨3|2+5|6]^2}{⟨17⟩[25]⟨7|2+6|5]s_{256}}+\nonumber\\
                                                       &&\frac{-2⟨25⟩⟨37⟩⟨2|3+5|4]-2⟨23⟩⟨27⟩[34]⟨35⟩}{⟨17⟩⟨26⟩⟨67⟩⟨7|2+6|5]}\phantom{+} \\
  A^{d\Wrad}_\tree (5_{\gamma}^{-}, 6_g^{+}, 7_g^{+} ) &=& \frac{-2⟨23⟩^2[34]⟨5|3+4|1+7|6⟩}{⟨17⟩⟨26⟩⟨67⟩⟨5|3+4|5]⟨6|1+7|5]}+\nonumber\\
                                                       &&\frac{⟨25⟩⟨35⟩(4⟨23⟩[34]-2⟨25⟩[45])}{⟨17⟩⟨26⟩⟨67⟩⟨5|3+4|5]}+\nonumber\\
                                                       &&\frac{2[17]^2⟨23⟩^2[34]}{[15]⟨26⟩⟨6|1+7|5]s_{157}}+\nonumber\\
                                                       &&\frac{-2⟨23⟩⟨5|2+3|4]⟨5|1+6+7|5]}{⟨17⟩⟨67⟩⟨6|1+7|5]s_{234}}+\nonumber\\
                                                       &&\frac{2⟨23⟩⟨56⟩⟨2|3+5|4]}{⟨17⟩⟨26⟩⟨67⟩⟨6|1+7|5]}\phantom{+} \\
  A^{e\Wrad}_\tree (5_{\gamma}^{-}, 6_g^{+}, 7_g^{+} ) &=& \frac{2⟨45⟩⟨2|3+5|4]^2}{⟨17⟩⟨26⟩[35]⟨67⟩⟨5|3+4|5]}
\end{eqnarray}

\subsubsection{Tree $5_{\gamma}^+, 6_g^-, 7_g^-$}
\begin{eqnarray}
  A^{e\Wrad}_\tree (5_{\gamma}^{+}, 6_g^{-}, 7_g^{-} ) &=& \frac{-2[45]⟨3|4+5|1]^2}{[17][26]⟨35⟩[67]⟨5|3+4|5]}
\end{eqnarray}

\subsection{Quark radiation}
\label{Sevenpointtrees:quarks}

The amplitudes presented in this section are for the real radiation of a quark-anti-quark pair (legs 2-7). They are labelled by $h_5$, the helicity of the photon, and $h_7$, the helicity of the radiated anti-quark, with the remaining helicities being fixed by the choice of electric charge of the $W$ boson and helicity conservation along quark lines
\begin{equation}
  A^{(0)} (5^{h_5}_{\gamma},7^{h_7}_{\bar{q}}) = A^{(0)} ( 1^+_{\bar{u}},2^{-h_7}_q,3^-_{\nu},4^+_{\lrad},5^{h_5}_{\gamma},6^-_{d},7^{h_7}_{\bar{q}}) \, .
\end{equation}
The relation to Eq.~\ref{fourquark} is a simple swap of legs 2 and 6.

Let us decompose the amplitude as
\begin{equation}
  A^{(0)} (5^{h_5}_{\gamma},7^{h_7}_{\bar{q}}) = A(5^{h_5}_{\gamma},7^{h_7}_{\bar{q}}) + B(5^{h_5}_{\gamma},7^{h_7}_{\bar{q}})
\end{equation}
where the $A$ represents $W\gamma$ radiation from the same quark line (1-6), and $B$ from different quark lines ($W$ from 1-6 and $\gamma$ from 2-7).

The latter case is easier and is simply given by
\begin{gather}
  B (5^{-}_{\gamma},7^{-}_{\bar{q}}) = \frac{2\,P(s_{34})}{[25][57]s_{257}} \Bigg[  \frac{⟨36⟩[12][2|5+7|3+6|4]}{s_{346}} - \frac{[14]⟨3|1+4|2]⟨6|5+7|2]}{s_{134}} \Bigg] \, ,  \\
  B (5^{+}_{\gamma},7^{-}_{\bar{q}}) = \frac{-2\,P(s_{34})}{⟨25⟩⟨57⟩s_{257}} \Bigg[ \frac{⟨67⟩[14]⟨3|1+4|2+5|7⟩}{s_{134}} + \frac{⟨36⟩⟨7|2+5|1]⟨7|3+6|4]}{s_{346}} \Bigg] \, ,  
\end{gather}
with the two remaining helicity configurations given by the following relation
\begin{equation}
  B ( 1^+_{\bar{u}},2^{h_7}_q,3^-_{\nu},4^+_{\lrad},5^{-h_5}_{\gamma},6^-_{d},7^{-h_7}_{\bar{q}}) = B ( 6^+_{\bar{u}},2^{-h_7}_q,4^-_{\nu},3^+_{\lrad},5^{h_5}_{\gamma},1^-_{d},7^{h_7}_{\bar{q}}) \, \bigl\{ \langle \rangle \leftrightarrow [] \bigr\},
\end{equation}
where for the sake of clarity we have reintroduced the suppressed indices.

Radiation from the same quark line is slightly more complicated, and resembles the gluon radiation case. As before, we eliminate double propagators $P(s_{34})P(s_{345})$ with the partial fraction relation of Eq.~\ref{eq:Wpropagatorspartialfraction}, to obtain
\begin{eqnarray}
  A (5^{h_5}_{\gamma},7^{h_7}_{\bar{q}})  &=&
                                              Q_{u} \, P(s_{34}) \, A^{u\Wrad}_\tree (5^{h_5}_{\gamma},7^{h_7}_{\bar{q}})   + \nonumber \\
                                          && Q_{d} \, P(s_{34}) \, A^{d\Wrad}_\tree  (5^{h_5}_{\gamma},7^{h_7}_{\bar{q}})   +  \nonumber \\
                                          && (Q_u-Q_d) P(s_{345}) \, A^{\lrad\Wrad}_\tree  (5^{h_5}_{\gamma},7^{h_7}_{\bar{q}}) . 
\end{eqnarray}
An analogous relation to the one for the gluon case holds
\begin{gather}
  A^{u\Wrad}_\tree  ( 1^+_{\bar{u}},2^{h_7}_q,3^-_{\nu},4^+_{\lrad},5^{-h_5}_{\gamma},6^-_{d},7^{-h_7}_{\bar{q}}) =  \nonumber \\
  - A^{d\Wrad}_\tree  ( 6^+_{\bar{u}},2^{-h_7}_q,4^-_{\nu},3^+_{\lrad},5^{h_5}_{\gamma},1^-_{d},7^{h_7}_{\bar{q}}) \, \bigl\{ \langle \rangle \leftrightarrow [] \bigr\}.
\end{gather}
A minimal complete set of expressions follows.

\subsubsection{Tree $5_{\gamma}^-, 7_{\bar{q}}^-$}

\begin{eqnarray}
  A^{u\Wrad}_\tree (5_{\gamma}^{-}, 7_{\bar{q}}^- ) &=& \frac{-2⟨34⟩⟨5|3+4|6]⟨7|1+2|4]^2}{⟨27⟩[56]⟨1|2+7|6]⟨5|3+4|5]s_{127}}+\nonumber \\
                                                    && \frac{-2[2|6+7|3+5|4](⟨3|1+4|2]⟨5|3+4|6]+[12]⟨15⟩⟨35⟩[56])}{[27][56]⟨1|2+7|6]⟨5|3+4|5]s_{267}}+\nonumber \\
                                                    && \frac{-2[14]⟨3|1+4|2]⟨5|6+7|2]}{[27][56]s_{134}s_{267}}\phantom{+} \\
  A^{d\Wrad}_\tree (5_{\gamma}^{-}, 7_{\bar{q}}^- ) &=&  \frac{[12]^2⟨6|3+5|4](-2⟨36⟩⟨5|3+4|1]-2[15]⟨35⟩⟨56⟩)}{[15][27]⟨5|3+4|5]⟨6|2+7|1]s_{127}}+\nonumber \\
                                                    && \frac{-2[14]^2⟨34⟩⟨67⟩^2⟨5|3+4|1]}{[15]⟨27⟩⟨5|3+4|5]⟨6|2+7|1]s_{267}}+\nonumber \\
                                                    && \frac{2[12]^2⟨36⟩⟨5|3+6|4]}{[15][27]s_{127}s_{346}}\phantom{+} \\
  A^{e\Wrad}_\tree (5_{\gamma}^{-}, 7_{\bar{q}}^- ) &=&  \frac{2[12]^2⟨45⟩⟨6|3+5|4]^2}{[27][35]⟨5|3+4|5]⟨6|2+7|1]s_{127}}+ \nonumber \\
                                                    && \frac{2[14]^2⟨45⟩⟨67⟩^2s_{345}}{⟨27⟩[35]⟨5|3+4|5]⟨6|2+7|1]s_{267}}
\end{eqnarray}

\subsubsection{Tree $5_{\gamma}^+, 7_{\bar{q}}^+$}

\begin{eqnarray}
  A^{e\Wrad}_\tree (5_{\gamma}^{+}, 7_{\bar{q}}^+ ) &=& \frac{-2[17]^2⟨36⟩^2[45]s_{345}}{[27]⟨35⟩⟨5|3+4|5]⟨6|2+7|1]s_{127}}+\nonumber\\
                                                    && \frac{-2⟨26⟩^2[45]⟨3|4+5|1]^2}{⟨27⟩⟨35⟩⟨5|3+4|5]⟨6|2+7|1]s_{267}}\phantom{+}
\end{eqnarray}

\subsubsection{Tree $5_{\gamma}^+, 7_{\bar{q}}^-$}

\begin{eqnarray}
  A^{u\Wrad}_\tree (5_{\gamma}^{+}, 7_{\bar{q}}^- ) &=&  \frac{-2[12]^2[34]⟨36⟩^2⟨6|3+4|5]}{[27]⟨56⟩⟨5|3+4|5]⟨6|2+7|1]s_{127}}+\nonumber\\
                                                    && \frac{⟨67⟩^2⟨3|4+5|1](-2[14]⟨6|3+4|5]+2[15][45]⟨56⟩)}{⟨27⟩⟨56⟩⟨5|3+4|5]⟨6|2+7|1]s_{267}}+\nonumber\\
                                                    && \frac{2[14]⟨67⟩^2⟨3|1+4|5]}{⟨27⟩⟨56⟩s_{134}s_{267}}\phantom{+} \\
  A^{d\Wrad}_\tree (5_{\gamma}^{+}, 7_{\bar{q}}^- ) &=& \frac{2[34][56]⟨3|4+5|1+2|7⟩^2}{⟨27⟩⟨1|2+7|6]⟨5|3+4|5]⟨5|3+4|6]s_{127}}+\nonumber\\
                                                    && \frac{-2[34]⟨1|3+4|5]⟨3|6+7|2]^2}{⟨15⟩[27]⟨1|2+7|6]⟨5|3+4|5]s_{267}}+\nonumber\\
                                                    && \frac{-2⟨34⟩⟨7|3+6|4]^2}{⟨15⟩⟨27⟩⟨5|3+4|6]s_{346}}\phantom{+} \\
  A^{e\Wrad}_\tree (5_{\gamma}^{+}, 7_{\bar{q}}^- ) &=&  \frac{2[12]^2⟨36⟩^2[45]s_{345}}{[27]⟨35⟩⟨5|3+4|5]⟨6|2+7|1]s_{127}}+\nonumber\\
                                                    && \frac{2[45]⟨67⟩^2⟨3|4+5|1]^2}{⟨27⟩⟨35⟩⟨5|3+4|5]⟨6|2+7|1]s_{267}}\phantom{+}
\end{eqnarray}

\subsubsection{Tree $5_{\gamma}^-, 7_{\bar{q}}^+$}

\begin{eqnarray}
  A^{e\Wrad}_\tree (5_{\gamma}^{-}, 7_{\bar{q}}^+ ) &=& \frac{-2[17]^2⟨45⟩⟨6|3+5|4]^2}{[27][35]⟨5|3+4|5]⟨6|2+7|1]s_{127}}+\nonumber\\
                                                    && \frac{-2[14]^2⟨26⟩^2⟨45⟩s_{345}}{⟨27⟩[35]⟨5|3+4|5]⟨6|2+7|1]s_{267}}\phantom{+}
\end{eqnarray}

\bibliography{main}
\bibliographystyle{JHEP}

\end{document}